\documentclass[11pt,twoside]{book} 
\usepackage{asp2008s}
\usepackage{times}
\usepackage{lscape}
\usepackage{epsf}

\usepackage{graphicx}
\usepackage{amssymb}
\usepackage{longtable}
\usepackage[figuresright]{rotating}
\usepackage{multirow}

\newcommand{\kms}{km~s{$^{-1}$}}

\newcommand{\msun}{M{$_{\odot}$}}

\newcommand{\simless}{\mathbin{\lower 3pt\hbox {$\rlap{\raise 5pt\hbox{$\char'074$}}\mathchar"7218$}}}

\mainmatter

\begin{document}

\title{The Nearest OB Association: Scorpius-Centaurus~(Sco\,OB2)}
\author{Thomas Preibisch}
 \affil{Max-Planck-Institut f\"ur Radioastronomie, Auf dem
 H\"ugel 69, D--53121 Bonn, Germany\\
  Universit\"ats-Sternwarte M\"unchen,
       Scheinerstr.~1, D-81679 M\"unchen, Germany}
\author{Eric Mamajek}
 \affil{Harvard-Smithsonian Center
 for Astrophysics, 60 Garden Street, MS-42, Cambridge, MA 02138, USA}
\begin{abstract}
We summarize observational results on the stellar population and star
formation history of the Scorpius-Centaurus OB Association (Sco OB2),
the nearest region of recent massive star formation.  It consists of
three subgroups, Upper Scorpius (US), Upper Centaurus-Lupus (UCL), and
Lower Centaurus-Crux (LCC) which have ages of about 5, 17, and $
16$~Myr.  While the high- and intermediate mass association members
have been studied for several decades, the low-mass population
remained mainly unexplored until rather recently.

In Upper Scorpius, numerous studies, in particular large multi-object
spectroscopic surveys, have recently revealed hundreds of low-mass
association members, including dozens of brown dwarfs.  The
investigation of a large representative sample of association members
provided detailed information about the stellar population and the
star formation history.  The empirical mass function could be
established over the full stellar mass range from $0.1\,M_\odot$ up to
$20\,M_\odot$, and was found to be consistent with recent
determinations of the field initial mass function.  A narrow range of
ages around 5~Myr was found for the low-mass stars, the same age as
had previously (and independently) been derived for the high-mass
members.  This supports earlier indications that the star formation
process in US was triggered, and agrees with previous conjectures that
the triggering event was a supernova- and wind-driven shock-wave
originating from the nearby UCL group.

In the older UCL and LCC regions, large numbers of low-mass members
have recently been identified among X-ray and proper-motion selected
candidates.  In both subgroups, low-mass members have also been
serendipitously discovered through investigations of X-ray sources in
the vicinity of better known regions (primarily the Lupus and TW Hya
associations). While both subgroups appear to have mean ages of
$\sim$16\,Myr, they both show signs of having substructure. Their
star-formation histories may be more complex than that
of the younger, more compact US group.

Sco-Cen is an important ``astrophysics laboratory'' for detailed
studies of recently formed stars. For example, the ages of the
sub-groups of 5~Myr and $\sim 16$~Myr are ideal for studying how
circumstellar disks evolve.  While no more than a few percent of the
Sco-Cen members appear to be accreting from a circumstellar disk,
recent {\it Spitzer} results suggest that at least $\sim$35\% still
have cold, dusty, debris disks.

\end{abstract}

\section{Star Formation in OB Associations \label{intro}}

OB associations were first recognized by \citet{Blaauw46} and
\citet{Ambartsumian47} as extended moving groups of blue luminous
stars.  They are defined as loose stellar systems (stellar mass
density of $\la 0.1\,M_\odot\,{\rm pc}^{-3}$) containing O- and/or
early B-type stars \citep[for a recent review see][]{Briceno06}. At
such low densities, the associations are unstable against Galactic
tidal forces, and therefore it follows from their definition that OB
associations must be young ($\la 30-50$~Myr) entities.  Most of their
low-mass members are therefore still in their pre-main sequence (PMS)
phase.

There are different models for the origin of OB associations. One
possibility is that they start as initially dense clusters, which get
unbound and expand quickly as soon as the massive stars expel the gas
\citep[see][]{Kroupa01}.  An alternative model assumes that OB
associations originate from {\em unbound} turbulent giant molecular
clouds \citep[see][]{Clark05}, i.e.~start already in a spatially
extended configuration and thus form in a fundamentally different way
than dense, gravitationally bound clusters.
Many well investigated OB associations show remarkably small internal
velocity dispersions (often $\leq 1.5$~km/sec), which are in some
cases much smaller than required to explain the large present-day size
(typically tens of parsecs) by expansion over the age of the
association. This excludes the expanding cluster model and provides
strong support for an origin as an extended unbound cloud for these
associations.

The considerable number of OB associations in the solar neighborhood
\citep[e.g.,][]{deZeeuw99} suggests that they account for a large,
maybe the dominant, fraction of the total Galactic star formation.  A
good knowledge of their stellar content is thus essential in order to
understand the nature of the star formation process not only in OB
associations but also on Galactic scales.

For many years star formation was supposed to be a bimodal process
\citep[e.g.][]{Larson86,Shu88} according to which high- and low-mass
stars should form in totally different sites.  Although it has been
long established that low-mass stars {\em can} form alongside their
high-mass siblings in nearby OB associations
\citep[e.g.~][]{Herbig62}, it is still not well known {\em what
quantities} of low-mass stars are produced in OB environments.  There
have been many claims that high-mass star forming regions have a
truncated initial mass function (IMF), i.e.~contain much smaller
numbers of low-mass stars than expected from the field IMF \citep[see,
e.g.,][]{Slawson92,Leitherer98,Smith01,Stolte05}.  Possible
explanations for such an effect are often based on the strong
radiation and winds from the massive stars.  For example, increased
radiative heating of molecular clouds may raise the Jeans mass;
lower-mass cloud cores may be completely dispersed by photoevaporation
before low-mass protostars can even begin to form; the radiative
destruction of CO molecules should lead to a change in the equation of
state of the cloud material, affecting the fragmentation processes and
ultimately leading to the formation of a few massive stars rather than
the ``normal'' IMF which is dominated by low-mass stars
\citep[e.g.][]{Li03}.  However, several well investigated massive star
forming regions show {\em no} evidence for an IMF cutoff \citep[see,
e.g.,][for the cases of NGC~3603, 30~Dor, and NGC~346,
respectively]{Brandl1999,Brandner01,Sabbi08}, and notorious difficulties in
IMF determinations of distant regions may easily lead to wrong
conclusions about IMF variations \citep[see, e.g., discussion
in][]{Selman05, Zinnecker93}.

If the IMF in OB associations is not truncated and similar to the
field IMF, it would follow that most $(\ga 60\%)$ of the total stellar
mass is found in low-mass $(< 2\,M_\odot)$ stars.  This would then
imply that most of the current Galactic star formation is taking place
in OB associations \citep[as initially suggested by][]{Miller78}, and
the {\em typical} environment for forming stars (and planets) would be
close to massive stars and not in isolated regions like Taurus.  The
presence of nearby massive stars affects the evolution of young
stellar objects and their protoplanetary disks in OB environments.
For example, photoevaporation by intense UV radiation can remove a
considerable amount of circumstellar material around young stellar
objects \citep[e.g.,][]{Bally98,Richling98}, and these objects will
therefore ultimately end up with smaller final masses than if they
were located in isolated regions \citep{Whitworth04}.  Although
generally considered to be a threat for forming planetary systems,
photoevaporation may actually help to form planets, as it seems to
play an important role in the formation of planetesimals
\citep{Throop05}.

OB associations provide excellent targets to investigate these effects
on the formation and evolution of low-mass stars (and their forming
planetary systems) during ages between a few Myr and a few ten
Myr. However, before one can study the low-mass members, one first has
to find them.  Although this statement sounds trivial, the major
obstacle on the way towards a reliable knowledge of the low-mass
population in OB associations is the problem to identify the
individual low-mass members.  Unlike stellar clusters, which can be
easily recognized on the sky, OB associations are generally very
inconspicuous: since they extend over huge areas in the sky (often
several hundred square-degrees for the nearest examples), most stars
in the area actually are unrelated foreground or background stars.
Finding the association members among these field stars is often like
finding needles in a haystack.  As the low-mass members are often too
faint for proper-motion studies, the only reliable sign to discern
between low-mass association members and unrelated, much older field
stars is the strength of the 6708~\AA\, lithium line in the stellar
spectrum: at ages of $\leq 30$~Myr, the low-mass association members
still have most of their initial Li preserved and show a strong Li
line, whereas the older foreground and background field stars do not
show this line since they have already depleted their primordial Li
\citep[e.g.,][]{D'Antona94}.  However, an accurate measurement of Li
line width requires at least intermediate resolution spectroscopy, and
thus the observational effort to identify the widespread population of
PMS stars among the many thousands of field stars is huge.  Many
empirical IMF determinations are therefore based on photometric data
only; while this strategy rather easily provides a complete spatial
coverage and allows one to work with large samples, photometry alone
cannot give completely reliable membership information.  Most studies
with spectroscopically identified member samples, on the other hand,
include only very small fractions of the total stellar population and
are strongly affected by small number statistics and the necessity of
using large extrapolation factors.  Therefore, most studies dealing
with OB associations have been restricted to estimating the
number-ratio of low-mass versus high-mass members
\citep[e.g.][]{Walter94,Dolan02,Sherry04}.

During the last years, new and very powerful multiple-object
spectrographs like 2dF at the Anglo-Australian Telescope
\citep[see][]{Lewis02} made large spectroscopic surveys for low-mass
PMS members feasible.  In combination with the Hipparcos results,
which allowed the complete identification of the high- and
intermediate-mass $(\ge 2\,M_\odot)$ stellar population in many nearby
associations \citep[see][]{deZeeuw99}, studies of the {\em complete}
stellar population in OB associations are now possible, enabling us to
investigate in detail the spatial and temporal relationships between
high- and low-mass members.

In this chapter, we review the state of knowledge regarding the
stellar populations of the nearest OB association: Scorpius-Centaurus
(Sco OB2).  We structure this chapter in the following manner.
Section~2 gives a general, and historical, overview of the association
and its subgroups. Sections 3, 4, and 5 discuss the surveys for the
members of the three primary subgroups of Sco-Cen: Upper Scorpius,
Upper Centaurus-Lupus, and Lower Centaurus-Crux, respectively. Readers
uninterested in the individual surveys may want to skip ahead to
Section~6, where we discuss the astrophysical ramifications, and
interpretations, of studies of the Sco-Cen members.

\section{The Scorpius-Centaurus OB Association \label{sco-cen}}

\subsection{Morphology and Nomenclature of the Subgroups \label{morph_nomen}}

\begin{figure}[!h]
\plotfiddle{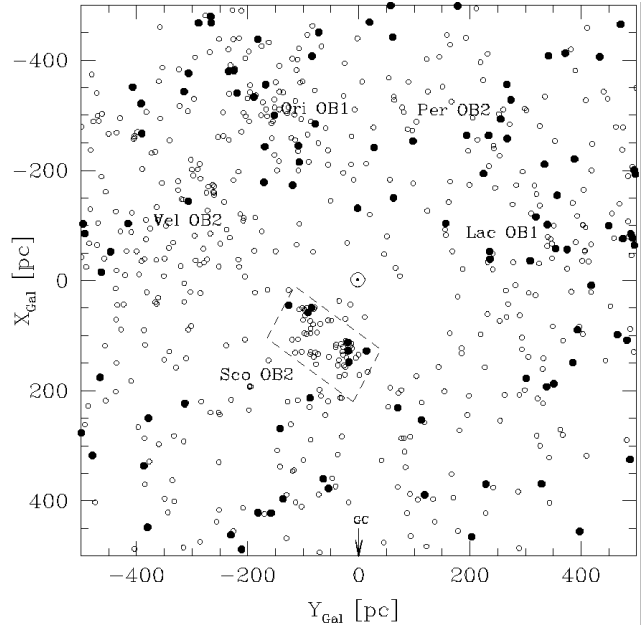}{10.9cm}{0}{60}{60}{-190}{-100}
\caption{ View of the local O- and early B-type stars, as looking down
upon the Galactic disk. The direction of the Galactic Center (GC) is
towards bottom, and the Sun is at the origin.  Stars with spectral
types B0 or hotter are {\it solid circles}, B1-B2 stars are {\it open
circles}.  Sco-Cen (Sco OB2) is the concentration (dashed box) of
early-B stars closest to the Sun.  Note that stars near the edge of
the plots typically have individual distance errors of $\sim$50\%, and
hence associations appear to be very stretched. Galactic positions
were calculated using {\it Hipparcos} \citep{Perryman97} celestial
coordinates and parallaxes.
\label{nearby_OB.fig}}
\end{figure}

\begin{figure}[!h]
\plotone{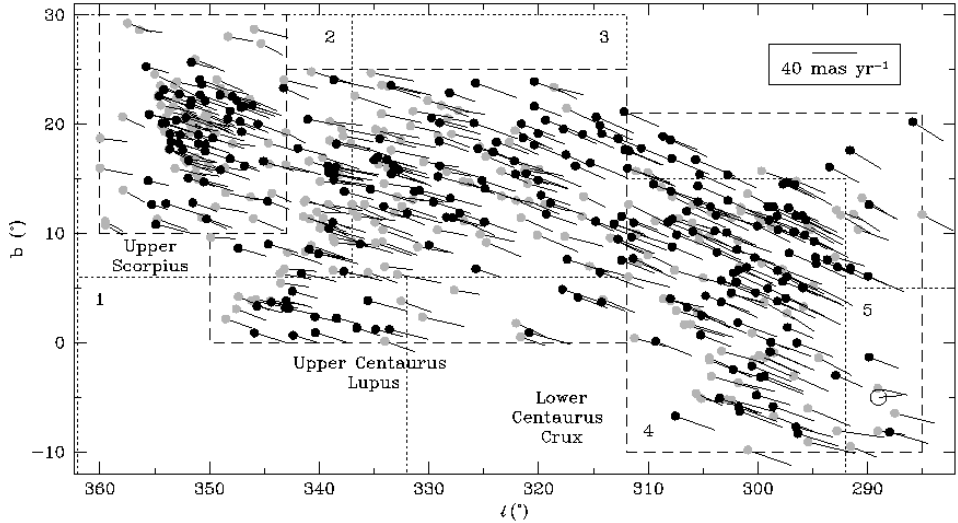}
\caption{Map of the Sco-Cen region, showing the positions and proper
motions of the Hipparcos members (adapted from \citet{deZeeuw99}
 and kindly provided
by Tim de Zeeuw).
\label{scocen_map.fig}}
\end{figure}

The Scorpius-Centaurus (Sco-Cen) association is the OB association
nearest to the Sun (see Fig.~\ref{nearby_OB.fig}).  It contains at
least $\sim$150 B stars which concentrate in the three subgroups Upper
Scorpius (US\footnote{As a historical note, in pre-1960s literature,
one also finds that Upper Scorpius was called \mbox{\it II Sco}
\citep{Morgan53}, {\it Collinder 302}, or the {\it Antares Moving
Group} \citep{Collinder31}. All three names have fallen into disuse.}),
Upper Centaurus-Lupus (UCL), and Lower Centaurus-Crux \citep[LCC;
Fig.~\ref{scocen_map.fig}; cf.][]{Blaauw64,Blaauw91,deZeeuw99}.  There
is one O-type star associated with Sco-Cen: the runaway O9V star
$\zeta$ Oph. The currently recognized Sco OB2 subgroups were defined
by \citet{Blaauw46}: Subgroup \#2 is US, \#3 is UCL, and \#4 is LCC.
\Citet{deGeus89} derived ages for the B-type stars in the different
subgroups from the main sequence turnoff in the HR diagram and found
that Upper Scorpius is the youngest subgroup ($\sim 5-6$~Myr), whereas
Lower Centaurus Crux ($\sim 11-12$~Myr) and Upper Centaurus-Lupus
($\sim 14-15$~Myr) are considerably older\footnote{Note that
\citet{Mamajek02} revised these ages to $\sim 16$~Myr for Lower
Centaurus-Crux and $\sim 17$~Myr for Upper Centaurus-Lupus.}.

\subsection{A Century of Sco-Cen Research \label{history}}

Although the focus of our review is the low-mass membership of
Sco-Cen, it is worth reviewing the history regarding the unveiling of
the group's membership\footnote{Much of the early research on Sco-Cen
is not yet electronically available through ADS
(http://adsabs.harvard.edu/).  The most valuable list of pre-1973
references on Sco-Cen is in the ``Alter'' files accessible through
Vizier (http://vizier.u-strasbg.fr/, catalogs VII/31B and VII/101A),
which complement the ``Catalogue of Star Clusters and Associations''
\citep{Ruprecht82,Ruprecht83}. Relevant Sco-Cen references are given
under ``Sco OB2'' and ``Collinder 302''}.  For most of the past
century ``Sco-Cen'' was studied as a single entity, and only the
highest mass members (earlier than B5-type) were called ``members'' with
any degree of confidence. The discovery of the low-mass population
(FGKM stars) in large numbers has only recently become technically
feasible. Many early studies of Sco-Cen involved isolating the members
from the field population through kinematic and photometric
means. Sco-Cen contained a large sample of B-type stars with large,
convergent proper motions, which provided a critical lower rung in the
Galactic (and cosmic) distance ladder \citep[e.g.][]{Morgan53}. Our
brief historical review here is not exhaustive, but complements the
recent summary by \citet[][their Sect.~4.1]{deZeeuw99}.

Around 1910, large compilations of proper motion and spectral type
data were becoming available for Galactic structure investigations --
primarily Lewis Boss's (1910) {\it Preliminary General Catalog} and
the spectral type compilations from the Henry Draper Memorial project
\citep[e.g.][]{Cannon01}. \citet{Kapteyn14} appears to have been the
first to make a convincing case that the B-type ``helium'' stars in
the Scorpius-Centaurus region demonstrate convergent proper motions,
and constituted a moving group.  Subsequent work on the association
during the early 20th century revolved around testing the reality of
the group, ascertaining its membership, and (most importantly for the
rest of the astronomical community) estimating the group's
distance. The importance of understanding, and exploiting, this
conspicuous group of bright B-type stars was not lost on
\citet{Kapteyn14}: ``{\it The real question of importance is this: is
the parallelism and equality of motion in this part of the sky
[Sco-Cen] of such a nature that we can derive individual
parallaxes?}''

The reality of Sco-Cen as a moving group was challenged, most notably,
by \citet{Smart36,Smart39}, and \citet{Petrie62}.  Some primary
objections raised by these authors were that (1) the convergent point
of the Sco-Cen proper motions was so close to the solar antapex that
it could be construed simply as solar reflex motion on unrelated
B-type stars in the field, (2) the group was so dispersed that it
should disintegrate on a short time scale, and (3) the Sco-Cen space
motion varied as one isolated different subsamples within the group,
enough so to cast doubt on the derived cluster parallaxes for the
constituent stars.  Many of the kinematic objections by Smart and
others were addressed by Adriaan \citet{Blaauw46} in his PhD thesis,
which concluded that Sco-Cen was indeed a true moving group.  Perhaps
Blaauw's (1946) most amusing retort was in telling the Regius Chair of
Astronomy at Glasgow to simply look up (p. 19): ``{\it Smart has not
paid attention to the fact that the existence of the cluster is
evident from the apparent distribution of the bright stars in the
sky.}''

In hindsight, the first objection is understandable as the young stars
in the solar neighborhood form from molecular gas which itself has
small peculiar motions with respect to the LSR \citep[][]{Stark89}.
The last two objections are symptomatic of OB associations in general,
when compared to the more coherent kinematic groups like the Hyades
cluster. As one of the few nearby OB associations with appreciable
proper motion, the kinematic studies of Blaauw, and others, were
critical to understanding the dynamical state of OB associations in
general.  The modern consensus \citep[e.g.][]{Blaauw64,deZeeuw99} is
that Sco-Cen constitutes a moving group, but that it has subgroups
isolated by position, age, and space motion, and that these structures
are young and unbound. Investigations of the expansion of the Sco-Cen
subgroups have been undertaken by \citet{Bertiau58}, \citet{Jones71},
and \citet{Madsen02}.  A modern study using the radial velocities of
the low-mass members is sorely needed to confirm, and build upon,
these findings.

The most recent investigation of the membership of high-mass stars in
Scorpius-Centaurus was carried out by \citet{deZeeuw99}.  They used
Hipparcos proper motions and parallaxes in conjunction with two moving
group methods in order to accurately establish the high-mass (and
sometimes intermediate mass) stellar content of 12 nearby OB
associations.  In Sco-Cen the membership of nearly 8000 Hipparcos
Catalogue stars was investigated. A total of 120 stars in US, 221
stars in UCL, and 180 stars in LCC were identified as high probability
members.  The mean distances of the subgroups, derived from the
Hipparcos parallaxes, are 145 pc for US, 140~pc for UCL, and 118~pc
for LCC.

\subsection{The Sco-Cen Complex \label{complex}}

The vicinity of Sco-Cen is rich with well-studied sites of current,
and recently terminated, star-formation. Besides the Ophiuchus and
Lupus star-forming regions, there are other dark clouds, T
associations, and somewhat older ``outlying'' stellar groups which
appear to be genetically related to Sco-Cen by virtue of their ages,
positions, and space motions \citep{Mamajek01}.  Several of these are
described elsewhere in this volume. These neighboring regions
demonstrate a broad evolutionary spectrum of star formation.  They
include dark clouds with little, if any, star-formation activity
(e.g. Musca, Coalsack, Pipe Nebula, Cha III), molecular cloud
complexes that are currently forming stars (e.g. Cha I \& II, CrA),
and several recently discovered groups of $\sim5-12$~Myr-old stars
with little or no trace of the dark clouds from which they formed
(e.g. the TW Hya, $\beta$ Pic, $\eta$ Cha, $\epsilon$ Cha groups).  In
an investigation of the origins of the $\eta$ Cha cluster,
\citet{Mamajek00} and \citet{Mamajek01} found that many nearby, young
groups, and isolated young stars, in the southern hemisphere within
$\sim$200\,pc (e.g. $\eta$ Cha, TW Hya, $\beta$ Pic, $\epsilon$ Cha,
CrA, etc.) are not only spatially close to the Sco-Cen OB association,
but {\it moving away from the subgroups}. The star-forming clouds
in Oph, CrA, and Cha I manifest head-tail morphologies, with the
star-forming ``heads'' on the side facing Sco OB2. More detailed
investigations of the space motions and star-formation histories of
the associations in this region are needed, however the preliminary
results suggest that star-formation in these small ``satellite''
groups near Sco-Cen may have been triggered by the massive
star-formation event in the primary Sco-Cen subgroups
\cite[see, e.g.~the comprehensive model scenario for the formation of the
Sco-Cen association and the
young stellar groups proposed by][]{Fernandez08}.

The Sco-Cen region, including the OB subgroups, molecular clouds, and
outlying associations, may be thought of as a small star-forming {\it
complex} \citep{Elmegreen00}.  The {\it Sco-Cen complex} has been
variously referred to as the {\it Oph-Sco-Cen association}
\citep[OSCA;][]{Blaauw91}, {\it Greater Sco-Cen} \citep{Mamajek01},
or, perhaps with tongue in cheek, the {\it Oph-Sco-Lup-Cen-Cru-Mus-Cha
star-formation region} \citep{Lepine03}.  Throughout this review, we
will mostly limit our discussions to the 3 subgroup regions outlined
by Fig. 2 (US, UCL, LCC), {\it but excluding the regions associated
with the Ophiuchus and Lupus molecular cloud complexes} (see chapters
by Wilking et al.~and Comer\'on). Both
complexes are within \citet{deZeeuw99} projected boundaries of US and
UCL, respectively, and are approximately co-distant with the subgroups
($d \simeq 140$~pc).

There is no evidence for ongoing star formation activity in the OB
subgroups of Sco-Cen itself.  This makes it an ideal target for an
investigation of the {\em outcome of the recently completed star
formation process}.  The area is essentially free of dense gas and
dust clouds, and the association members show only very moderate
extinctions ($A_V \la 2$~mag). This is probably the consequence of the
massive stellar winds and several supernova explosions, which have
cleared the region from diffuse matter and created a huge system of
loop-like H\,I structures around the association. These loop structures
have a total mass of about $3 \times 10^5\,M_\odot$ and seem to be
the remnants of the original giant molecular cloud in which the
OB subgroups formed
\citep[cf.][]{deGeus92}.

\subsection{Other Proposed Subgroups \label{poorly_studied}}
\smallskip

For historical completeness, we mention some candidate stellar groups
in the vicinity of Sco-Cen which have been proposed, but later
refuted. \Citet{deZeeuw99} was unable to verify the existence of
kinematic groups in Blaauw's (1946) areas \#1 (CrA region) and \#5
(Car-Vel region).  Blaauw's areas \#6 and \#7 correspond to the
modern-day Vel OB2 and Collinder 121 associations, however they were
sufficiently detached from groups \#2-\#4 in position, distance, and
velocity, that \citet{Blaauw46} did not include them in his final
census of Sco-Cen groups.

\citet{Makarov00} claimed to have discovered a moving group of X-ray
bright stars adjacent to LCC in Carina-Vela, in essentially the same
region as Blaauw's subgroup \#5. They claimed that the new group
is a ``near extension of the Sco-Cen complex'' and that the open
cluster IC 2602 was part of this group. The status of Car-Vel as a
coherent group is very unlikely, let alone any relation to Sco-Cen.
\citet{Makarov00} show that the inferred ``kinematic'' parallaxes for
their proposed Car-Vel membership show a disturbing ``finger-of-god''
effect with distances ranging from $\sim 30-500$~pc, with a large gap
in the distribution.  Closer examination of the kinematic and
spectroscopic data for these stars by \citet[][ and in
prep.]{Jensen04} show that the objects appear to constitute a
heterogeneous sample of low-mass members of IC 2602 and the $a$~Car
cluster (= Platais 8), and probable Gould Belt stars with a large
range of distances. \citet{Zuckerman04} claim that a subsample
of the Car-Vel stars at d $\simeq 30$~pc constitute a previously
unknown nearby $\sim200$~Myr-old group, which they dub
``Carina-Near''. Regardless, the Carina stars appear to be unrelated
to Sco-Cen. The consensus from studies of the high mass and low mass
populations appears to be that the western ``edge'' of Sco-Cen lies
near Galactic longitude 290$^{\circ}$.

Recently, \citet{Eggen98} and \citet{Sartori03} suggested that an OB
association spatially contiguous to LCC might exist to the south in
Chamaeleon. While there are a total of 4 known B stars associated with
the Cha I (2), $\epsilon$ Cha (1), and $\eta$ Cha (1) kinematic groups
in Chamaeleon, \citet{Mamajek03} argued that the kinematic and stellar
density data are inconsistent with the idea of an OB association in
Cha, at least of the size proposed by \citet[][21 B-type
stars]{Sartori03}. Their sample appears to be dominated by field stars
completely unrelated to the Chamaeleon molecular clouds or Sco-Cen.

\section{Upper Scorpius (US) \label{us}}

The Upper Scorpius association is the best studied part of the Sco-Cen
complex. Despite its rather young age ($\sim 5$~Myr) and the
neighborhood to the $\rho$ Oph molecular cloud (see chapter by Wilking
et al.~in this book), which is located in front of the southeastern edge of
US and is well known for its strong star formation activity, there are
no indications for ongoing star formation in US.
Below we will make a distinction between the high- and the low-mass
stellar population. The former refers to stars of spectral types F and
earlier ($M \ge 2\,M_\odot$) whose membership of US was established
using Hipparcos data \citep{deZeeuw99}.  The latter refers to G, K,
and M stars in the mass range $0.1\!-\!2.0$~\msun\, for which
membership was established from their PMS character.

\subsection{The High-Mass Stellar Population \label{us_high_mass}}
\label{highmass_sec}

\Citet{deZeeuw99} investigated the membership of 1215 stars in US
listed in the Hipparcos Catalogue; 120 of these were identified as
genuine members. The spectral types of the members on the
(pre-)main-sequence range from B0.5V to G5V, and there are some
evolved stars with giant luminosity classes (including the M1.5
supergiant Antares [$\alpha$~Sco]).  The most massive star in US was
presumably a $\sim 50\,M_\odot$ O5--O6 star, which exploded as a
supernova about 1.5~Myr ago.  \citet{Hoogerwerf01} suggested that the
pulsar PSR~J1932+1059 is the remnant of this supernova and that
the runaway star $\zeta$ Oph was the previous binary companion of the
supernova progenitor and was ejected by the explosion.  However, the
new parallax of PSR~J1932+1059 determined by \citet{Chatterjee04}
challenged this scenario. The new data suggest that the pulsar was
probably {\em not} the former binary companion of $\zeta$ Oph, but it
is still possible that PSR~J1932+1059 was created in US $\sim 1-2$~Myr
ago.

The 120 kinematic members of US cover an area of about 150~deg$^2$
on the sky. The large intrinsic size of the association
suggests that the spread of individual stellar distances cannot be
neglected. The projected diameter on the sky is $\sim 14\deg$ which at
the distance of US corresponds to $\sim 35$~pc.
However, while the Hipparcos data allow the determination of a very
accurate mean distance of $145 \pm 2$~pc for US \citep{deZeeuw99}, the
errors on the trigonometric parallaxes ($\sim 1$ mas) are too large to
resolve the internal spatial structure.  The only conclusion that can
be drawn directly from the Hipparcos parallaxes is that the
line-of-sight depth of US cannot be much larger than $\sim 70$~pc.
This prompted \citet{deBruijne99} to carry out a more detailed
investigation of the Hipparcos members of Scorpius-Centaurus by
performing a careful kinematic modeling of the proper motion and
parallax data. He used a maximum likelihood scheme based on a
generalized moving-cluster method to derive secular (or `kinematically
improved') parallaxes for the association members. He showed that the
method is robust and that the secular parallaxes are a factor of $\sim
2$ more precise than the Hipparcos trigonometric parallaxes. The
secular parallaxes for the members of US show a much reduced
line-of-sight dispersion, confirming that the dominant part of the
scatter in the Hipparcos distances is caused by the trigonometric
parallax errors and not by a large intrinsic dispersion of individual
stellar distances. The distribution of secular parallaxes for US is
not resolved, which means that the distance spread cannot be larger
than $\sim 50$ pc.
Hence one can assume that US has a roughly spherical shape, i.e.~that
the intrinsic spread of distances is about $\pm\,20$ pc from the mean
value of 145 pc.

Previous investigations of the IMF of US focused on the high- to
intermediate mass stellar content.  \Citet{deGeus89} established the
membership of stars in US using Walraven multi-color photometry, and
determined their physical parameters ($\log g$ and $\log
T_\mathrm{eff}$), from which they derived stellar masses.
\citet{Brown98} used the preliminary results on membership from the
Hipparcos data to determine the luminosity function for the high-mass
stars using the Hipparcos parallaxes, and then transformed a smoothed
version of this into a mass function using the mass-luminosity
relation listed in \citet{Miller79}.  Adopting a conservative
completeness limit of $V\approx7$, which corresponds to masses of
about $2.8\,M_\odot$, he concluded that down to this mass limit the
IMF is consistent with a single power-law ${\rm d}N/{\rm d}M \propto
M^{\alpha}$ with slope $\alpha \approx-2.9$.  There are, however,
several problems with this determination of the IMF, most importantly
the fact that a mass-luminosity relation had been used that might not
be appropriate for young stars (see discussion in Brown 1998).

Finally, \citet{Hoogerwerf00} attempted to extend the completeness of
the kinematic studies of the membership of US toward lower masses by
making use of the TRC and ACT astrometric catalogs
\citep{Hoeg98,Kuzmin99,Urban98}, which are believed to be complete to
$V\approx10.5$.  He selected some 250 candidate members with $V\approx
7-11$ from these catalogs by searching for stars with proper motions
consistent with those of the Hipparcos association members, and which
lie $\le 1$ mag below and $\le 1.5$ mag above the main sequence in the
color-magnitude diagram.  However, we note that the second selection
criterion actually excludes most of the late-type association members,
which, at an age of about 5 Myr, lie well above the main sequence.

\subsection{Searches for Low-Mass Members \label{us_low_mass}}

Numerous studies have tried to reveal low-mass stars in US.  Most of
these, however, focused on very small subregions of the
association. For example, \citet{Meyer93} studied IRAS sources in a
$\sim 2$~deg$^2$ field near $\sigma$ Sco and found 4 young
stars. \citet{Sciortino98} used deep pointed ROSAT X-ray observations
to search for PMS stars in a $\sim 4$~deg$^2$  area and found
several candidates for PMS stars. \citet{Martin98b} analyzed pointed
ROSAT observations in the vicinity of the $\rho$~Oph star forming
region and found a number of additional PMS stars in this area.

The first systematic search for low-mass members covering a
significant part of US was performed by \citet{Walter94}, who obtained
spectroscopy and photometry for the optical counterparts of X-ray
sources detected in 7 individual EINSTEIN fields. They classified 28
objects as low-mass PMS stars and placed them into the
HR-diagram. They found a remarkably small dispersion in stellar
ages\footnote{Mart\'in~(1998) suggested that the spread in Li line
widths in these stars is an indication of a large age spread.
However, this interpretation would be very difficult to reconcile with
the locations of the stars in the HR diagram and would require a much
larger spread in the individual stellar distances of the low-mass
stars than the best estimate of the line-of-sight depth of about
$\pm\,20$~pc derived from the Hipparcos data for the massive members.}
and interpreted this as an indication that the formation of these
stars was triggered by some external event.

In another study, M.~Kunkel investigated the optical counterparts of
more than 200 ROSAT All Sky Survey (RASS) X-ray sources in a
$\sim\!60$~deg$^2$ area in US and UCL \citep[see][for a list of
these stars]{Kohler00}.  32 objects in this sample that are located in
US can be classified as new low-mass members \citep[cf.][]{PZ99}.

A deep search for very-low mass PMS star and brown dwarf candidates in
US was presented by \citet{Ardila00}. Their photometric survey covered
an area of 14~deg$^2$ and yielded some 100 candidate members.
Low-resolution spectroscopy for some of these candidates led to the
classification of 20 stars with strong H$\alpha$ emission as potential
association members.  For eleven of these candidates
\citet{Mohanty04a} and \citet{Mohanty04b} performed high-resolution
optical spectroscopy and derived stellar parameters.  They showed that
five of these objects have masses $<0.075$\,\msun, i.e.~are brown
dwarfs.  \citet{Martin04} presented low-resolution optical
spectroscopy of further candidate very low-mass members of US.  Their
analysis indicated that 28 of these objects are most likely members of
US, and 18 objects have spectral types in the range M6.5--M9, i.e.~are
likely young brown dwarfs.

\citet{Argiroffi06} analyzed deep {\em XMM-Newton} X-ray observations
of two fields in US. Among the 224 detected X-ray sources they
identified 22 stars as photometric member candidates, 13 of which were
not known to be association members before.  \citet{Slesnick06}
presented a wide-field optical/near-infrared photometric survey of
US. Follow-up spectroscopy of selected stars led to the identification
of 43 new low-mass members with estimated masses in the $0.02 -
0.2\,M_\odot$ range, 30 of which are likely new brown dwarf members of
US.  Finally, \citet{Lodieu06} identified about a dozen additional new
likely low-mass members of US from an analysis of UKIRT Infrared Deep
Sky Survey Early Data Release data of a 9.3~deg$^2$ field in
US and follow-up observations. In continuation of this work,
\citet{Lodieu07} used UKIDSS Galactic Cluster Survey data of a 6.5~deg$^2$
region in US and identified 129 members by photometric
and proper motion criteria.  The estimated masses of these objects are
in the range between $0.3$ and $0.007\,M_\odot$ and they conclude that
the sample contains a dozen new brown dwarf candidates below 15
Jupiter masses.

\subsection{Multi-Object Spectroscopic Surveys for Low-Mass Members \label{mos}}

\begin{figure}
\plotone{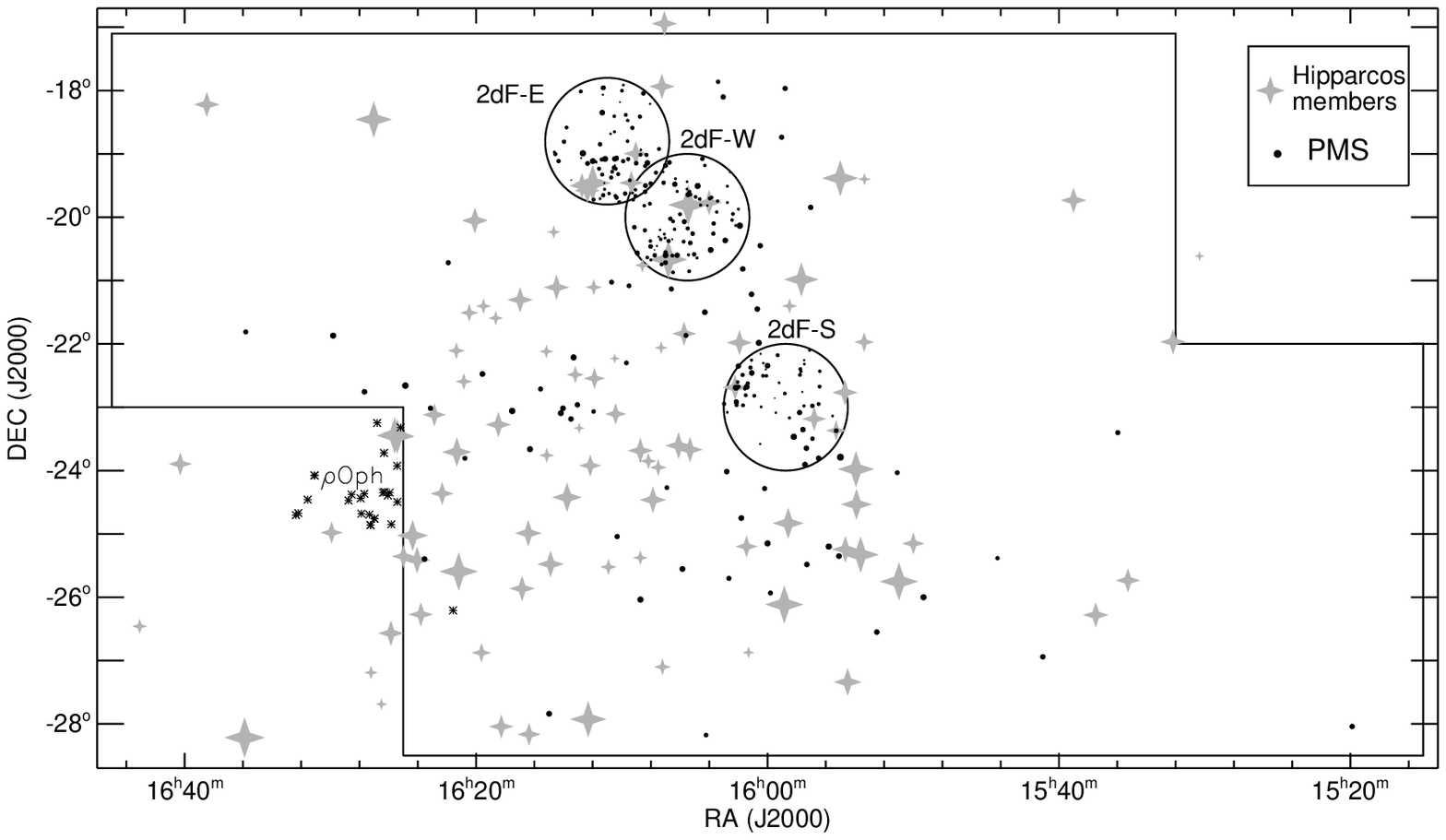}
\caption{Map of the Upper Scorpius region.  The 160~deg$^2$ area
of the FLAIR survey \citep{Preibisch98} is marked by the thick solid
line.  The three fields of the 2dF survey
\citep{Preibisch01,Preibisch02} are marked by circles.  Hipparcos
members are shown as asterisks and low-mass PMS stars as dots, with
symbol size proportional to the magnitude of the stars.  T Tauri stars
associated with the $\rho$~Oph star forming region listed in the
Herbig-Bell Catalogue \citep{Herbig88} are shown as small
asterisks. \label{usco_map.fig}}
\end{figure}

During the last couple of years, extensive spectroscopic surveys for
low-mass members of US were performed with wide-field multi-object
spectrographs at the Anglo-Australian Observatory.  The first step was
a survey with the wide-field multi-object spectrograph FLAIR at the
1.2\,m United Kingdom Schmidt Telescope to reveal PMS stars among
ROSAT All Sky Survey X-ray sources in a 160~deg$^2$ area
\citep{Preibisch98}.  In this spatially complete, but flux-limited
survey covering nearly the full area of the association, 39 new PMS
stars were found. \citet{PZ99} investigated the star formation history
in US.  In a detailed analysis of the HR diagram, properly taking into
account the uncertainties and the effects of unresolved binaries, they
found that the low-mass PMS stars have a mean age of about 5 Myr and
show no evidence for a large age dispersion. The PMS sample of
\citet{PZ99} (see Table~\ref{PZ99tab}) is statistically complete for
stars in the mass range $\sim 0.8$\,\msun\, to $\sim 2$\,\msun.

The next step was to reveal the {\em full} population of low- and
very-low mass $(\sim 0.1\!-\!0.8$\,\msun) stars in a representative
area of US, in order to allow a direct determination of the full IMF
of this OB association.  The multi-object spectrograph 2dF at the
3.9\,m Anglo-Australian-Telescope was used to obtain intermediate
resolution spectra of more than 1000 stars with magnitudes $R =
12.5\!-\!18.0$ in a 9~deg$^2$ area. Among these, 166 new PMS
stars were found, nearly all of them M-type stars, by their strong Li
absorption lines.  The results of these observations were reported in
\citet{Preibisch01} and \citet{Preibisch02}, and the newly revealed
low-mass members are listed in Table~\ref{2dfpmstab}.  Combining these
results with the earlier investigation yielded a sample of 250 PMS
stars in the mass range $\sim 0.1\,$\msun\, to $\sim 2$\,\msun.  A map
of the survey region showing the locations of the low-mass PMS stars
as well as the high-mass Hipparcos members is shown in
Fig.~\ref{usco_map.fig}.  One can see that the low-mass members are
spatially coincident with the early type members of the US
association.

\subsection{The HR-Diagram for Upper Scorpius}

\citet{Preibisch02} studied the properties of the full stellar
population in US on the basis of a large sample of 364 association
members. This sample was composed of the following parts:

(1) the 114 Hipparcos members constitute a complete sample of all
members with masses above $2\,M_\odot$ and an (incomplete) sample of
lower-mass members with masses down to $\sim 1\,$\msun, covering the
full spatial extent of the association.

(2) the 84 X-ray selected PMS stars from \citet{PZ99} provide a
statistically complete sample of the $\sim 0.8\!-\!2\,$\msun\, members
(plus some lower mass stars) in a 160~deg$^2$ area.

(3) the 166 low-mass PMS stars identified with 2dF
 \citep{Preibisch01,Preibisch02} constitute a statistically complete,
 unbiased sample of the member population in the $\sim
 0.1\!-\!0.8\,$\msun\, mass range, which covers a 9~deg$^2$
 area.

\begin{figure}[!htb]
\plotone{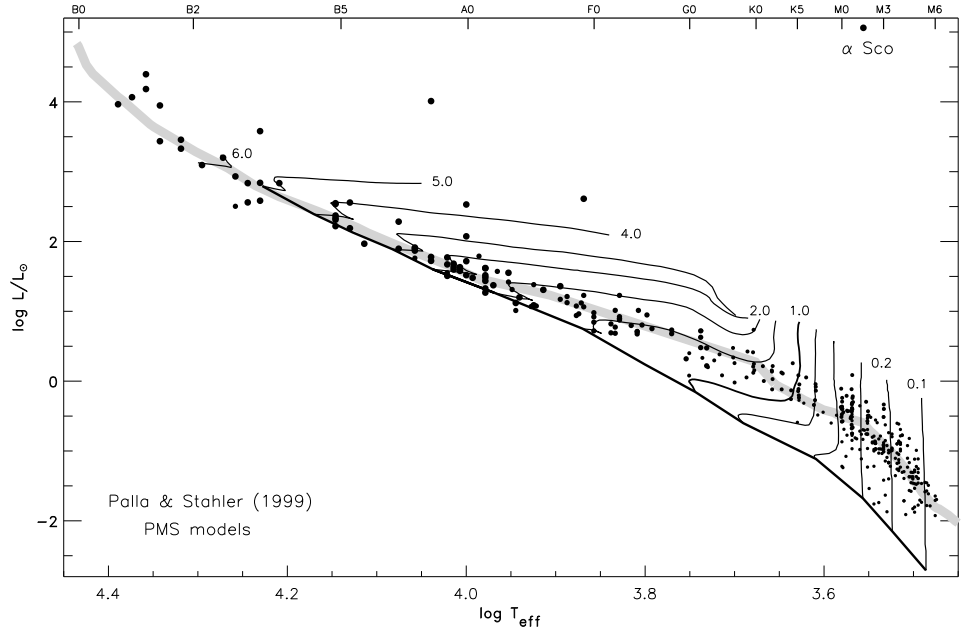}
\caption{HR diagram for the Upper Scorpius members
described in \citet{Preibisch02}.  The lines show the evolutionary
tracks from the \citet{Palla99} PMS models, some labeled by their
masses in solar units.  The thick solid line shows the main sequence.
The 5 Myr isochrone is shown as the thick grey line; it was composed
from the high-mass isochrone from \citet{Bertelli94} for masses
$6\!-\!30\,$\msun, the \citet{Palla99} PMS models for
$1\!-\!6\,$\msun, and the \citet{Baraffe98} PMS models for
$0.02\!-\!1\,$\msun. \label{hrd_full.fig} }
\end{figure}

\noindent Figure~\ref{hrd_full.fig} shows the HR-diagram with all
these US members.

\subsection{Ages of the Upper Scorpius Stars \label{us_ages}}

From Fig.~\ref{hrd_full.fig} one can see that not only the majority of
the low-mass stars, but also most of the intermediate- and high-mass
stars lie close to or on the 5 Myr isochrone. There clearly is a
considerable scatter around this isochrone that may seem to suggest a
considerable spread of stellar ages. However, it is very important to
be aware of the fact that the masses and especially the ages of the
individual stars read off from their position in the HR-diagram are
generally not identical to their true masses and ages. For the case of
US, the most important factor is the relatively large spread of
individual stellar distances \citep[$\sim \pm\,20$~pc around the mean
value of 145~pc;][and priv.~comm.]{deBruijne99} in this very nearby
and extended region, which causes the luminosities to be either over-
or under-estimated.  Another important factor is the presence of
unresolved binary companions, which cause over-estimates of the
luminosity. Further factors include photometric errors and
variability, and the uncertainties in the calibrations used to derive
bolometric luminosities and effective temperatures.  Detailed
discussions and simulations of these effects can be found in
\citet{PZ99} and  \citet{Hill08}.  The net effect of the uncertainties
is that in the observed HR diagram a (hypothetical) perfectly coeval
population of stars will {\em not} populate just a single line
(i.e.~the corresponding isochrone), but will always display a finite
spread, mimicking an age spread.  \citet{PZ99} and \citet{Preibisch02}
found via statistical modeling of these effects
that the observed HR-diagram for the low-mass stars in US is
consistent with the assumption of a {\em common stellar age of about 5
Myr; there is no evidence for an age dispersion, although small age
spreads of $\sim 1-2$~Myr cannot be excluded by the data}.
\citet{Preibisch02} showed that the derived age is also robust when
taking into account the uncertainties of the theoretical PMS models.
The mean age of 5~Myr and the absence of a
significant age spread among the US stars
has been confirmed in the independent studies
of \citet{Allen03} and \citet{Slesnick07}.

It is remarkable that the age derived for the low-mass stars is very
well consistent with previous independent age determinations based on
the nuclear and kinematic ages of the massive stars \citep{deGeus89},
which also yielded 5 Myr.  This very good agreement of the {\em
independent\/} age determinations for the high-mass and the low-mass
stellar population shows that {\em low- and high-mass stars are
coeval} and thus have formed together.  Furthermore, the absence of a
significant age dispersion implies that all stars in the association
have formed more or less simultaneously. This means that the
star-formation process must have started rather suddenly and
everywhere at the same time in the association, and also must have
ended rather suddenly after at most a few Myr.  The star formation
process in US can thus be considered as a {\em burst of star
formation\/}.  This will be discussed in more detail in Sect.~\ref{sf_in_usco}

\subsection{The Full Initial Mass Function of Upper Scorpius \label{us_imf}}

\begin{figure}[!b]
\plotfiddle{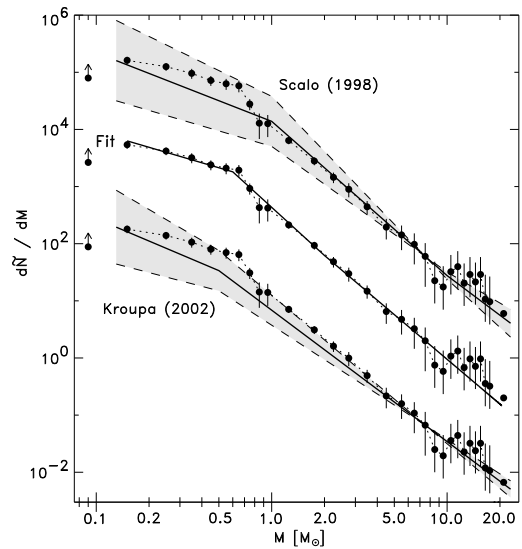}{7.0cm}{0}{60}{60}{-150}{-220}
\caption{Comparison of the derived mass function for Upper Scorpius
with different representations of the field IMF.  The US mass function
is shown three times by the solid dots connected by the dotted lines,
multiplied by arbitrary factors. The middle curve shows the original
mass function, the solid line is our multi-part power-law fit.  The
upper curve shows the US mass function multiplied by a factor of 30
and compared to the \citet{Scalo98} field IMF representation (solid
line); the grey shaded area delimited by the dashed lines represents
the range allowed by the errors of the model.  The lower curve shows
the US mass function multiplied by a factor of 1/30 and compared to
the \citet{Kroupa02} field IMF representation (solid line).
\label{imf2.fig}
}
\end{figure}

Upper Scorpius is an ideal target for an investigation of the initial
mass function (IMF) for several reasons.  First, the star formation
process is completed and thus stars of all masses are already present;
since the molecular cloud has already been dispersed, the members are
also rather easily observable. Second, the age of $\sim5$~Myr is young
enough that nearly all members are still present; only very few of the
most massive members have already evolved away from the main-sequence,
and these stars can be accounted for individually.  Therefore, {\em
the present day mass function} (with the addition of the one member
that already exploded as a supernova) {\em is identical to the initial
mass function.}

Figure \ref{imf2.fig} shows the empirical mass function for US as
derived in \citet{Preibisch02}.  The best-fit multi-part power law
function for the probability density distribution is given by

\begin{equation}
\frac{{\rm d}N}{{\rm d}M} \propto   \left\{ \begin{array}{lcrcr}
     M^{-0.9\pm0.2} & \mbox{for} &0.1\!&\le M/M_\odot <&\!0.6\\
     M^{-2.8\pm0.5} & \mbox{for} &0.6\!&\le M/M_\odot <&\!2\\
     M^{-2.6\pm0.5} & \mbox{for} &2\!&\le M/M_\odot <&\!20\end{array}\right.
 \label{imf_usco}
\end{equation}
or, in shorter notation,
$\alpha[0.1\!-\!0.6] = -0.9\pm0.2$,
$\alpha[0.6\!-\!2.0] = -2.8\pm0.5$,
$\alpha[2.0\!-\!20] = -2.6\pm0.3$.
The derived US mass function is compared to two different
representations of the field IMF.  The first one is the field IMF
representation suggested by \citet{Scalo98}, given by
$\alpha[0.1\!-\!1] = -1.2\pm0.3$, $\alpha[1\!-\!10] = -2.7\pm0.5$,
$\alpha[10\!-\!100] = -2.3\pm0.5$, the second is the parameterization
for the average Galactic field star IMF in the solar neighborhood
derived by \citet{Kroupa02}, which is given by $\alpha[0.02\!-\!0.08]
= -0.3\pm0.7$, $\alpha[0.08\!-\!0.5] = -1.3\pm0.5$,
$\alpha[0.5\!-\!100] = -2.3\pm0.3$.  Although the observed US mass
function is not identical to either the \citet{Scalo98} or
\citet{Kroupa02} field IMF representations, it is well within the
ranges of slopes derived for similar mass ranges in other young
clusters or associations, as compiled in \citet{Kroupa02}.  Therefore,
it can be concluded that, within the uncertainties, the general shape
of the US mass function is consistent with recent field star and
cluster IMF determinations.

The {\em total stellar population} of US in the $0.1\,$\msun\, to
$20\,$\msun\, mass range can be described by the `best fit' mass
function (eqn.~\ref{imf_usco}).  The integration of this function
yields 2525 stars with a total mass of $1400\,$\msun.  75\% of all
stars have masses below $0.6\,$\msun\, and contribute 39\% of the
total mass. Only 3\% of all stars have masses above $2\,$\msun, but
they contribute 33\% of the total mass.  All these numbers are based
on the primary star mass function.  For a reasonable estimate of the
total stellar mass one has to take into account that most of the stars
are probably in multiple systems.  The binary frequency
\citep[i.e.~the probability that a given object is multiple;
cf.][]{Reipurth93} for late type stars is at least about 50\%
\citep[c.f.][see K\"ohler et al.~2000 for the case of US]{Duquennoy91,
Fischer92} and even higher for early type stars
\citep{Mason98,Abt90,Preibisch99}.  The total mass of the companions
in the multiple systems can be estimated as follows: assuming a binary
frequency of 100\% and random pairing of secondaries from the same
underlying mass function (eqn.\ref{imf_usco}) as the primaries, the
total mass of all companions in the $0.1\!-\!20\,$\msun\, mass range
is 40\% of the total mass of all primaries.  For the estimate of the
total stellar mass one also has to include the most massive stars,
that have already evolved away from the main-sequence: Antares ($\sim
22\,M_\odot$) and its B2.5 companion ($\sim 8\,M_\odot$), the
supernova progenitor ($\sim 50\,M_\odot$), and $\zeta$ Oph ($\sim
20\,M_\odot$).  The total stellar mass is then: $ 1.4 \times
1400\,M_\odot + 22\,M_\odot + 8\,M_\odot + 50\,M_\odot + 20\,M_\odot =
2060\,M_\odot$.

\section{Upper Centaurus-Lupus (UCL) \label{ucl}}

While US has been the focus of many investigations over the past
century, its older siblings -- the UCL and LCC regions -- have
received much less attention. There appear to be several biases which
made these two groups more difficult to study, or perhaps more easy to
ignore, at least historically.  These are: (1) UCL and LCC are not as
concentrated as US, (2) a significant fraction of their memberships
are within 10$^{\circ}$--15$^{\circ}$ of the Galactic plane, making
separating their membership from the Galactic population more
difficult, (3) most of UCL and LCC are inaccessible to telescopes in
the northern hemisphere, (4) and early investigations of PMS stars
centered on dark and reflection nebulae, which are in abundance in and
near US, but more lacking or absent in UCL and LCC. Only over the past
decade has the availability of high quality astrometry ({\it
Hipparcos}, Tycho, etc.)  and the {\it ROSAT} All-Sky Survey, enabled
the efficient identification of low-mass UCL and LCC members. Given
the numbers of B-type stars in UCL and LCC, there are probably {\it
thousands} of low-mass members awaiting discovery.

\subsection{The High-Mass Stellar Population of UCL \label{ucl_high_mass}}

There appear to be no post-MS members of UCL, however there is a
fairly well-defined main sequence turn-off near spectral type B1.5
\citep{deGeus89,Mamajek02}.  The notable turn-off stars are, in
approximate order of mass: $\mu^1$ Sco (B1.5IV), $\alpha$ Lup
(B1.5III), $\beta$ Lup (B2III), $\delta$ Lup (B1.5IV), $\nu$ Cen
(B2IV), $\mu^2$ Sco (B2IV), and $\eta$ Lup \citep[B1.5V; where MK
spectral types are from ][]{Hiltner69}.  The positions of these
massive stars are plotted and labeled in the UCL map in
Fig.~\ref{ucl_map.fig}.  All of these stars were selected as {\it
Hipparcos} members of UCL by \citet{deZeeuw99} except for $\mu^1$ Sco
and $\beta$ Lup.  \citet{Hoogerwerf00} has noted that the longer
baseline ACT and TRC proper motions for these stars are more conducive
to UCL membership.  The Tycho-2 proper motion for $\beta$ Lup is
well-pointed toward the UCL convergent point defined by
\citet{deBruijne99}, however its magnitude is somewhat larger than
most other UCL members.

Using the \citet{Bertelli94} evolutionary tracks, \citet{Mamajek02}
estimated the main sequence turn-off age of UCL to be $\sim$17\,Myr.
It was \citet{deGeus92} who first calculated the number of supernovae
to have exploded in UCL ($\sim$6\,$\pm$\,3), and showed that the sum
of the total kinetic energy predicted to be imparted by the supernovae
and stellar winds of the deceased UCL members ($\sim$10$^{51}$~erg) is
roughly consistent with the kinetic energy of the $\sim$100\,pc-radius
expanding shell of H\,{\sc I} centered on UCL.  This satisfactory
agreement supports the original scenario by \citet{Weaver79}, that the
massive stars in the oldest Sco-Cen subgroups (UCL and LCC) destroyed
the proto-Sco-Cen molecular cloud complex, and radially dispersed most
of the gas into what is seen today as large, expanding, loop-like H~I
structures centered on Sco-Cen. A modern estimate by the authors of
the number of exploded supernovae in UCL ($\sim$7 SNe), using the
\citet{deZeeuw99} membership list for UCL, and adopting a
\citet{Kroupa02} initial mass function, corroborates de Geus's (1992)
prediction.

UCL contains a Herbig Ae star (HD 139614) and two accreting Fe
stars (AK Sco and HD 135344).  Both HD 139614 and HD 135344 are close
to the western edge of the Lupus clouds near $\ell$ $\simeq$
332$^{\circ}$ (see Fig. 8 of the Lupus chapter; Comer\'on, this volume),
and so may be younger than the mean age for UCL.

\bigskip

\subsection{The Low-Mass Stellar Population of UCL \label{ucl_low_mass}}
\medskip

\noindent{\em The Pre-{\it ROSAT} Era:}\smallskip

\noindent A small number of PMS members of UCL were found before the
 arrival of the {\it ROSAT} X-ray surveys.  In the course of a radial
 velocity survey of B-type members of Sco-Cen, \citet{Thackeray66}
 found that many of their visual companions shared similar radial
 velocities, i.e. they probably constituted physical binaries.  Some
 of these objects were of AFGK spectral types, and include HD 113703B
 (K0Ve; LCC), HD 113791B (F5V; LCC), HD 143099 (G0V; UCL), and HD
 151868 (F6V; UCL). The high-mass primaries are still considered
 secure members of US, UCL, and LCC \citep{deZeeuw99}.
 \citet{Thackeray66} proclaimed ``{\it [t]hese observations do in fact
 represent the first to establish the presence of stars later than
 type A0 in the group. They include one K-dwarf which is presumably no
 older than the group to which \citet{Blaauw64} assigns an age of 20
 million years}''.  \citet{Catchpole71} noted strong Li in the
 spectrum of HD 113703B, confirming its extreme youth.  The first two
 stars, along with HD 129791B (K5Ve; UCL) and HD 143939B (K3Ve; UCL),
 are included in the well-studied ``Lindroos'' sample of post-T Tauri
 companions to massive stars \citep[e.g.][]{Lindroos86,Huelamo00}.

In an effort to tie the absolute magnitude calibration of B-type stars
to that of later type stars, as well as explore the luminosity
function of Sco-Cen, \citet{Glaspey72} identified 27 candidate A- and
F-type stars in a $uvby$ survey of UCL. Of these 27 candidates, 24
have {\it Hipparcos} astrometry, and \citet{deZeeuw99} retained 14 of
these as kinematic members of UCL.\bigskip

\noindent{\em The {\it ROSAT} Era: The Lupus Region Surveys:}\smallskip

\noindent Most of the low-mass members of UCL which have been
 identified over the past decade have been due to {\it ROSAT} X-ray
 pointed observations and its all-sky survey.  The focus of most of
 these surveys was not UCL, but the Lupus molecular cloud complex.  In
 a wide-field survey of X-ray-luminous stars in a 230~deg$^2$
 region around the Lupus clouds, \citet{Krautter97} identified 136
 candidate T Tauri stars. While 47 of these new T Tauri stars (TTS)
 were found in pointed {\it ROSAT} observations of the Lupus clouds,
 the majority (89) were found scattered over a wider region with the
 {\it ROSAT} All-Sky Survey (RASS).  \citet{Wichmann97} found that if
 the ``off-cloud'' T Tauri stars were co-distant with the Lupus
 clouds, then their mean isochronal age was significantly older than
 the ``on-cloud'' Lupus T Tauri stars ($\sim$7\,Myr
 vs. $\sim$1-3\,Myr, respectively). \citet{Wichmann97} hypothesized
 that the off-cloud TTS formed in the Lupus clouds, but were dispersed
 either due to a large intrinsic velocity dispersion in the clouds, or
 due to ejection \citep{Sterzik95}.

\begin{sidewaysfigure}[!p]
\centering
\includegraphics[draft=False,height=9.8cm]{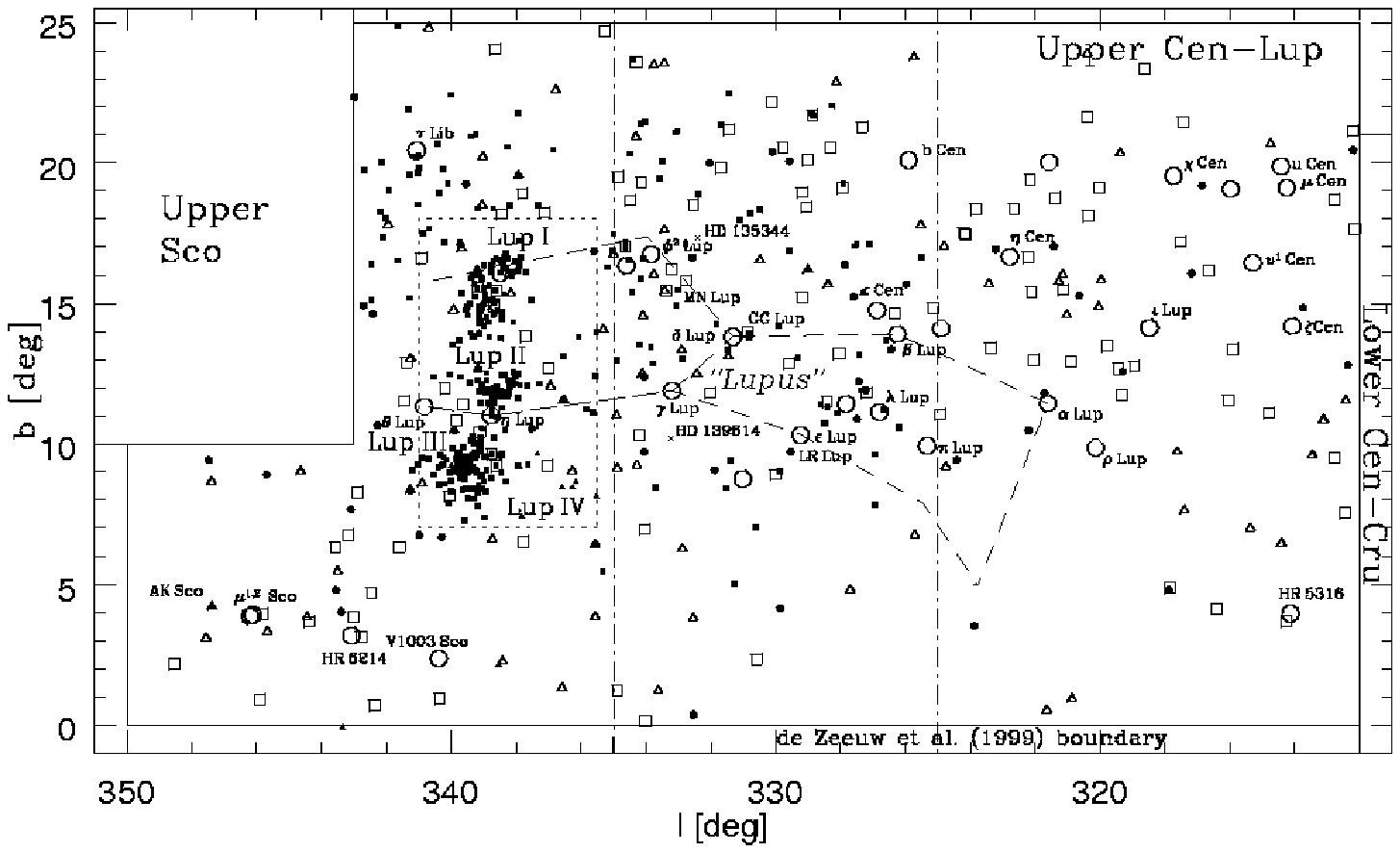}
\caption{The Upper Centaurus-Lupus (UCL) region.  The {\it solid line}
is the UCL border from \citet{deZeeuw99}. The {\it dotted line} is the
box enclosing the Lupus star-forming clouds (Sect.~\ref{ucl_low_mass}; the
cloud positions are labeled).  The {\it dash-dotted line} is the
region investigated by \citet{Wichmann97GB}. The {\it long-dashed}
lines outline the Lupus constellation. UCL members from the {\it
Hipparcos} study of \citet{deZeeuw99} are plotted as follows: B0-B5
stars ({\it large open circles}), B6-A9 stars ({\it medium open
squares}), FGKM stars ({\it small open triangles}).  PMS stars from
X-ray surveys are plotted as follows: \citet{Mamajek02} ({\it filled
circles}), \citet{Wichmann97GB} and \citet{Kohler00} ({\it filled
squares}), \citet{Krautter97} and \citet{Wichmann97} ({\it tiny filled
squares}).  PMS stars from the Herbig-Bell Catalog \citep{Herbig88}
are plotted as {\it small black triangles}. Two ``isolated'' Herbig
Ae/Be stars (HD 135344 \& 139614) are plotted with {\it
Xs}. \label{ucl_map.fig}}
\end{sidewaysfigure}

\citet{Wichmann97GB} also conducted a spectroscopic survey of RASS
sources west of the Lupus clouds in a strip between 325$^{\circ}$ $<$
$\ell$ $<$ 335$^{\circ}$ and $-5^{\circ}$ $<$ $b$ $<$ $+50^{\circ}$
({\it dotted line} in Fig. \ref{ucl_map.fig}).  They identified 48
Li-rich stars, most of which were concentrated between Galactic
latitudes +8$^{\circ}$ and +22$^{\circ}$. \citet{Wichmann97GB}
hypothesized that the majority of these objects were ``Gould Belt''
members, with ages of $<$60\,Myr.

The idea that there are older, dispersed RASS TTSs near the Lupus
clouds becomes somewhat less surprising when it is appreciated that
the Lupus clouds are adjacent to the US and UCL subgroups of Sco-Cen
(with ages of 5 and $\sim$17~Myr, respectively, and thousands of
predicted members).  Sco-Cen and UCL are not mentioned in the
\citet{Krautter97}, \citet{Wichmann97}, or \citet{Wichmann97GB}
surveys, although in this region of sky the ``Gould Belt'' is
essentially {\it defined} by the high-mass stars of UCL
\citep[e.g.][]{Lesh72} and the gas associated with the Lupus
clouds. Two of the clumps of TTSs seen by Wichmann et al.~(1997b)
seem to be co-spatial with the over-densities of high-mass UCL
members, associated with $\lambda$ Lup and $\phi^2$ Lup. A more
detailed kinematic study of the Krautter-Wichmann TTSs is needed to
disentangle the star-formation of the region, explore the relation
between the modern-day Lupus clouds and the ``completed''
star-formation in UCL.

Recently, \cite{Makarov07} investigated the kinematics of the
\citet{Krautter97} TTS in the Lupus region. After correcting for the
distribution of individual stellar distances, the color-magnitude
diagram revealed two separate stellar populations with clearly
different ages: a young ($\sim 1$ Myr) population of TTS which are
closely concentrated at the Lupus dark clouds, and an older ($\sim
5-27$ Myr) population of stars which are much more widely dispersed in
the area.  Based on kinematic arguments, Makarov concludes that ``it
is unlikely that the T Tauri members were born in the same
star-forming cores as the more compactly located classical T Tauri
stars.'' We agree with this assessment, but add that the positions,
motions, and {\it mean} age of the outlying members are very well
consistent with the idea that the \citet{Krautter97} stars are in fact
low-mass members of the US and UCL subgroups of Sco-Cen. Rather than
subsume all of the RASS stars into a ``Lupus Association'' as
\citet{Makarov07} elects, it is probably wise for astrophysical
studies to separate the on- or near-cloud Lupus members from the
off-cloud UCL/US members. We have visually attempted to do this with a
dashed line box in Fig.~6, but clearly more study is warranted.

There is good kinematic evidence that the \citet{Wichmann97GB} RASS
stars are mostly UCL members.  A preliminary kinematic analysis by
E.M. suggests that for the 33 \citet{Wichmann97GB} {\it ROSAT} stars
with either Tycho-2 \citep{Hoeg00} or UCAC2 \citep{Zacharias04} proper
motions, {\it all but four have motions consistent with UCL
membership}.  The {\it ROSAT} X-ray stars RX J1412.2-1630,
J1419.3-2322, J1509.3-4420, and J1550.1-4746 can all be rejected as
kinematic UCL members. The first two objects are also well outside of
the UCL region defined by \citet{deZeeuw99}, so their rejection is
perhaps unsurprising.  The inferred cluster parallax distances for the
rest\footnote{Cluster parallax distances were calculated using either
the UCAC2 or Tycho-2 proper motions, and the UCL space motion vector
from \citet{deBruijne99}.  See \citet{Mamajek02} for details on the
technique.}  range from $\sim$90\,pc to $\sim$200\,pc, suggesting
considerable depth to the UCL subgroup. Of note for future Li
depletion studies of UCL members, we find that the M-type Wichmann et
al. stars RX J1512.8-4508B (M1), RX J1505.4-3716 (M0), and RX
J1457.3-3613 all have proper motions consistent with membership in
UCL.  The components of the wide M-type binary RX J1511.6-3249A and B
(M1.5+M1.5; $48''$ separation) also have proper motions consistent with
UCL membership.

\begin{figure}[!ht]
\plotfiddle{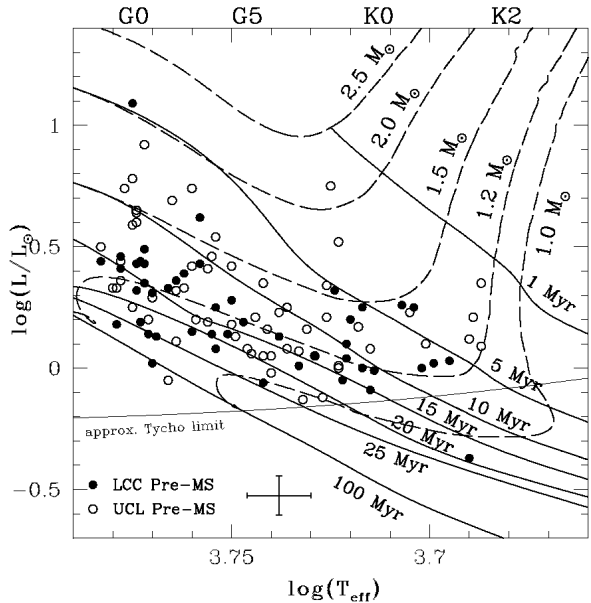}{9.3cm}{0}{50}{50}{-150}{-78}
\caption{Theoretical H-R diagram for PMS members of UCL
(open circles) and LCC (filled circles) from the survey
of \citet[][their Fig. 6]{Mamajek02}. PMS evolutionary
tracks are from \citet{D'Antona97}. \label{hrd_ucl_lcc.fig}}
\end{figure}
\medskip

\noindent{\em The ROSAT Era: The Mamajek et al. Survey:} \smallskip

\noindent
\citet{Mamajek02} conducted the first wide-field, spectroscopic survey
searching explicitly for PMS GK-type members of UCL. They selected 56
UCL candidates by cross-referencing proper motion-selected stars from
the kinematic study of \citet[][but with color-magnitude constraints
more amenable to identifying PMS G- and K-type members from the ACT
and TRC astrometric catalogs]{Hoogerwerf00} with X-ray sources from
the {\it ROSAT} All-Sky Survey Bright Star Catalog \citep{Voges99}.
They also measured optical spectra of 18 GK-type {\it Hipparcos} stars
selected as probable kinematic members by \citet{deZeeuw99}. Blue and
red optical spectra of the candidates were taken with the DBS
spectrograph on the Siding Springs 2.3-m telescope, with resolution of
$\sim$2.8\,\AA\, in the blue, and $\sim$1.3\,\AA\, in the red.  Stars
were classified as PMS by virtue of having strong Li (stronger than
the loci of $\sim$30--100 Myr-old clusters) and subgiant luminosity
classes (measured via the Sr II $\lambda$4077 line). Between UCL and
LCC, the X-ray and proper motion selection was $\sim$93\% efficient at
selecting PMS stars, while 73\% of the kinematic candidates selected
by \citet{deZeeuw99} were PMS (well in line with the $\sim$30\%
contamination rate predicted by those
investigators). \citet{Mamajek02} confirmed 12 out of 18 of the
\citet{deZeeuw99} proper motion-selected UCL candidates as PMS, and
found 50 Li-rich UCL members from the X-ray and proper motion-selected
sample.  None of the UCL PMS stars identified could be considered
classical T Tauri candidates (i.e.~accreting), however a few
demonstrated small H$\alpha$ emission excesses (all $<$2\,\AA),
probably due to enhanced chromospheric activity. Although
\citet{Mamajek02} found no new classical T Tauri UCL members, there
are two known F-type accretors \citep[AK Sco and HD 135344;
e.g.][]{Alencar03}.  The mean age of the PMS UCL members in the
\citet{Mamajek02} survey depends, unsurprisingly, on one's choice of
evolutionary tracks, but ranged from 15--22~Myr (see
Fig.~\ref{hrd_ucl_lcc.fig}). Using the evolutionary tracks of
\citet[][where appropriate]{Baraffe98}, and accounting for the
magnitude limit of the \citet{Mamajek02} survey (see Sect.~7.1 of their
paper), one finds a median age for the UCL PMS stars of 16~Myr.

While Mamajek et al.'s survey of UCL was very broad, admittedly it was
not very deep, and more members could be easily identified with
existing catalog data (e.g.~RASS Faint Source Catalog, UCAC2, SACY,
etc.).  \citet{Mamajek02} claim that the majority ($\sim$80\%) of the
$\sim$1.1--1.4\,\msun\, members have probably been identified, but
that $\sim$2000 $<$1\,\msun\, members likely await discovery. The
lowest-mass UCL candidates which have been found (and which are not in
the immediate vicinity of the Lupus star-forming clouds) are the
\citet{Krautter97} Li-rich early-M-type stars RX\,J1514.0-4629A (LR
Lup; M2) and RX\,J1523.5-3821 (MN Lup; M2). The equivalent widths of
the Li\,I $\lambda$6707.7 line for these M2 stars
($\sim$0.38\AA), as measured in low-resolution spectra by
\citet{Wichmann97}, {\it suggests that modest Li depletion is taking
place among the early-M stars in UCL}.  A high resolution
spectroscopic survey of Li-rich M-type members could place independent
constraints on the age of UCL, as well as allow an interesting
comparison between Li depletion ages and main sequence turn-off ages.
\medskip

\noindent{\em A List of Low-Mass UCL Members:}\smallskip

\noindent
We have assembled a membership list of candidate low-mass members of
UCL from the previously mentioned literature (Table
\ref{table_ucl_list}).  The list is meant to be rather exclusive in
that it selects only those low-mass (GKM) stars that are Li-rich {\it
and} have kinematics consistent with group membership. All are known
X-ray sources, except for a small number of proper motion candidates
from \citet{deZeeuw99} which were confirmed to be Li-rich by
\citet{Mamajek02}. The F stars in UCL have not been thoroughly
investigated spectroscopically. Given the very high efficiency of
selecting PMS stars amongst X-ray and proper motion selected GK stars
\citet[93\%]{Mamajek02}, we list only the proper motion-selected
F-type candidate members from the literature which have ROSAT All-Sky
Survey X-ray counterparts within 40'' of their optical positions. The
sample is not meant to be complete, by any means, but should represent
a relatively clean sample of $\la$2\,\msun\, UCL members.  The
incidence of non-members is probably well below $\la$5\%. Two
confirmed PMS companions to high mass UCL members (HD 129791B and HD
143939B) were included, using data from \citet{Huelamo00},
\citet{Pallavicini92}, \citet{Lindroos83}, and distances (for the
primaries) from \citet{Madsen02}.  Although the majority of
\citet{Krautter97} and \citet{Wichmann97GB} stars appear to be bona
fide UCL members (by virtues of their positions, proper motions,
appropriate HR-D positions, etc.), we have not included them in our
membership list, but refer the reader to those papers.

In Table \ref{table_ucl_list} we have flagged the UCL members which
are near the Lupus molecular clouds. We have defined a box around the
Lupus clouds that are actively forming stars: 335.5$^{\circ}$ $<$
$\ell$ $<$ 341$^{\circ}$ and +7$^{\circ}$ $<$ $b$ $<$
+18$^{\circ}$. {\it The population of stars in the off-cloud region in
that box may represent a mix of stars that have recently formed in the
Lupus clouds, along with older UCL PMS stars}. The HRD positions of
the flagged UCL stars in Table \ref{table_ucl_list} are consistent
with having ages of $\sim$7--23 Myr, so they are probably UCL members
projected against (or behind?) the Lupus clouds.

\section{Lower Centaurus-Crux (LCC) \label{lcc}}

\begin{figure}[!h]
\plotone{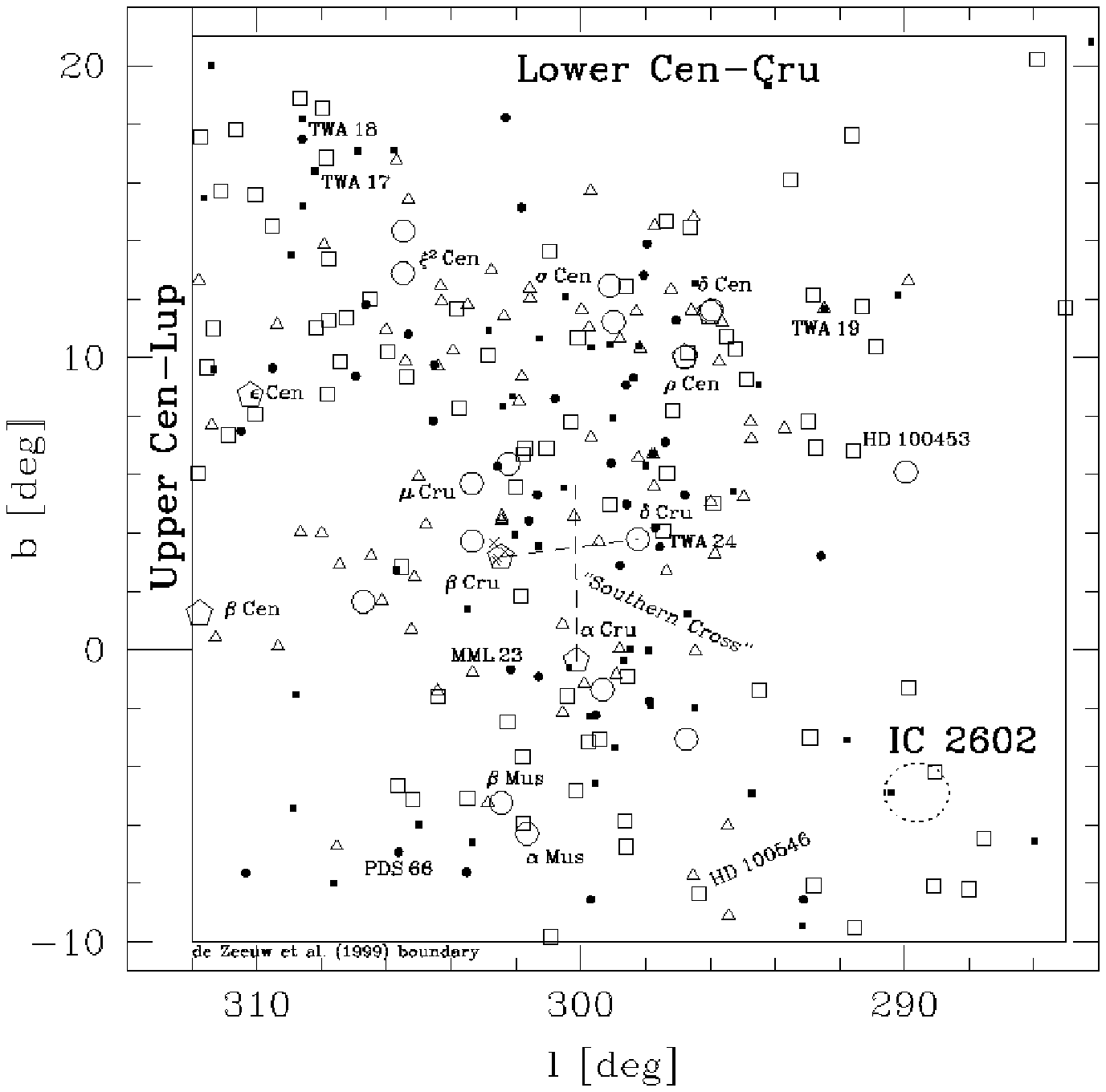}
\caption{The Lower Centaurus-Crux region.  The {\it solid line} is the
LCC border from \citet{deZeeuw99}. LCC members from the {\it
Hipparcos} study of \citet{deZeeuw99} are plotted as follows: B0-B5
stars ({\it large open circles}), B6-A9 stars ({\it medium open
squares}), FGKM stars ({\it small open triangles}).  The ``super
Cen-Crux six'' early B-type stars (Sect.~\ref{lcc_high_mass}) are {\it
large open pentagons}.  PMS stars are plotted from the surveys of
\citet{Mamajek02} ({\it filled circles}), \citet{Park96} ({\it Xs};
near $\beta$ Cru), and TWA stars listed as probable LCC members by
\citet[][{\it filled squares}]{Mamajek05}. Li-rich SACY stars from
\citet{Torres06} selected as LCC members in this review are {\it
filled squares}. The position of the nearby cluster IC 2602 is
plotted, although its motion and older age are very distinct from that
of LCC \citep{deZeeuw99}. Three of the four stars that outline the
``Southern Cross'' ({\it dashed line}) are probable LCC members.
\label{lcc_map.fig}
}
\end{figure}

LCC straddles the Galactic equator in Crux, stretching from Galactic
latitude \mbox{$\sim -10^{\circ}$} to $+20^{\circ}$ between Centaurus, Crux,
and Musca (see Fig.~\ref{lcc_map.fig}).  Although LCC is the closest
recognized OB association subgroup to the Sun ($\langle D \rangle =
118$~pc), it is the least studied of the Sco-Cen regions. There
is some hint of substructure in the group, and it appears that the
northern part of the group is somewhat more distant, older, and richer
($\sim$17~Myr, 120~pc) than the southern part of the group ($\sim$12~Myr,
110~pc).

\subsection{The High-Mass Stellar Population of LCC \label{lcc_high_mass}}

The upper main sequence of LCC is poorly defined, and so the group has
the least secure turn-off age of the Sco-Cen subgroups.  In
approximate order of mass, the LCC turnoff stars with fairly secure
kinematic membership \citep{deZeeuw99} are: $\delta$ Cru (B2IV),
$\sigma$ Cen (B2V), $\alpha$ Mus (B2IV-V), $\mu^1$ Cru (B2IV-V), and
$\xi^2$ Cen \citep[B1.5V; MK types from][]{Hiltner69}.  There are
other massive stars in this region, whose membership in LCC has been
debated, but whose {\it Hipparcos} proper motions were {\it
inconsistent} with membership \citep{deZeeuw99}. Among these systems
are the ``{\it super Cen-Crux six}'': $\alpha$ Cru (B0.5IV+B1V+B4IV),
$\beta$ Cru (B0.5III), $\beta$ Cen (B1III), $\delta$ Cen (B2IVne),
$\epsilon$ Cen (B1III), and $\rho$ Cen \citep[B3V;][]{Hiltner69}.  As
noted by \citet{Mamajek02} for the most massive subset of these,
``{\it These stars are $\sim$10--20\,\msun, with inferred ages of
$\sim$5--15 Myr and distances of $\sim$100--150\,pc.  Such stars are
extremely rare, and their presence in the LCC region appears to be
more than coincidental.''} As noted by \citet{deZeeuw99}, all of these
objects are flagged as having unusual motions due to either binarity
or variability in the {\it Hipparcos} catalog, however they only
consider $\delta$ Cen, $\beta$ Cru, and $\rho$ Cen to be probable LCC
members whose {\it Hipparcos} proper motions excluded them from
kinematic selection.  In a follow-up paper, \citet{Hoogerwerf00}
demonstrated that the long-baseline proper motions for $\rho$ Cen (ACT
catalog) and $\beta$ Cru (TRC catalog) are more consistent with LCC
membership, however they did not address the remainder of the
``six''. A more thorough kinematic investigation of the center-of-mass
motions of these systems is needed.  Although the Sco-Cen subgroup
memberships seem to have a $\sim$1--1.5\,\kms\, internal velocity
dispersion \citep{deBruijne99}, if significant dynamical evolution has
taken place among the massive multiple systems over the past
$\sim$10--15~Myr (i.e.~due to supernovae, or ejections from
short-lived trapezia), then the resultant kicks could explain the
presence of these massive multiple systems with deviant motions in the
vicinity of the Sco-Cen subgroups. The results from a kinematic study
of these ``super six'', as well as the runaway star and pulsar
candidates in the vicinity \citep[e.g.][]{Hoogerwerf01} could have
important implications for the IMF and star-formation history of
LCC.

LCC contains two known Herbig Ae/Be stars: HD 100453 (A9Ve) and
HD 100546 (B9Ve). Although sometimes labeled ``isolated'' HAEBE stars,
their positions, proper motions, and distances ($d \simeq 100-110$~pc)
are all consistent with LCC
membership. \Citet{vandenAncker98} estimated an age of $>$10 Myr for
HD 100546, consistent with other LCC members. Circumstellar PAH
emission was recently resolved around both stars using VLT
\citep{Habart06}, while the disk for HD 100546 was resolved in the
thermal IR at Magellan with MIRAC \citep{Liu03}. \citet{Chen06}
recently identified a faint candidate companion to HD 100453 at $1''$
separation consistent with being an M-type PMS star, however it has
not been spectroscopically confirmed. HD 100546 has a gap in its disk
consistent with the presence of a substellar object at $\sim 6$~AU
\citep{Acke06}.\bigskip

\subsection{The Low-Mass Stellar Population of LCC \label{lcc_low_mass}}
\medskip

The low-mass population of LCC has been investigated even less than
that of UCL. Low-mass members were identified serendipitously through
surveys of the regions near $\beta$ Cru \citep{Park96} and the TW Hya
association \citep{Zuckerman01}, which is near the northwest corner of
the LCC box defined by \citet{deZeeuw99}, but at roughly half the
distance \citep[$\sim$50\,pc;][]{Mamajek05} as LCC ($\sim$118\,pc).
\citet{Mamajek02} conducted the only systematic survey, thus far,
whose goal was to identify low-mass members over the whole LCC region.

\bigskip

\noindent{\em The Park-Finley {\it ROSAT} Stars near $\beta$ Cru:}\medskip

\noindent \citet[][]{Park96} identified 6 unknown, variable X-ray
sources near the B0.5III star $\beta$ Cru in a {\it ROSAT} PSPC
pointing.  By virtue of their X-ray spectral fits, variability, and
X-ray to optical flux ratios, \citet{Park96} conjectured that the 6
objects were T Tauri stars\footnote{The stars are indexed 1 through 6
with the acronym {\it [PF96]} (SIMBAD) or {\it Cru}
\citep{Feigelson97,Alcala02}. We adopt the SIMBAD nomenclature.}, and
that they had discovered a previously unknown star-forming region
centered on $\beta$ Cru. A low-resolution spectroscopic survey by
\citet{Feigelson97} found that the X-ray stars were Li-rich, late-type
stars, consistent with classification as weak-lined T Tauri
stars. After rejecting several hypotheses regarding the origins of
these young stars, \citet{Feigelson97} argued that Park \& Finley
serendipitously uncovered the first known ``isolated'' low-mass
members of LCC. \citet{Alcala02} conducted a high-resolution
spectroscopic study of the Park-Finley stars, and confirmed that 4 of
the 6 are sufficiently Li-rich to be classified as
PMS. \citet{Alcala02} also find that the radial velocities and
isochronal ages ($\sim$5--10 Myr)\footnote{The distances to the PF96
stars are unknown. We follow \citet{Alcala02}, and assume $d$ =
110\,pc (the distance to $\beta$ Cru). The reason for the discrepancy
in ages between the Park-Finley sample and the Mamajek et al.  sample
is unknown, but could be due to the unknown distances to the
Park-Finley stars.}  for the 4 most Li-rich stars are roughly
consistent with LCC membership.  The 4 PMS stars are K5-M4 in spectral
type, and have Li abundances intermediate between those of T Tauri
stars and the $\sim$50-Myr-old IC 2602 cluster. {\it The discovery of
more M-type Sco-Cen members may prove to be an interesting means of
testing PMS evolutionary tracks and Li depletion models}.
\medskip

\noindent{\em TWA {\it ROSAT} Stars in LCC:}\smallskip

\noindent
In their attempt to identify new members of the nearby TW Hya
association (TWA; age $\simeq$ 10\,Myr; $D$ $\simeq$ 50\,pc),
\citet{Zuckerman01} conducted a spectroscopic survey of RASS BSC X-ray
sources in the vicinity of the famous debris disk star HR 4796. They
identified eight new T Tauri stars, however they were different from
the rest of the TWA members thus far found: their optical, infrared,
and X-ray fluxes were significantly dimmer. \citet{Zuckerman01}
claimed that the new TWA stars (\#14-19)
 were further away from the
original TWA \#1-13 sample, with distances of perhaps
70-100\,pc. As
pointed out in \citet{Mamajek01}, the positions of TWA 14-19 overlap
with the LCC region defined by \citet{deZeeuw99}. \citet{Lawson05}
have presented evidence that the distribution of rotational periods of
TWA 1-13 and 14-19 are very different, suggesting
two different
populations.  In a detailed kinematic investigation of `TWA'' stars in
the literature, \citet{Mamajek05} concluded that the TWA stars are
dominated by two populations: a group of two dozen stars with
distances of $\sim$49\,$\pm$\,12\,pc (what probably constitutes the
true ``TW Hya association''), and a subset of objects with distances
of $\sim$100--150\,pc, which are likely LCC members (partially
corroborating the Lawson \& Crause findings). \citet{Mamajek05} claims
that TWA 12,
17,
18,
19, and
 24 are very likely LCC members, with
TWA 14 a borderline case. TWA 19 (= HIP 57524) was identified as an LCC
member by \citet{deZeeuw99}, as was TWA 24 (= MML 5) by
\citet{Mamajek02}.
\medskip

\noindent{\em The Mamajek et al. ROSAT Survey:}\smallskip

\noindent
As discussed at length in Sect.~4.2, \citet{Mamajek02} conducted a search
for $\sim$1\,\msun\, members of UCL and LCC amongst an X-ray and
proper motion-selected sample. They identified 37 LCC members, and
confirmed the youth of 10 (out of 12) of the \citet{deZeeuw99} {\it
Hipparcos} G-type candidates they observed. \citet{Mamajek02} claimed
that the classical T Tauri star PDS 66 \citep[][]{Gregorio-Hetem92} is
actually a LCC member, and appears to be the only known accreting
low-mass star in LCC. The mean age of the PMS LCC members in the
\citet{Mamajek02} survey ranged from 17 to 23~Myr, depending on the
choice of evolutionary tracks. Using the evolutionary tracks of
\citet[][where appropriate]{Baraffe98}, and accounting for the
magnitude limit of the Mamajek et al. survey, one finds a median age
for the LCC PMS stars of 18 Myr.
\medskip

\noindent{\em Torres et al. ``SACY'' ROSAT Survey:}\smallskip

\noindent
\citet{Torres06} presented results for a high resolution spectroscopic
survey of 1151 stars in the southern hemisphere with {\it ROSAT}
All-Sky Survey X-ray counterparts. Results for the ``SACY'' (Search
for Associations Containing Young Stars) survey regarding nearby young
low-density stellar groups are reported elsewhere in this volume
(Torres et al.). There are numerous Li-rich late-type SACY stars in
the Sco-Cen region, and most have proper motions and radial velocities
suggestive of membership to the Sco-Cen groups. For this review, we
have only attempted to assign membership of SACY stars to the Lower
Centaurus-Crux group, although the SACY catalog no doubt contains many
new UCL and US members as well. The selection of LCC members from the
SACY catalog will be discussed in more detail by Mamajek (in prep.).

To construct a sample of LCC members in the SACY catalog, we start
with the 138 SACY stars that lie within the \citet{deZeeuw99} boundary
for LCC. Of these objects, 45 are previously known LCC members found
either by \citet{deZeeuw99} or \citet{Mamajek02}.  We further prune
the SACY sample by removing giants, and selecting only those that are
Li-rich (EW(Li 6707\AA) $>$ 100m\AA) and that have proper motions
within 25 mas\,yr$^{-1}$ of the \citet{deBruijne99} mean value for
LCC. Lastly we run a convergent point algorithm from \citet{Mamajek05}
on the full sample of \citet{deZeeuw99} and \citet{Mamajek02} members
along with the remaining 49 SACY objects. Following
\citet{deBruijne99}, we assume an internal velocity dispersion of 1.14
km\,s$^{-1}$, and calculate cluster parallax distances using the
velocity vector of \citet{deBruijne99}. In total we identify 45 SACY
stars as probable new members of LCC. The median RV for the SACY stars
selected as LCC members is $-13$\,\kms, which is nearly identical to the
subgroup mean RV from \citet{deZeeuw99} ($-12$\,\,\kms).

\medskip

\noindent{\em A List of Low-Mass LCC Members:}\smallskip

\noindent
Table \ref{table_lcc_list} provides a modern catalog of low-mass LCC
members, and was constructed in a similar manner as that for UCL
(Sect.~\ref{ucl_low_mass}). The positions of known LCC members are plotted
in Fig.~\ref{lcc_map.fig}.  Individual cluster parallax distances were
adopted from \citet{Mamajek02} or \citet{Madsen02}. Where no published
cluster parallax distance was available, we calculate new values using
the \citet{deBruijne99} motion vector for LCC, while using the proper
motions for the TWA objects listed in \citet{Mamajek05} and from the
SACY catalog \citep{Torres06}.

\section{Sco-Cen as an Astrophysics Laboratory \label{lab}}

\subsection{Implications on the Star Formation Process in
Upper Scorpius \label{sf_in_usco}}

The US group has been particularly well-studied, and can give us some
quantitative insight into the star formation history as well as
constraints on the mechanism responsible for triggering the star
formation.
As described in detail in Sect.~\ref{us_ages}, the populations of the
high-mass as well the low-mass stars both have a common age of 5~Myr.
There is no evidence for a significant age spread, and the data are
consistent with the idea that all stars have formed within a period of
no more than $\sim 1\!-\!2$ Myr.

Another important aspect is the initial configuration of the region at
the time when the stars formed.  Today, the bulk (70\%) of the
Hipparcos members (and thus also the low-mass stars) lie within an
area of 11 degrees diameter on the sky, which implies a characteristic
size of the association of 28~pc.  \Citet{deBruijne99} showed that the
internal 1D velocity dispersion of the Hipparcos members of US is only
1.3~km/s.  In combination with the well determined age of US and the
present-day size, this strikingly small velocity dispersion clearly
shows that US cannot have originated in a compact cluster
configuration that expanded later, but must have been in a spatially
extended configuration from the beginning. US seems to have formed in
an extended, unbound giant molecular cloud, similar to the models
considered by \citet{Clark05}.

The initial size of the association can be estimated by assuming that
the stars expanded freely from their initial positions. With a
Gaussian velocity distribution characterized by the measured velocity
dispersion, a single point in space would have expanded to a size of
about 13~pc in 5~Myr.  Subtracting this number from the current
characteristic size of 28~pc in quadrature leads to an initial size of
25~pc.

This implies an lateral stellar crossing time of 25~pc~/~1.3~km/s
$\sim$~20~Myr in the initial configuration.  This large crossing time
is in remarkable contrast to the upper limit on any possible age
spread among the association members of only $\sim 1\!-\!2$ Myr.  The
fact that the lateral stellar crossing time is much (about an order of
magnitude) larger than the age spread of the association members
clearly suggests that some external agent must have coordinated the
onset of the star formation process over the full spatial extent of
the association. In order to account for the small spread of stellar
ages, the triggering agent must have crossed the initial cloud with a
velocity of at least $\sim 20$ km/s.
Also, some mechanism must have terminated the star formation process
about 1 Myr after it started.  Finally, we note that the US region
does no longer contain significant amounts of molecular cloud
material.  The original molecular cloud, in which the stars formed,
has been nearly completely dispersed.  Today, most of this material
appears to be situated in an expanding, (mostly) atomic H\,{\small I}
superbubble centered on US.  \Citet{deGeus92} estimated the mass of
this superbubble to be $\sim 80\,000\,M_\odot$. Comparing this to the
estimated total mass of all stars in US ($\sim 2060\,M_{\odot}$; see
Sect.~\ref{us_imf}) suggests that only a small fraction of the initial
cloud mass was transformed into stars.

\medskip

These findings can be understood as consequences of
the feedback from massive stars.  High-mass stars, above about ten
solar masses, profoundly affect their environment in several ways.
Their strong ionizing radiation can photo\-evaporate molecular cloud
clumps (e.g.~see Hester et al.~1996 and McCaughrean \& Andersen 2002
for the case of M16) and circumstellar matter around young stellar
objects \citep[see][]{Bally98,Richling98,Richling00}.  Their powerful
stellar winds deposit considerable amounts of momentum and kinetic
energy into the surrounding medium.  Finally, supernova explosions
cause strong shock waves that transfer typically some $10^{51}$~erg of
kinetic energy into the ambient interstellar medium.  The supernova
blast wave will initially expand within the wind-blown bubble formed
by the supernova progenitor; as it catches up with the bubble shock
front, it will accelerate the expansion of the bubble \citep[see
e.g.][]{Oey95}, further disrupt the parental molecular cloud
\citep[see e.g.][]{Yorke89} and sweep up a massive shell of dust and
gas.

In general, massive stars have a very destructive effect on their
nearby environment; they can disrupt molecular clouds very quickly and
therefore prevent further star formation in their surroundings.  At
somewhat larger distances, however, the wind- and supernova-driven
shock waves originating from massive stars can have a constructive
rather than destructive effect by driving molecular cloud cores into
collapse.  Several recent numerical studies
\citep[e.g.][]{Boss95,Foster96,Foster97,Vanhala98,Fukuda00} have found
that the outcome of the impact of a shock wave on a cloud core mainly
depends on the type of the shock and its velocity: In its initial,
adiabatic phase, the shock wave is likely to destroy ambient clouds;
the later, isothermal phase, however, is capable of triggering cloud
collapse if the velocity is in the right range.  Shocks traveling
faster than about 50 km/s shred cloud cores to pieces, while shocks
with velocities slower than about 15 km/s usually cause only a slight
temporary compression of cloud cores.  Shock waves with velocities in
the range of $\sim 15 - 45$ km/s, however, seem to be well able to
induce collapse of molecular cloud cores.  A good source of shock
waves with velocities in that range are
expanding superbubbles driven by the winds and
supernova explosions of massive stars at distances\footnote{In the
immediate vicinity of a supernova,
the shock wave is so strong and fast that it will destroy clouds;
at larger distances, the supernova shock wave will accelerate
the expansion of the (pre-supernova) wind-driven superbubble to
velocities in the suitable range.}
between $\sim 20$~pc and $\sim 100$~pc
\citep[see][]{Oey95}.
Observational evidence for star forming
events triggered by shock waves from massive stars has, for example,
been discussed in \citet{Carpenter00}, \citet{Walborn99},
\citet{Yamaguchi01}, \citet{Efremov98}, \citet{Oey95}, \citet{Oey05},
\citet{Reach04}, \citet{Cannon05}, and \citet{Gorjian04};
see also the discussions in \citet{Elmegreen98} and \citet{PZ07}.

\subsection{A Triggered Star Formation Scenario for Upper Scorpius \label{us_trigger}}

For the star burst in US, a very suitable trigger is the shock-wave of
the expanding superbubble around the UCL group, which is driven by the
winds of the massive stars and several supernova explosions that
started to occur about 12 Myr ago.  The structure and kinematics of
the large H~I loops surrounding the Scorpius-Centaurus association
suggest that this shock wave passed through the former US molecular
cloud just about 5 Myr ago \citep{deGeus92}.  This point in time
agrees very well with the ages found for the low-mass stars as well as
the high-mass stars in US.  Furthermore, since the distance from UCL
to US is about 60~pc, this shock wave probably had just about the
right velocity ($\sim 25$~km/s) that is required to induce star
formation according to the modeling results mentioned above.  Thus,
the assumption that this wind- and supernova-driven shock wave
triggered the star formation process in US provides a self-consistent
explanation of all observational data.

\begin{figure}[!ht]
\centering
\includegraphics[width=0.49\textwidth,draft=False]{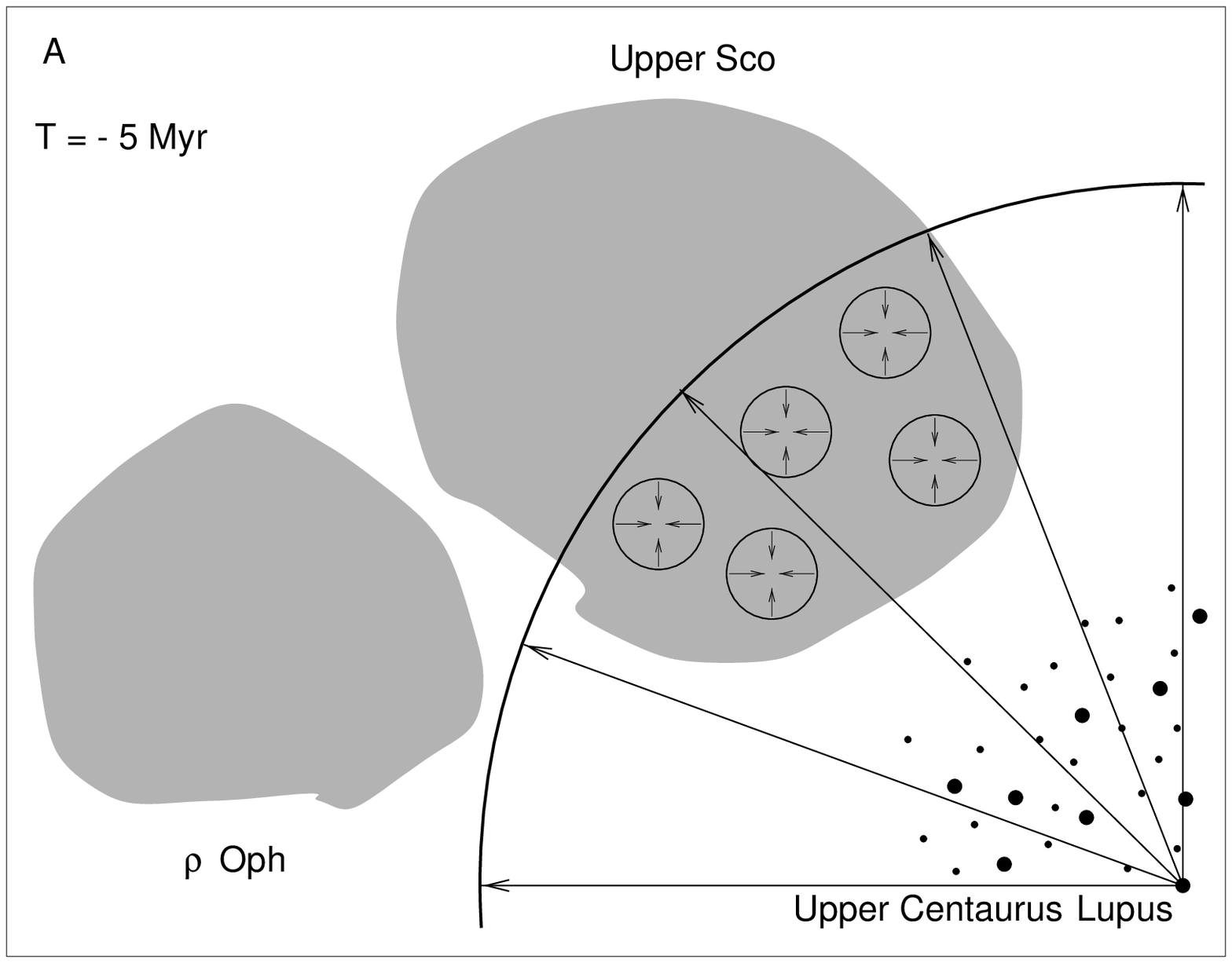}
\includegraphics[width=0.49\textwidth,draft=False]{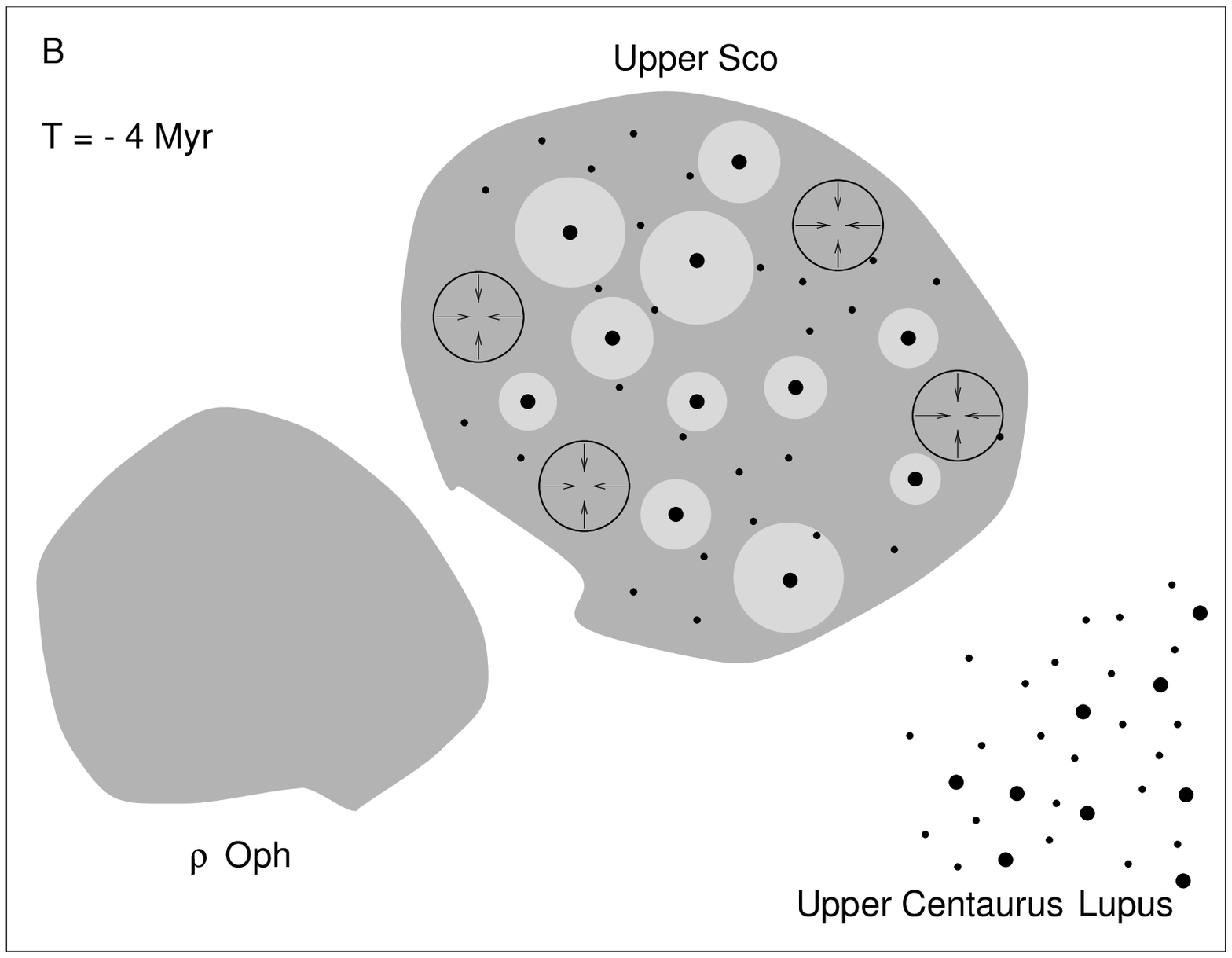}\\
\includegraphics[width=0.49\textwidth,draft=False]{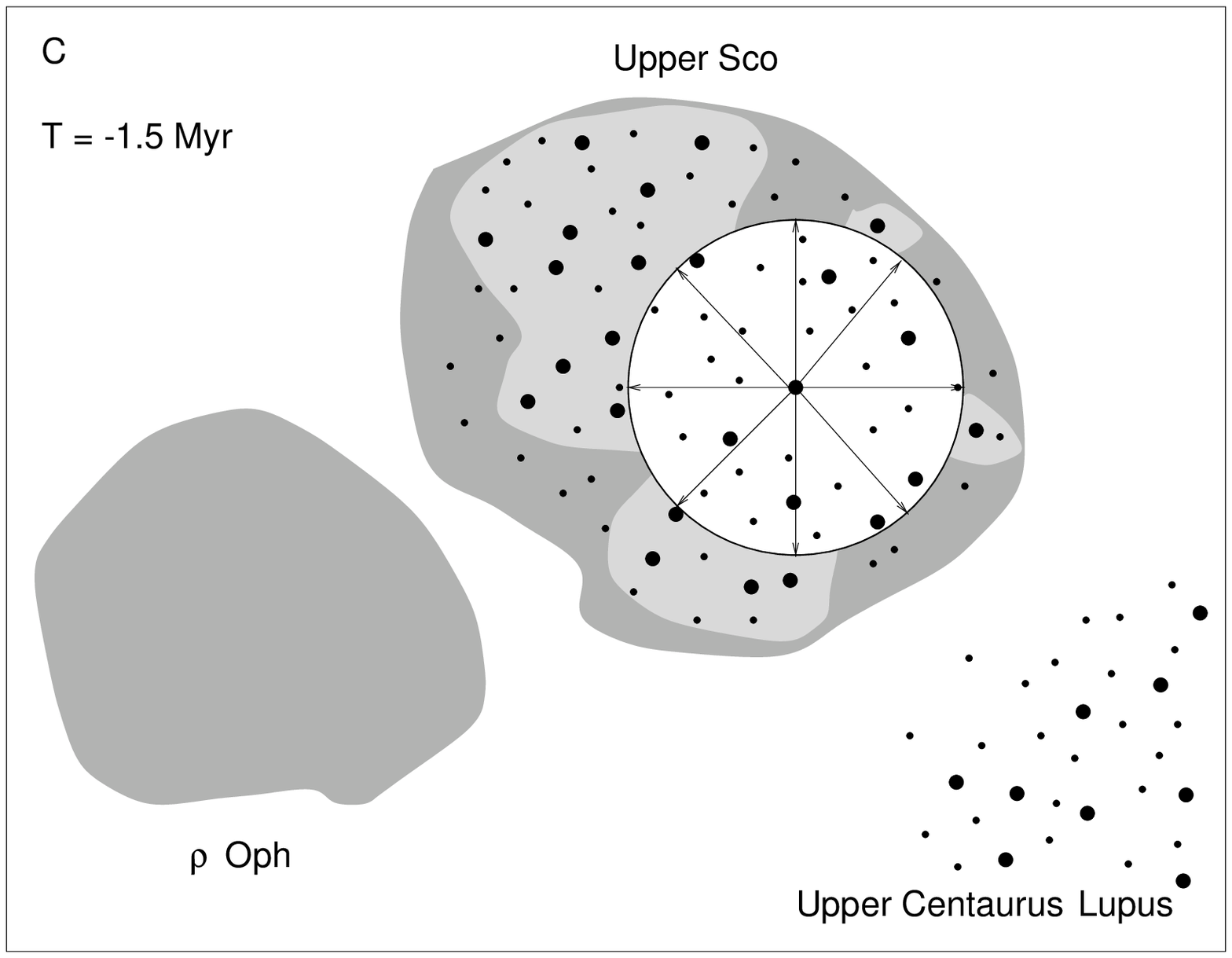}
\includegraphics[width=0.49\textwidth,draft=False]{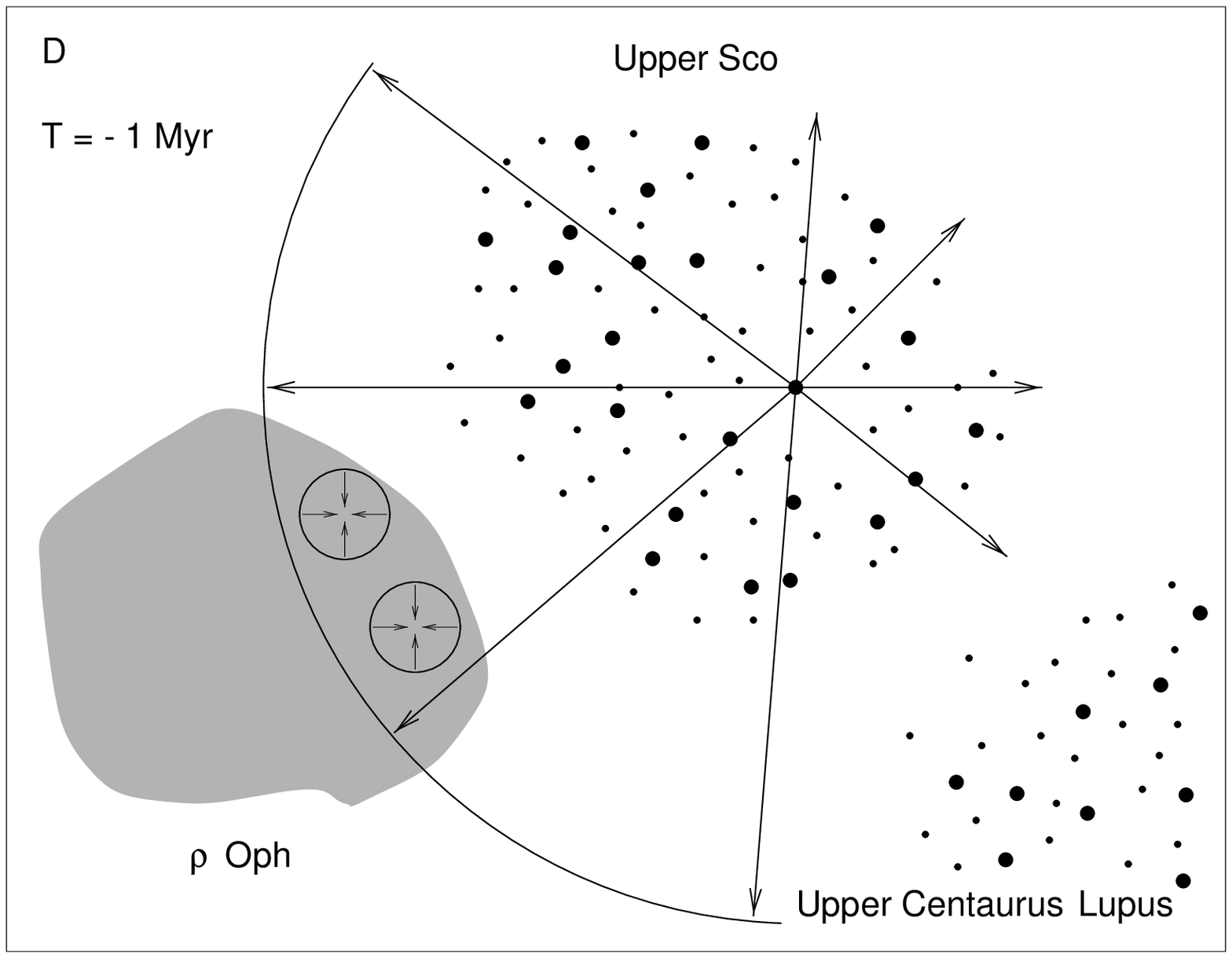}

\caption{Schematic view of the star formation history in the
Scorpius-Centaurus association. Molecular clouds are shown as dark
regions, high-mass and low-mass stars as large resp.~small dots. For
further details on the sequence of events see text.}
\label{sfm}
\end{figure}

A scenario for the star formation history of US consistent with the
observational results described above is shown in Fig.~\ref{sfm}.  The
shock-wave crossing US about 5~Myr ago initiated the formation of some
2500 stars, including 10 massive stars upwards of $10\,M_\odot$.  When
the new-born massive stars `turned on', they immediately started to
destroy the cloud from inside by their ionizing radiation and their
strong winds. This affected the cloud so strongly that after a period
of $\la 1$ Myr the star formation process was terminated, probably
simply because all the remaining dense cloud material was
disrupted. This explains the narrow age distribution and why only
about 3\% of the original cloud mass was transformed into stars.

About 1.5 Myr ago, the most massive star in US, presumably the
progenitor of the pulsar PSR~J1932+1059
\citep[see][]{Hoogerwerf01,Chatterjee04}, exploded as a supernova and
created a strong shock wave, which fully dispersed the US molecular
cloud and removed basically all the remaining diffuse material.

Finally, it is interesting to note that this shock wave must have
crossed the $\rho$ Oph molecular cloud within the last 1 Myr
\citep{deGeus92}.  The strong star formation activity we witness right
now in the L~1688 cloud of the $\rho$ Oph region might therefore be
triggered by this shock wave \citep[see][]{Motte98} and would
represent the third generation of sequential triggered star formation
in the Scorpius-Centaurus-Ophiuchus complex.  Furthermore, we note
that the Lupus 1 \& 2 dark clouds are also located near the edge of
the shell around US.  These two dark clouds (see chapter by Comer\'on
in this book) contain numerous young stellar objects with estimated
ages of $\la 1$~Myr; their recent star formation activity may also
well have been triggered by the passage of the shock related to the
expanding shell around US \citep[see, e.g.,][]{Tachihara01}.
\medskip

\noindent{\em An updated model for triggered cloud and star
formation:}\smallskip

\noindent
While the scenario described above provides a good explanation of the
star formation history, a potential problem is its implicit assumption
that the US and $\rho$ Oph molecular clouds existed for many Myr
without forming stars before the triggering shock waves arrived
(otherwise one should see large age spreads in the stellar
populations, contrary to the observational evidence).
Numerous recent studies \citep[e.g.,][]{Elemegreen00,Hartmann01,Hartmann03,BH07,Elmegreen07}
have provided increasing and convincing evidence that the lifetime
of molecular clouds is much shorter than previously thought, and the
whole process of molecular cloud formation, star formation, and cloud
dispersal (by the feedback of the newly formed stars) occurs on
timescales of only a few $(\la 5$)~Myr.  It is now thought that
molecular clouds form by the interaction of flows in the interstellar
medium that accumulate matter in some regions.  As soon as the column
density gets high enough for self-shielding against the ambient UV
radiation field, the atomic gas transforms into molecular clouds.
This new paradigm for the formation and lifetime of molecular clouds
seems to invalidate the idea of a shock wave that hit a pre-existing
molecular cloud and triggered star formation.  Nevertheless, the basic
scenario for the sequence of processes in Sco-Cen may still be valid.
As pointed out by \citet{Hartmann01}, wind and supernova shock waves
from massive stars are an important kind of driver for ISM flows, and
are especially well suited to create {\em coherent large-scale flows}.
Only large-scale flows are able to form large molecular clouds, in
which whole OB associations can be born.
An updated scenario for Sco-Cen could be as follows: Initially, the
winds of the OB stars created a superbubble around UCL.  The
interaction of this expanding superbubble with flows in the ambient
ISM started to sweep up clouds in some places.  When supernovae
started to explode in UCL (note that there were presumably some 6
supernova explosions in UCL up to today), these added energy and
momentum to the wind-blown superbubble and accelerated its expansion.
The accelerated shock wave (now with $v \sim 20 - 30$~km/sec) crossed
a swept-up cloud in the US area, and the increased pressure due to
this shock triggered star formation.
This scenario does not only explain the temporal sequence of events in
a way consistent with the ages of the stars and the kinematic
properties of the observed HI shells.  The following points provide
further evidence: (1) The model \citep[see Fig.~3 in][]{Hartmann01}
predicts that stellar groups triggered in swept-up clouds should move
away from the trigger source.  A look at the centroid space motions of
the Sco-Cen subgroups \citep{deBruijne99} actually shows that US moves
nearly radially away from UCL with a velocity of $\sim 5
(\pm3)$~km/sec.  Furthermore, (2) a study by \citet{Mamajek01}
revealed that several young stellar groups, including the $\eta$~Cha
cluster, the TW Hydra association, and the young stars associated with
the CrA cloud, move away from UCL at velocities of about 10 km/sec;
tracing their current motions back in time shows that these groups
were located near the edge of UCL 12 Myr ago (when the supernova
exploded). (3) The model also predicts that molecular clouds are most
efficiently created at the intersection of two expanding bubbles.  The
Lupus~1 cloud, which is located just between US and UCL, seems to be a
good example of this process.  Its elongated shape is very well
consistent with the idea that it was swept up by the interaction of
the expanding superbubble around US and the (post-SN) superbubble
created by the winds of the remaining early B stars in UCL.

\subsection{Ages and Star-Formation History of UCL and LCC \label{ucl_lcc}}

In the UCL and LCC subgroups, \citet{Mamajek02} found that the HR
diagram positions of the stars in both groups are consistent with 95\%
of their low-mass star-formation occurring within a $<8\!-\!12$~Myr span.
The {\it upper limit} on the duration of star-formation is relatively
large due to the fact that most of the stars from the
\citet{Mamajek02} survey were on the radiative portion of the PMS
evolutionary tracks (where the isochrones are packed closely
together), and the binarity of these stars is unstudied.  Although
\citet{Mamajek02} did account for the spread in distances of the
subgroup members by using cluster parallaxes, the previously mentioned
effects conspire to make it difficult to be more precise about the
group star-formation history at this time.  The mean ages of the PMS
populations, using the \citet{Baraffe98} tracks (16 Myr for UCL, 18
Myr for LCC), and other sets of models \citep[15-22 Myr for UCL, 17-23
Myr for LCC;][]{D'Antona97,Palla01,Siess00} agree fairly well with the
revised turn-off ages \citep[17 Myr for UCL, 16 Myr for LCC; using
tracks of ][]{Bertelli94}.

We note two age trends in LCC that are worthy of future
investigation. First, there appears to be a mass-dependence on the
derived age for LCC. The turn-off age for the B-stars is $\sim$16 Myr
(but see discussion in Mamajek et al. 2002), while the median ages of
the cool members listed in Table 4 are 17 Myr for the F stars, 17 Myr
for the G stars, 12 Myr for the K stars, and 4 Myr for the M stars.
The 8 M-type members of LCC listed in Table 4 all have isochronal ages
in the range 2-7 Myr. The isochronal age for the M-type members is
inconsistent with the positions of the main sequence turn-on
($\sim$F5) and turn-off points ($\sim$B1). We do not believe that this
is evidence for non-coeval formation for the low mass vs.~high mass
stars, but instead attribute the difference to errors with the
evolutionary tracks and/or incompleteness of the member sample
among the faintest (i.e.~apparently older) low-mass members.

There also appears to be a trend in median age vs.~Galactic latitude
in LCC, which is probably due to substructure. The median age of the
low-mass stars south of the Galactic equator is 12\,$\pm$\,2 Myr,
while the stars north of the equator have a median age of 17\,$\pm$\,1
Myr. The observed dispersion in the ages, characterized by the 68\%
confidence intervals, is $\sim$7 Myr, however much of this is probably
due to observational uncertainties and binarity.  The {\it Hipparcos}
parallax data for the \citet{deZeeuw99} members of LCC show that the
mean distance to the southern part of the group is closer
(109\,$\pm$\,1 pc) than the northern part (123\,$\pm$2
pc). Immediately south of LCC at the same distance ($\sim$110 pc) is
the $\epsilon$ Cha region which formed stars as recently as $\sim$3-6
Myr ago.  Could we be seeing evidence for a north-south axis of
triggered star-formation? In this picture, star-formation would have
started $\sim$17 Myr ago in northern LCC near $\sigma$, $\rho$, and
$\delta$ Cen, progressed southward through the Southern Cross forming
$\alpha$ and $\beta$ Cru and the southern part of LCC some $\sim$12
Myr ago, and progressing further southward to form the $\epsilon$ and
$\eta$ Cha groups some $\sim$6 Myr ago.

In comparison to the more elegant picture of star-formation in the
smaller Upper Sco region, UCL and LCC probably have more complex
histories. The spatial distributions of their members suggest that
they contain substructure (and may warrant further subdivision), and
are inconsistent with being the evaporating, expanding remnants of
{\it two} massive embedded clusters.  Instead, we suspect that the
bulk of star-formation in UCL and LCC proceeded $\sim$10-20 Myr ago in
a series of multiple embedded clusters and filaments, containing tens
to hundreds of stars each, in a dynamically unbound giant molecular
cloud complex.

\subsection{Stellar Multiplicity}

Members of the Sco-Cen association have been the target of several
studies concerning the multiplicity of the stars.  \citet{Brandner98}
and \citet{Kohler00} carried out K-band speckle observations of more
than 100 X-ray-selected weak-line T Tauri stars in Sco-Cen.  The
derived multiplicity of the stars they observed is $\sim 1.6$ times
higher than typical for main-sequence stars; almost all of the
brighter M-type PMS stars are binaries.  \citet{Shatsky02} examined
115 B-type stars in Sco-Cen for the existence of visual companions in
the near-infrared.  They derived a $\sim 1.6$ times higher binary
fraction than found for low-mass dwarfs in the solar neighborhood and
in open clusters in the same separation range.

\citet[][2007]{Kouwenhoven05} performed near-infrared adaptive
optics surveys of 199 A-type and late B-type Hipparcos members of
Sco-Cen with the aim to detect close companions.  They found 176
visual companions among these stars and conclude that 80 of these are
likely physical companion stars.  The estimated masses of these
companions are in the range from $0.03\,M_\odot$ to $1.2\,M_\odot$.
The comparison of their results with visual, spectroscopic, and
astrometric data on binarity in Sco-Cen suggests that each system
contains on average at least 0.5 companions; this number is a strict
lower limit due to the likely presence of further, yet undiscovered
companion stars.

\citet{Kraus05} performed a high-resolution imaging survey of 12 brown
dwarfs and very low-mass stars in US with the Advanced Camera for
Surveys on HST. This survey discovered three new binary systems, and
lead to an estimated binary fraction of $33\%\pm 17\%$, which is
consistent with that inferred for higher mass stars in US.
\citet{Costado05} searched for planetary-mass objects and brown dwarfs
around nine low-mass members of US, but found no convincing cases of
such companions.
\citet{Luhman05} discovered a wide ($\sim 140$~AU), low-mass binary
system in US with estimated masses of $\sim 0.15\,M_\odot$ for both
components.  He concluded that this new system further establishes
that the formation of low-mass stars and brown dwarfs does not require
ejection from multiple systems and that wide, low-mass binaries can
form in OB associations as well as in smaller clusters.
\citet{Bouy06} performed an adaptive-optics imaging survey for
multiple systems among 58 M-type members of US and resolved nine pairs
with separations below $1''$. One of these systems is probably a young
brown dwarf binary system.

\subsection{Modeling of Binaries}

There appear to be several Sco-Cen objects which may be of interest in
the broader context of testing stellar evolution theory, as well as
refining estimates for the age and metallicity of the subgroups.  The
most interesting objects in this regard are well-characterized binary
systems (mainly eclipsing systems), and low-mass stars which have
demonstrated Li depletion.  UCL has at least two eclipsing binary
systems: $\mu^1$ Sco and GG Lup.  The eclipsing binary $\mu^1$ Sco is
a well-studied massive binary
\citep[12.8\,M$_{\odot}$+ 8.4\,M$_{\odot}$;
B1.5V+B3;][]{Schneider79,Giannuzzi83,Stickland96}, and has been
recognized as a Sco-Cen member for many decades.  We will briefly
discuss some low-mass binaries in more detail.

Although the eclipsing binary GG Lup \citep[B7V+B9V;][]{Andersen93}
was assigned UCL membership by \citet{deZeeuw99}, its membership has
not been recognized in the binary modeling literature.  Within
0.5$^{\circ}$ of GG Lup, and having nearly identical proper motions
(also suggestive of UCL membership), are the {\it ROSAT} T Tauri stars
TTS 26 and 27 \citep[G8 and K0 types;][]{Krautter97}, the A8/9V star
HD 135814, and the MS turn-off star $\delta$ Lup (B2IV).  By comparing
the parameters of GG Lup to the evolutionary tracks of
\citet{Claret92}, \citet{Andersen93} claim that the system must have a
somewhat sub-solar metallicity (Z $\simeq$ 0.15), with an age of
$\sim$20\,Myr.  \citet{Pols97} derived ages of 15\,$\pm$\,6~Myr (no
convective overshoot) and 17$^{+7}_{-5}$~Myr (with overshoot) for the
system, however work by \citet{Lastennet02} was only able to constrain
the age to be $<$50\,Myr, with a wide range of possible metallicities
(Z $\in$ 0.014-0.037).  The agreement between the model ages for GG
Lup with the quoted mean age of the low-mass PMS UCL members
\citep[15--22~Myr;][]{Mamajek02}, and recent estimates of the UCL
turn-off age \citep[$\sim$17~Myr;][]{Mamajek02}, are surprisingly
good.  Further observations and modeling of GG Lup and its UCL
neighbors could further improve our knowledge of the age and
metallicity of UCL.

Another well-constrained binary, however non-eclipsing, is AK Sco
\citep{Andersen89}. \citet{Alencar03} use the {\it lack} of observed
eclipses for this equal mass system (two F5 stars), along with the
{\it Hipparcos} parallax ($\varpi$ = 6.89\,$\pm$\,1.44\,mas), to
constrain the masses of the components (1.35\,$\pm$\,0.07
$M_{\odot}$). The HR-D positions for the AK Sco A and B, individually,
correspond to (log\,T$_{\rm eff}$, log\,L/L$_{\odot}$ = 3.807,
0.76). Using the evolutionary tracks of \citet{Palla01}, this
corresponds to an age of 11~Myr and 1.5\,M$_{\odot}$. The mass is some
2$\sigma$ higher than Alencar et al.'s dynamic estimate, but still
statistically consistent.  All the more remarkable, AK Sco
demonstrates spectroscopic and photometric evidence for a
circumstellar accretion disk.  AK Sco, along with PDS 66, appear to be
rare examples of stars which can retain accretion disks for
$>$10\,Myr.

\citet{Reiners05} reported the discovery of a low-mass spectroscopic
binary, UScoCTIO~5 \citep[spectral type M4;][]{Ardila00}, in US.  The
lower limit to the system mass derived from the orbit fit is higher
than the mass suggested by PMS models based on the luminosity and
effective temperature, suggesting either problems of the theoretical
PMS evolutionary models or uncertainties in the empirical spectral
type and/or temperature determination.

\subsection{Circumstellar Disk Evolution}

Another important aspect is how circumstellar disks evolve and whether
their evolution is affected by environmental conditions.  Some
interesting results on the time scales for mass accretion and disk
dissipation have been obtained for Sco-Cen members. First, it was
found that the fraction of accreting stars (as traced by strong
H$\alpha$ emission) is $\sim 10\%$ in US \citep{Preibisch01}, but
$<$few\% in UCL and LCC \citep{Mamajek02}.  A recent {\it Spitzer}
study by \citet{Carpenter06} of US members across the whole stellar
mass spectrum has found that circumstellar disk evolution appears to
be very mass dependent.  Roughly 20\% of the $<1.2\,M_{\odot}$
(K/M-type) members of US were found to have infrared SEDs consistent
with primordial accretion disks, whereas {\it none} of the $\sim
1.2\!-\!1.8\,M_{\odot}$ (F/G-type) stars, and only one (of 61) of the
massive A/B-type, showed signs of having accretion disks. A small
fraction ($\sim$20\%) of the A/B-type massive stars did show evidence
for small excesses at 16\,$\mu$m, which could be due to cold, dusty
debris disks. A {\it Spitzer} study by \citet{Chen05} investigated
infrared excesses tracing dust at $\sim$~3--40~AU from F/G-type stars
from \citet{deZeeuw99}.  The MIPS observations of 40 F- and G-type
proper motion members of Sco-Cen detected 14 objects that possess
$24\,\mu$m fluxes $\ge30\%$ larger than their predicted photospheres,
tentatively corresponding to a disk fraction of $\ge35\%$, including
seven objects that also possess $70\,\mu$m excesses $\ge100$ times
larger than their predicted photospheres.  20\% of the targets in US,
9\% in UCL, and 46\% in LCC display $24\,\mu$m excesses.

\subsection{Lithium Depletion}

Another critical area of stellar evolution theory which studies of
Sco-Cen membership could address is Li depletion.  Li depletion in
low-mass stars is predicted to take place the fastest for stars of
mass $\sim$0.6\,M$_{\odot}$ \citep{Chabrier97}, which corresponds to
spectral type $\sim$M2 for $\sim$10-20\,Myr-old objects.  Only a small
sample of early-M-type stars are known in the UCL and LCC subgroups,
primarily among the Krautter-Wichmann and Park-Finley {\it ROSAT} TTS
samples.  Despite the evidence for a short duration of star-formation
in US \citep[e.g.][]{Preibisch02}, \citet{Palla05} have recently
suggested that the spread in Li abundances among the low-mass members
of US may be interpreted as evidence for a significant age spread. A
high-resolution spectroscopic study of low-mass Sco-Cen members may be
able to address the following questions: Is there a demonstrable
spread in Li abundances for stars of a given mass? Does the spread
correlate with any fundamental parameters (i.e.~rotation, binarity,
etc.)? Is there evidence for a spread in ages amongst the low-mass
members? Larger samples of low-mass Sco-Cen members are sorely needed
to address these topics.

\section{The Future \label{future}}

If the IMF predictions are any indication, then the surface of the
census of low-mass Sco-Cen members has barely been scratched.  Due to
its intermediate age, between that of well-studied T associations and
nearby ZAMS clusters, as well as its close proximity, the nearest OB
association promises to be an important region for studying
star-formation up close, and understanding of the PMS evolution of
low-mass stars and their circumstellar disks. The Sco-Cen region is a
promising hunting ground for future ground-based, high-cadence,
photometric and astrometric survey facilities (e.g., Pan-STARRS,
SkyMapper) as well as the GAIA astrometric mission. By combining the
photometric and astrometric capabilities of these facilities with a
focused spectroscopic survey, it is possible that we could construct
an accurate 7-dimensional picture (position, velocity, age) of
the complete stellar census of our nearest ``starburst'' within the
next decade or two.

\vspace{10mm}
\acknowledgements

We dedicate this review to Adriaan Blaauw, whose Ph.D. thesis 60 years
ago, and series of influential papers over the past six decades, laid
the foundation for much of our understanding of Sco-Cen, and the other
OB associations.  We are grateful to Hans Zinnecker and Ronnie
Hoogerwerf for valuable comments on the manuscript, and Tim de Zeeuw
for providing a figure. Eric Mamajek is supported by a Clay
Postdoctoral Fellowship from the Smithsonian Astrophysical
Observatory.  Thomas Preibisch would like to thank Hans Zinnecker for
many years of motivation and advice in studying Upper Scorpius.  This
work made extensive use of NASA's Astrophysics Data System
Bibliographic Services and the SIMBAD database (CDS, Strasbourg,
France).

\clearpage

{ \footnotesize
\begin{longtable}{@{}l@{\hskip11pt}l@{\hskip11pt}r@{\hskip11pt}r@{\hskip11pt}r@{\hskip11pt}r@{\hskip11pt}r@{\hskip11pt}r@{\hskip11pt}r@{\hskip11pt}r@{}}

\caption[]{PMS stars in Upper Sco used in the study of Preibisch \& Zinnecker (1999).
 The source name in the first column consists of the
 J2000 coordinates of the star.
 The second column lists the spectral type, the third column tells
 whether the star was detected in the ROSAT All Sky Survey
  (``R'') and whether it was selected as a
 candidate for a PMS star (``C'') or as a non-PMS candidate (``nC'').
 The other colums give photometric data and stellar parameters.
}\\
  \hline
 USco   &  SpT &RASS&  $I$ &$R-I$&$A_V$&$\log T_{\rm eff}$&
 $\log \frac{\displaystyle L}{\displaystyle L_\odot}$&$M_\star$& age\rule[-2mm]{0mm}{7mm} \\
        &      &    & [mag]&[mag]&[mag]&[K]&   &[$M_\odot$]& [Myr]\rule[-2mm]{0mm}{5mm}\\\hline
\endfirsthead

\caption[]{$-$ continued}\\
  \hline
 USco   &  SpT &RASS& $I$ &$R-I$&$A_V$&$\log T_{\rm eff}$&
 $\log \frac{\displaystyle L}{\displaystyle L_\odot}$&$M_\star$& age\rule[-2mm]{0mm}{7mm} \\
        &      &    & [mag]&[mag]&[mag]&[K]&   &[$M_\odot$]& [Myr]\rule[-2mm]{0mm}{5mm}\\\hline
\endhead

 \hline
 \endfoot

 \hline
 \endlastfoot

 153557.8$-$232405& K3: &RC & 10.88& 0.86&1.4 & 3.649& $-0.118$&0.90& 3.0 \\
 154106.7$-$265626& G7  &RC &  9.98& 0.77&1.6 & 3.701& $ 0.258$&1.42& 3.0 \\
 154413.4$-$252258& M1  &RC & 11.33& 1.13&0.6 & 3.564& $-0.431$&0.38& 1.0 \\
 154920.9$-$260005& K0  &RC &  9.71& 0.64&0.7 & 3.676& $ 0.208$&1.10& 1.8 \\
 155106.6$-$240218& M2  &RC & 12.08& 1.19&0.4 & 3.551& $-0.759$&0.35& 2.5 \\
 155231.2$-$263351& M0  &RC & 10.88& 0.75&0.0 & 3.576& $-0.372$&0.43& 1.0 \\
 155459.9$-$234718& G2  &RnC&  8.14& 0.41&0.2 & 3.738& $ 0.722$&1.82& 2.8 \\
 155506.2$-$252109& M1  &RnC& 10.57& 0.86&0.0 & 3.564& $-0.241$&0.33& 0.6 \\
 155517.1$-$232216& M2.5&   & 12.17& 1.11&0.0 & 3.545& $-0.863$&0.30& 2.9 \\
 155548.7$-$251223& G3  &RC &  9.44& 0.42&0.2 & 3.731& $ 0.200$&1.34& 9.0 \\
 155702.3$-$195042& K7: &RC & 10.24& 0.77&0.2 & 3.609& $-0.080$&0.50& 0.8 \\
 155716.6$-$252918& M0  &   & 10.94& 0.85&0.0 & 3.576& $-0.396$&0.44& 1.2 \\
 155720.0$-$233849& M0  &   & 10.95& 0.92&0.0 & 3.576& $-0.397$&0.44& 1.2 \\
 155725.8$-$235422& M0.5&   & 10.94& 1.16&0.9 & 3.570& $-0.210$&0.34& 0.5 \\
 155734.4$-$232111& M1  &RC & 11.37& 1.18&0.8 & 3.564& $-0.401$&0.38& 0.9 \\
 155750.0$-$230508& M0  &RC & 11.46& 0.95&0.1 & 3.576& $-0.574$&0.50& 2.9 \\
 155812.7$-$232835& G2  &RnC&  9.25& 0.47&0.4 & 3.738& $ 0.333$&1.47& 8.0 \\
 155847.8$-$175800& K3  &RC & 10.46& 0.94&1.8 & 3.649& $ 0.123$&0.78& 1.0 \\
 155902.1$-$184414& K6  &RC & 10.40& 0.94&1.3 & 3.620& $ 0.053$&0.53& 0.7 \\
 155950.0$-$255557& M2  &RC & 11.89& 1.31&1.0 & 3.551& $-0.573$&0.35& 1.5 \\
 160000.0$-$222037& M1  &   & 11.08& 1.13&0.6 & 3.564& $-0.331$&0.36& 0.8 \\
 160000.7$-$250941& G0  &RnC&  9.69& 0.36&0.0 & 3.750& $ 0.080$&1.10&18.0 \\
 160013.3$-$241810& M0  &   & 11.68& 1.04&0.6 & 3.576& $-0.580$&0.50& 3.0 \\
 160031.3$-$202705& M1  &   & 11.24& 1.22&1.0 & 3.564& $-0.313$&0.35& 0.7 \\
 160040.6$-$220032& G9  &RC &  9.95& 0.56&0.5 & 3.685& $ 0.053$&1.22& 4.0 \\
 160042.8$-$212737& K7  &   & 10.97& 0.88&0.8 & 3.609& $-0.272$&0.60& 1.8 \\
 160105.2$-$222731& M3  &   & 11.27& 1.33&0.5 & 3.538& $-0.398$&0.27& 0.5 \\
 160108.0$-$211318& M0  &RC & 10.86& 0.86&0.0 & 3.576& $-0.364$&0.42& 1.0 \\
 160125.7$-$224040& K1  &RC & 10.16& 0.63&0.6 & 3.667& $ 0.002$&1.05& 3.0 \\
 160147.4$-$204945& M0  &RnC& 10.92& 1.08&0.8 & 3.576& $-0.239$&0.38& 0.8 \\
 160151.4$-$244524& K7  &RC & 10.54& 0.84&0.6 & 3.609& $-0.136$&0.53& 1.0 \\
 160158.2$-$200811& G5  &RC &  9.29& 0.58&0.8 & 3.717& $ 0.383$&1.60& 3.3 \\
 160200.3$-$222123& M1  &RC & 11.23& 1.23&1.1 & 3.564& $-0.300$&0.35& 0.8 \\
 160208.5$-$225457& M1  &   & 11.91& 1.13&0.6 & 3.564& $-0.663$&0.43& 2.8 \\
 160210.4$-$224128& K5  &RC &  9.89& 0.71&0.4 & 3.630& $ 0.083$&0.60& 0.7 \\
 160239.1$-$254208& K7  &RC & 10.80& 0.73&0.0 & 3.609& $-0.341$&0.65& 2.7 \\
 160251.2$-$240156& K4  &RC & 10.82& 0.71&0.6 & 3.639& $-0.257$&0.90& 4.0 \\
 160253.9$-$202248& K7  &RC & 10.58& 0.96&1.1 & 3.609& $-0.043$&0.50& 0.8 \\
 160302.7$-$180605& K4  &RC & 10.78& 0.83&1.1 & 3.639& $-0.132$&0.80& 2.2 \\
 160323.7$-$175142& M2  &RC & 11.06& 1.25&0.7 & 3.551& $-0.296$&0.30& 0.4 \\
 160354.9$-$203137& M0  &RC & 11.10& 1.10&0.9 & 3.576& $-0.293$&0.40& 0.9 \\
 160357.6$-$203105& K5  &RC & 10.92& 0.81&0.9 & 3.630& $-0.237$&0.78& 2.9 \\
 160420.9$-$213042& M2  &RC & 12.13& 1.57&2.2 & 3.551& $-0.431$&0.33& 0.9 \\

 160421.7$-$213028& K2  &RC & 10.64& 0.73&1.0 & 3.658& $-0.118$&1.00& 3.7 \\
 160447.7$-$193023& K2  &RC &  9.82& 0.65&0.6 & 3.658& $ 0.137$&0.90& 1.2 \\
 160527.3$-$193846& M1  &   & 11.99& 1.07&0.3 & 3.564& $-0.750$&0.44& 3.5 \\
 160539.1$-$215230& M1  &   & 11.93& 1.18&0.8 & 3.564& $-0.625$&0.43& 2.5 \\
 160542.7$-$200415& M2  &RC & 11.76& 1.34&1.1 & 3.551& $-0.494$&0.34& 1.0 \\
 160550.5$-$253313& G7  &RC &  9.99& 0.47&0.2 & 3.701& $-0.019$&1.20& 8.5 \\
 160612.5$-$203647& K5  &RC & 11.27& 1.00&1.8 & 3.630& $-0.203$&0.75& 2.5 \\
 160621.9$-$192844& M0.5&RC & 11.33& 1.09&0.6 & 3.570& $-0.430$&0.41& 1.1 \\
 160637.4$-$210840& M1  &RC & 11.67& 1.33&1.5 & 3.564& $-0.384$&0.37& 0.9 \\
 160639.9$-$200128& M3  &   & 12.79& 1.43&1.0 & 3.538& $-0.915$&0.27& 2.8 \\
 160654.4$-$241610& M3  &RC & 10.87& 1.08&0.0 & 3.538& $-0.335$&1.05& 3.0 \\
 160703.5$-$203626& M0  &RC & 10.22& 0.90&0.0 & 3.576& $-0.108$&0.33& 0.5 \\
 160703.6$-$204308& M1  &   & 12.15& 1.22&1.0 & 3.564& $-0.677$&0.43& 2.8 \\
 160703.9$-$191132& M1  &RC & 11.81& 1.23&1.1 & 3.564& $-0.532$&0.40& 1.7 \\
 160814.7$-$190833& K2  &RC & 10.28& 0.87&1.6 & 3.658& $ 0.154$&0.87& 1.2 \\
 160831.4$-$180241& M0  &RC & 11.26& 0.93&0.1 & 3.576& $-0.512$&0.47& 2.0 \\
 160843.4$-$260216& G7  &RC &  9.30& 0.71&1.3 & 3.701& $ 0.476$&1.50& 1.8 \\
 160856.7$-$203346& K5  &RC & 10.86& 0.92&1.4 & 3.630& $-0.113$&0.70& 1.8 \\
 160930.3$-$210459& M0  &RC & 11.14& 0.81&0.0 & 3.576& $-0.476$&0.47& 2.0 \\
 160941.0$-$221759& M0  &RC & 11.06& 0.84&0.0 & 3.576& $-0.444$&0.45& 1.8 \\
 161019.1$-$250230& M1  &RC &  9.76& 1.01&0.0 & 3.564& $ 0.087$&0.28& 0.1 \\
 161021.7$-$190406& M1  &RnC& 12.26& 1.45&2.1 & 3.564& $-0.511$&0.40& 1.8 \\
 161028.5$-$190446& M3  &   & 11.42& 1.27&0.2 & 3.538& $-0.513$&0.28& 0.9 \\
 161042.0$-$210132& K5  &RC & 10.74& 0.94&1.5 & 3.630& $-0.046$&0.66& 1.1 \\
 161108.9$-$190446& K2  &RC & 10.17& 0.89&1.7 & 3.658& $ 0.216$&0.84& 1.0 \\
 161120.6$-$182054& K5  &RnC& 10.73& 0.83&1.0 & 3.630& $-0.143$&0.70& 1.8 \\
 161156.3$-$230404& M1  &RC & 10.91& 1.18&0.8 & 3.564& $-0.217$&0.32& 0.5 \\
 161159.2$-$190652& K0  &RC & 10.18& 0.75&1.3 & 3.676& $ 0.120$&1.12& 2.5 \\
 161220.9$-$190903& M2.5&   & 12.68& 1.36&0.9 & 3.545& $-0.888$&0.30& 3.0 \\
 161240.5$-$185927& K0  &RC &  9.38& 0.72&1.1 & 3.676& $ 0.413$&1.01& 0.8 \\
 161302.7$-$225744& K4  &RC & 10.38& 1.08&2.3 & 3.639& $ 0.257$&0.64& 0.5 \\
 161318.6$-$221248& G9  &RC &  9.25& 0.66&0.9 & 3.685& $ 0.425$&1.18& 1.0 \\
 161329.3$-$231106& K1  &RC & 10.29& 0.92&2.0 & 3.667& $ 0.215$&0.95& 1.2 \\
 161402.1$-$230101& G4  &RC & 10.06& 0.81&2.0 & 3.724& $ 0.297$&1.50& 5.0 \\
 161411.0$-$230536& K0  &RC &  9.19& 0.99&2.4 & 3.676& $ 0.736$&1.02& 0.3 \\
 161459.2$-$275023& G5  &RC & 10.11& 0.61&1.0 & 3.717& $ 0.083$&1.28& 9.0 \\
 161534.6$-$224241& M1  &RC & 10.38& 1.06&0.3 & 3.564& $-0.115$&0.29& 0.3 \\
 161618.0$-$233947& G7  &RC &  9.55& 0.67&1.1 & 3.701& $ 0.339$&1.47& 2.2 \\
 161731.4$-$230334& G0  &RC &  9.14& 0.47&0.5 & 3.750& $ 0.400$&1.43& 9.0 \\
 161933.9$-$222828& K0  &RC & 10.09& 0.75&1.3 & 3.676& $ 0.156$&1.10& 2.0 \\
 162046.0$-$234820& K3  &RC & 10.70& 1.03&2.2 & 3.649& $ 0.109$&0.78& 1.0 \\
 162307.8$-$230059& K2  &RC & 10.18& 0.69&0.8 & 3.658& $ 0.029$&0.93& 2.0 \\
 162948.6$-$215211& K0  &RC &  9.77& 0.71&1.1 & 3.676& $ 0.248$&1.08& 1.8 \\
 \label{PZ99tab}
 \end{longtable}
}
\normalsize

{\footnotesize

\begin{longtable}{@{}l@{\hskip15pt}l@{\hskip15pt}l@{\hskip15pt}r@{\hskip15pt}r@{\hskip15pt}r@{\hskip15pt}r@{\hskip15pt}r@{}}

  \caption{PMS stars in Upper Sco detected in the 2dF survey of Preibisch et~al.~(2002).
The source name in the first column consists
of the J2000 coordinates of the star. The following columns give the $R$-band
magnitude, the equivalent width of the 6708\,\AA\, Li line and the
${\rm H\alpha}$ line, the spectral type,
the extinction, the adopted effective temperature and the bolometric luminosity.
}\\
  \hline
USco-&$R$ &$W({\rm Li})$&$W({\rm H\alpha})$&SpT&$A_V$&$\log T_{\rm eff}$&$\log
\frac{\displaystyle L}{\displaystyle L_\odot}$ \\
&[mag]&[\AA]&[\AA]&&[mag]&[K]& \\ \hline
\endfirsthead

\caption[]{ -- continued}\\
  \hline
USco-&$R$ &$W({\rm Li})$&$W({\rm H\alpha})$&SpT&$A_V$&$\log T_{\rm eff}$&$\log
\frac{\displaystyle L}{\displaystyle L_\odot}$ \\
&[mag]&[\AA]&[\AA]&&[mag]&[K]& \\ \hline
\endhead

 \hline
 \endfoot

 \hline
 \endlastfoot

 155532.4-230817 &    16.8 &    0.15 &       0.5 &    M1 &     5.7 &    3.569 &    -0.75\\
 155624.8-222555 &    15.5 &    0.78 &      -5.4 &    M4 &     1.7 &    3.516 &    -1.12\\
 155625.7-224027 &    14.6 &    0.53 &      -4.2 &    M3 &     1.8 &    3.532 &    -0.85\\
 155629.5-225657 &    14.3 &    0.54 &      -3.1 &    M3 &     0.9 &    3.533 &    -0.98\\
 155655.5-225839 &    13.1 &    0.54 &      -1.9 &    M0 &     0.7 &    3.578 &    -0.69\\
 155706.4-220606 &    16.2 &    0.52 &      -3.6 &    M4 &     2.0 &    3.513 &    -1.31\\
 155728.5-221904 &    16.7 &    0.40 &      -6.2 &    M5 &     2.3 &    3.505 &    -1.37\\
 155729.2-221523 &    17.1 &    0.29 &      -4.0 &    M5 &     2.1 &    3.503 &    -1.58\\
 155729.9-225843 &    15.8 &    0.55 &      -7.0 &    M4 &     1.4 &    3.511 &    -1.30\\
 155737.2-224524 &    16.7 &    0.15 &       0.0 &    M2 &     1.6 &    3.544 &    -1.76\\
 155742.5-222605 &    15.7 &    0.15 &      -2.9 &    M3 &     1.8 &    3.526 &    -1.22\\
 155744.9-222351 &    14.3 &    0.13 &       0.9 &    M2 &     0.7 &    3.558 &    -1.11\\
 155746.6-222919 &    14.7 &    0.69 &      -2.1 &    M3 &     1.3 &    3.533 &    -1.01\\
 155829.8-231007 &    15.4 &    0.41 &    -250 &    M3 &     0.0 &    3.532 &    -1.63\\
 155848.6-224657 &    14.5 &    0.19 &      -0.5 &    M0 &     0.7 &    3.579 &    -1.24\\
 155912.5-223650 &    16.4 &    0.89 &     -10.5 &    M5 &     0.3 &    3.493 &    -1.70\\
 155918.4-221042 &    14.7 &    0.98 &      -1.0 &    M4 &     1.3 &    3.515 &    -0.90\\
 155925.9-230508 &    17.0 &    0.65 &     -19.5 &    M6 &     1.5 &    3.485 &    -1.57\\
 155930.1-225125 &    17.6 &    0.70 &      -5.0 &    M4 &     2.5 &    3.511 &    -1.71\\
 160004.3-223014 &    14.9 &    0.20 &      -0.4 &    M3 &     0.2 &    3.528 &    -1.40\\
 160007.2-222406 &    16.8 &    0.27 &      -4.4 &    M4 &     0.6 &    3.509 &    -1.91\\
 160017.4-221810 &    16.5 &    0.50 &      -3.9 &    M6 &     0.0 &    3.481 &    -1.79\\
 160018.4-223011 &    14.7 &    0.30 &    -150 &    M3 &     1.5 &    3.539 &    -0.97\\
 160026.3-225941 &    16.8 &    1.00 &     -10.0 &    M5 &     2.5 &    3.495 &    -1.25\\
 160028.5-220922 &    16.8 &    0.65 &       0.0 &    M6 &     0.0 &    3.476 &    -1.93\\
 160030.2-233445 &    17.2 &    0.50 &     -15.8 &    M6 &     1.0 &    3.476 &    -1.71\\
 160054.5-224908 &    14.7 &    0.10 &       0.0 &    M3 &     1.0 &    3.530 &    -1.09\\
 160106.0-221524 &    15.9 &    0.80 &      -8.0 &    M5 &     1.0 &    3.501 &    -1.38\\
 160110.4-222227 &    14.2 &    0.80 &      -9.4 &    M4 &     1.0 &    3.513 &    -0.78\\
 160121.5-223726 &    14.2 &    0.80 &      -8.4 &    M4 &     0.4 &    3.507 &    -0.92\\
 160129.8-224838 &    14.9 &    0.80 &      -4.4 &    M4 &     1.6 &    3.515 &    -0.90\\
 160132.9-224231 &    14.6 &    0.20 &      -1.4 &    M0 &     0.9 &    3.580 &    -1.22\\
 160140.8-225810 &    14.0 &    0.45 &    -120 &    M3 &     0.0 &    3.528 &    -1.16\\
 160142.6-222923 &    13.5 &    0.33 &      -0.6 &    M0 &     0.7 &    3.580 &    -0.87\\
 160158.9-224036 &    14.4 &    0.80 &      -7.6 &    M4 &     0.7 &    3.515 &    -0.97\\
 160159.7-195219 &    16.3 &    0.47 &      -2.7 &    M5 &     2.0 &    3.497 &    -1.22\\
 160202.9-223613 &    14.7 &    0.18 &      -1.5 &    M0 &     1.6 &    3.575 &    -1.08\\
 160207.5-225746 &    13.7 &    0.30 &      -3.2 &    M1 &     0.9 &    3.565 &    -0.81\\
 160210.9-200749 &    16.4 &    0.65 &      -3.5 &    M5 &     1.6 &    3.501 &    -1.40\\
 160222.4-195653 &    14.7 &    0.40 &      -6.6 &    M3 &     1.6 &    3.533 &    -0.95\\
 160226.2-200241 &    15.2 &    0.47 &      -4.9 &    M5 &     0.0 &    3.495 &    -1.39\\
 160236.2-191732 &    17.3 &    0.40 &       0.7 &    M3 &     2.5 &    3.539 &    -1.74\\
 160245.4-194604 &    15.9 &    0.30 &       0.7 &    M2 &     1.4 &    3.550 &    -1.54\\
 160245.4-193037 &    16.4 &    0.51 &      -1.1 &    M5 &     1.8 &    3.495 &    -1.31\\
 160245.7-230450 &    16.7 &    1.00 &    -100 &    M6 &     1.5 &    3.485 &    -1.46\\
 160258.5-225649 &    13.9 &    1.00 &      -9.4 &    M2 &     0.8 &    3.546 &    -0.88\\
 160325.6-194438 &    15.4 &    0.20 &      -1.4 &    M2 &     1.6 &    3.544 &    -1.22\\
 160329.4-195503 &    15.7 &    0.40 &      -4.9 &    M5 &     1.0 &    3.503 &    -1.30\\
 160341.8-200557 &    13.5 &    0.30 &      -2.2 &    M2 &     0.9 &    3.555 &    -0.70\\
 160343.3-201531 &    13.6 &    0.49 &      -3.5 &    M2 &     0.9 &    3.558 &    -0.76\\
 160350.4-194121 &    15.0 &    0.59 &      -4.9 &    M5 &     0.6 &    3.505 &    -1.17\\
 160357.9-194210 &    14.5 &    0.35 &      -3.0 &    M2 &     1.7 &    3.546 &    -0.87\\
 160407.7-194857 &    16.8 &    0.22 &      -4.0 &    M5 &     0.7 &    3.499 &    -1.80\\
 160418.2-191055 &    15.4 &    0.76 &      -0.5 &    M4 &     2.3 &    3.524 &    -0.99\\
 160428.0-190434 &    15.8 &    0.55 &      -3.4 &    M4 &     2.4 &    3.516 &    -1.05\\
 160428.4-190441 &    13.5 &    0.68 &      -5.0 &    M3 &     1.0 &    3.532 &    -0.63\\

 160435.6-194830 &    16.1 &    0.63 &      -7.2 &    M5 &     1.6 &    3.499 &    -1.29\\
 160439.1-194245 &    15.1 &    0.61 &      -3.8 &    M4 &     0.9 &    3.518 &    -1.18\\
 160449.9-203835 &    16.0 &    0.65 &      -7.8 &    M5 &     1.0 &    3.497 &    -1.37\\
 160456.4-194045 &    15.0 &    0.54 &      -5.2 &    M4 &     0.9 &    3.507 &    -1.10\\
 160502.1-203507 &    13.6 &    0.57 &      -5.9 &    M2 &     1.8 &    3.551 &    -0.53\\
 160508.3-201531 &    13.3 &    0.58 &      -4.3 &    M4 &     0.0 &    3.524 &    -0.80\\
 160508.5-201532 &    14.0 &    0.46 &      -5.5 &    M4 &     0.4 &    3.518 &    -0.90\\
 160516.1-193830 &    15.6 &    0.50 &      -2.4 &    M4 &     1.1 &    3.520 &    -1.35\\
 160517.9-202420 &    12.9 &    0.41 &      -5.2 &    M3 &     0.6 &    3.542 &    -0.56\\
 160521.9-193602 &    13.9 &    0.50 &      -1.9 &    M1 &     1.2 &    3.563 &    -0.83\\
 160522.7-205111 &    14.8 &    0.10 &      -4.5 &    M4 &     1.1 &    3.518 &    -1.03\\
 160525.5-203539 &    16.1 &    0.48 &      -6.1 &    M5 &     1.5 &    3.499 &    -1.30\\
 160528.5-201037 &    14.1 &    0.55 &      -1.9 &    M1 &     1.4 &    3.569 &    -0.85\\
 160531.3-192623 &    16.7 &    0.58 &      -8.8 &    M5 &     1.8 &    3.497 &    -1.48\\
 160532.1-193315 &    16.5 &    0.61 &     -26.0 &    M5 &     0.0 &    3.499 &    -1.88\\
 160545.4-202308 &    14.4 &    0.31 &     -35.0 &    M2 &     1.4 &    3.560 &    -0.98\\
 160600.6-195711 &    14.9 &    0.80 &      -7.5 &    M5 &     1.7 &    3.505 &    -0.82\\
 160611.9-193532 &    16.0 &    0.63 &      -8.2 &    M5 &     0.9 &    3.495 &    -1.39\\
 160619.3-192332 &    16.4 &    0.52 &      -5.5 &    M5 &     1.9 &    3.505 &    -1.37\\
 160622.8-201124 &    15.4 &    0.53 &      -6.0 &    M5 &     0.0 &    3.499 &    -1.47\\
 160628.7-200357 &    15.0 &    0.58 &     -30.0 &    M5 &     0.6 &    3.499 &    -1.12\\
 160629.0-205216 &    15.1 &    0.68 &      -6.2 &    M5 &     1.2 &    3.497 &    -0.99\\
 160632.1-202053 &    17.3 &    0.69 &      -5.1 &    M5 &     1.5 &    3.495 &    -1.78\\
 160643.8-190805 &    12.7 &    0.65 &      -3.8 &    K6 &     1.9 &    3.623 &    -0.31\\
 160647.5-202232 &    13.9 &    0.41 &      -3.2 &    M2 &     1.6 &    3.557 &    -0.70\\
 160700.1-203309 &    14.1 &    0.65 &      -0.6 &    M2 &     1.8 &    3.551 &    -0.69\\
 160702.1-201938 &    16.3 &    0.39 &     -30.0 &    M5 &     1.7 &    3.501 &    -1.37\\
 160704.7-201555 &    16.2 &    0.61 &      -4.2 &    M4 &     1.7 &    3.511 &    -1.39\\
 160707.7-192715 &    14.0 &    0.57 &      -5.0 &    M2 &     2.2 &    3.548 &    -0.52\\
 160708.7-192733 &    15.8 &    0.52 &      -4.0 &    M4 &     1.7 &    3.516 &    -1.26\\
 160710.0-191703 &    16.6 &    0.90 &      -9.0 &    M2 &     3.6 &    3.551 &    -1.23\\
 160716.0-204443 &    14.7 &    0.68 &      -3.6 &    M4 &     1.2 &    3.524 &    -0.97\\
 160719.7-202055 &    15.3 &    0.57 &      -2.3 &    M3 &     2.7 &    3.542 &    -0.89\\
 160722.4-201158 &    17.2 &    0.50 &     -14.0 &    M5 &     2.3 &    3.491 &    -1.47\\
 160726.8-185521 &    14.1 &    0.33 &      -1.9 &    M1 &     1.1 &    3.569 &    -0.97\\
 160727.5-201834 &    17.3 &    0.90 &     -13.0 &    M5 &     2.4 &    3.487 &    -1.45\\
 160735.5-202713 &    17.2 &    0.75 &      -4.9 &    M5 &     2.3 &    3.505 &    -1.57\\
 160739.4-191747 &    14.0 &    0.52 &      -2.3 &    M2 &     1.6 &    3.560 &    -0.72\\
 160744.5-203602 &    13.6 &    0.50 &      -4.8 &    M4 &     0.8 &    3.516 &    -0.61\\
 160745.8-203055 &    17.3 &    0.30 &      -2.0 &    M3 &     1.9 &    3.530 &    -1.87\\
 160800.5-204028 &    16.3 &    0.34 &      -5.2 &    M5 &     1.9 &    3.501 &    -1.31\\
 160801.4-202741 &    12.8 &    0.83 &      -2.3 &    K8 &     1.5 &    3.601 &    -0.45\\
 160801.5-192757 &    14.0 &    0.52 &      -3.6 &    M4 &     1.0 &    3.524 &    -0.76\\
 160802.4-202233 &    15.3 &    0.60 &      -6.1 &    M5 &     0.5 &    3.499 &    -1.27\\
 160803.6-181237 &    17.3 &    0.35 &      -7.7 &    M4 &     3.9 &    3.515 &    -1.23\\
 160804.3-194712 &    14.0 &    0.91 &      -5.2 &    M4 &     0.0 &    3.516 &    -1.17\\
 160815.3-203811 &    15.1 &    0.57 &      -1.9 &    M3 &     2.1 &    3.542 &    -0.98\\
 160818.4-190059 &    14.1 &    0.72 &      -3.7 &    M3 &     0.3 &    3.528 &    -1.03\\
 160822.4-193004 &    12.9 &    0.29 &      -3.0 &    M1 &     1.2 &    3.571 &    -0.43\\
 160823.2-193001 &    13.1 &    0.49 &      -6.0 &    K9 &     1.5 &    3.586 &    -0.49\\
 160823.5-191131 &    14.1 &    0.59 &      -4.1 &    M2 &     1.5 &    3.544 &    -0.79\\
 160823.8-193551 &    13.2 &    0.72 &      -2.1 &    M1 &     1.5 &    3.569 &    -0.47\\
 160825.1-201224 &    13.7 &    0.41 &      -2.0 &    M1 &     1.3 &    3.574 &    -0.77\\
 160827.5-194904 &    15.6 &    0.57 &     -12.3 &    M5 &     1.4 &    3.495 &    -1.11\\
 160841.7-185610 &    16.9 &    0.70 &     -11.0 &    M6 &     1.1 &    3.483 &    -1.61\\
 160843.1-190051 &    14.9 &    0.57 &      -5.8 &    M4 &     1.3 &    3.515 &    -1.00\\
 160845.6-182443 &    13.7 &    0.70 &       0.5 &    M3 &     0.2 &    3.533 &    -0.93\\
 160854.0-203417 &    14.8 &    0.57 &      -3.4 &    M4 &     1.5 &    3.518 &    -0.90\\
 160900.0-190836 &    15.9 &    0.62 &     -15.4 &    M5 &     0.7 &    3.503 &    -1.45\\
 160900.7-190852 &    12.8 &    0.53 &     -12.7 &    K9 &     0.8 &    3.592 &    -0.60\\
 160903.9-193944 &    14.6 &    0.68 &      -7.2 &    M4 &     0.6 &    3.520 &    -1.10\\
 160904.0-193359 &    15.5 &    0.80 &      -2.2 &    M4 &     2.5 &    3.524 &    -0.96\\
 160908.4-200928 &    13.4 &    0.66 &      -4.0 &    M4 &     0.7 &    3.518 &    -0.60\\

 160913.4-194328 &    14.9 &    0.54 &      -1.6 &    M3 &     1.8 &    3.537 &    -0.97\\
 160915.8-193706 &    16.1 &    0.78 &      -4.4 &    M5 &     0.7 &    3.499 &    -1.53\\
 160916.8-183522 &    13.8 &    0.55 &      -3.0 &    M2 &     1.0 &    3.546 &    -0.80\\
 160926.7-192502 &    14.9 &    0.40 &       0.0 &    M3 &     1.0 &    3.528 &    -1.16\\
 160933.8-190456 &    14.1 &    0.57 &      -3.5 &    M2 &     1.4 &    3.557 &    -0.83\\
 160935.6-182822 &    15.5 &    0.75 &      -4.2 &    M3 &     1.8 &    3.528 &    -1.17\\
 160936.5-184800 &    15.0 &    0.57 &     -18.0 &    M3 &     1.2 &    3.532 &    -1.14\\
 160943.8-182302 &    15.0 &    0.48 &      -5.8 &    M4 &     1.9 &    3.516 &    -0.86\\
 160946.4-193735 &    13.5 &    0.49 &      -1.6 &    M1 &     1.6 &    3.569 &    -0.57\\
 160953.6-175446 &    16.6 &    1.30 &     -22.0 &    M3 &     4.1 &    3.539 &    -1.03\\
 160954.4-190654 &    13.5 &    0.70 &      -3.1 &    M1 &     0.9 &    3.569 &    -0.78\\
 160959.4-180009 &    14.7 &    0.58 &      -4.0 &    M4 &     0.2 &    3.518 &    -1.26\\
 161007.5-181056 &    17.8 &    0.44 &     -13.3 &    M6 &     1.9 &    3.474 &    -1.70\\
 161010.4-194539 &    14.7 &    0.49 &      -5.6 &    M3 &     1.4 &    3.533 &    -1.01\\
 161011.0-194603 &    16.2 &    0.70 &      -4.4 &    M5 &     0.3 &    3.487 &    -1.58\\
 161014.7-191909 &    14.4 &    0.62 &      -2.3 &    M3 &     1.0 &    3.535 &    -0.97\\
 161021.5-194132 &    14.0 &    0.40 &      -4.3 &    M3 &     2.0 &    3.541 &    -0.59\\
 161024.7-191407 &    14.9 &    0.55 &      -3.7 &    M3 &     1.5 &    3.528 &    -0.99\\
 161026.4-193950 &    15.0 &    0.62 &      -4.4 &    M4 &     1.9 &    3.516 &    -0.87\\
 161028.1-191043 &    17.3 &    0.65 &     -11.4 &    M4 &     2.1 &    3.511 &    -1.72\\
 161030.0-183906 &    16.1 &    0.54 &      -6.5 &    M4 &     2.0 &    3.507 &    -1.23\\
 161030.9-182422 &    15.1 &    0.55 &      -3.4 &    M3 &     1.8 &    3.535 &    -1.04\\
 161031.9-191305 &    12.6 &    0.50 &      -2.3 &    K7 &     1.1 &    3.607 &    -0.48\\
 161039.5-191652 &    14.5 &    0.53 &      -4.3 &    M2 &     1.5 &    3.551 &    -0.96\\
 161043.9-192225 &    14.0 &    0.67 &      -2.3 &    M3 &     1.1 &    3.539 &    -0.83\\
 161046.3-184059 &    16.8 &    0.51 &      -7.2 &    M4 &     3.1 &    3.516 &    -1.28\\
 161052.4-193734 &    15.4 &    0.93 &      -3.9 &    M3 &     2.3 &    3.526 &    -1.00\\
 161110.9-193331 &    16.2 &    0.63 &      -6.3 &    M5 &     1.1 &    3.501 &    -1.46\\
 161112.3-192737 &    17.3 &    0.30 &     -50.0 &    M5 &     1.4 &    3.501 &    -1.83\\
 161115.3-175721 &    13.1 &    0.55 &      -2.4 &    M1 &     1.6 &    3.574 &    -0.42\\
 161116.6-193910 &    13.9 &    0.47 &      -3.4 &    M4 &     0.6 &    3.513 &    -0.77\\
 161118.1-175728 &    13.9 &    0.55 &      -4.8 &    M4 &     0.9 &    3.515 &    -0.69\\
 161118.2-180358 &    17.5 &    0.90 &     -20.0 &    M6 &     1.6 &    3.476 &    -1.65\\
 161120.4-191937 &    14.1 &    0.51 &      -5.5 &    M2 &     0.6 &    3.544 &    -1.00\\
 161123.0-190522 &    14.4 &    0.40 &      -6.9 &    M3 &     1.7 &    3.528 &    -0.74\\
 161129.4-194224 &    16.4 &    0.80 &     -13.0 &    M6 &     0.8 &    3.483 &    -1.49\\
 161133.6-191400 &    14.4 &    0.56 &      -3.7 &    M3 &     1.8 &    3.537 &    -0.79\\
 161146.1-190742 &    16.2 &    0.47 &      -6.3 &    M5 &     1.6 &    3.503 &    -1.37\\
 161156.2-194323 &    14.1 &    0.60 &      -6.3 &    M3 &     0.7 &    3.542 &    -0.97\\
 161247.2-190353 &    17.4 &    0.88 &     -10.6 &    M6 &     1.3 &    3.485 &    -1.78\\
 161248.9-180052 &    14.6 &    0.52 &      -3.8 &    M3 &     1.4 &    3.530 &    -0.92\\
 161328.0-192452 &    17.8 &    0.40 &     -17.0 &    M5 &     2.2 &    3.491 &    -1.73\\
 161347.5-183459 &    14.6 &    0.52 &      -3.5 &    M2 &     1.5 &    3.544 &    -0.98\\
 161358.1-184828 &    13.7 &    0.57 &      -1.9 &    M2 &     1.0 &    3.553 &    -0.76\\
 161420.2-190648 &    13.2 &    0.37 &     -52.0 &    K5 &     1.8 &    3.630 &    -0.59\\
 161433.6-190013 &    14.2 &    0.64 &     -26.0 &    M2 &     2.2 &    3.551 &    -0.60\\
 161437.5-185823 &    15.7 &    0.13 &      -3.6 &    M3 &     1.4 &    3.533 &    -1.42 \\
\label{2dfpmstab}
\end{longtable}
}
\normalsize

\begin{landscape}
\setlength{\LTcapwidth}{8.3in}

{\footnotesize
\begin{longtable}{@{}l@{\hskip12pt}llrrrrrrrrrl@{}}

\caption[]{PMS stars in UCL. The name (first column) reflects the mode of selection: HIP stars
from {\it Hipparcos} survey by \citet{deZeeuw99}, MML stars from \citet{Mamajek02},
and HD stars from \citet{Thackeray66}
and \citet{Lindroos86}. Stars with asterisks (*) next to their names are close to the Lupus
clouds. The following columns
provide the 2MASS identifier (proxy for accurate J2000 position), spectral type,
V magnitude, reddening, distance, X-ray luminosity, X-ray-to-bolometric flux
ratio, effective temperature, luminosity, mass, age, and X-ray catalog
counterpart. Masses and ages are inferred from the evolutionary tracks of
\citet{Baraffe98} for $M$ $<$ 1.4\,M$_{\odot}$, and from \citet{Palla01}
otherwise \citep[see tests of tracks in ][]{Hillenbrand04}.}\\
\hline
Name & 2MASS J & SpT & $V$ & A$_V$ & Dist & log\,$L_X$ & log\,$\frac{L_X}{L_{\odot}}$ & log\,$T_{eff}$ & log\,$\frac{L}{L_{\odot}}$ & Age & $M_{\star}$ & X-ray\\
     &         &     & [mag] & [mag]   & [pc]   & [erg/s]      &                              & [K]              &                            & [Myr] & [$M_{\odot}$] & counterpart\\
\hline
\endfirsthead

\caption[]{ -- continued}\\
\hline
Name & 2MASS J & SpT & $V$ & A$_V$ & Dist & log\,$L_X$ & log\,$\frac{L_X}{L_{\odot}}$ & log\,$T_{eff}$ & log\,$\frac{L}{L_{\odot}}$ & Age & $M_{\star}$ & X-ray\\
     &         &     & [mag] & [mag]   & [pc]   & [erg/s]      &                              & [K]              &                            & [Myr] & [$M_{\odot}$] & counterpart\\ \hline
\endhead

\hline
\endfoot

\hline
\endlastfoot

MML 36 & 13375730-4134419 & K0IV & 10.08 & -0.07 & 98 & 30.4 & -3.2 & 3.72 & 0.00 & 19 & 1.1 & 1RXS J133758.0-413448 \\
MML 38 & 13475054-4902056 & G8IVe & 10.82 & 0.45 & 148 & 30.5 & -3.2 & 3.74 & 0.05 & 22 & 1.1 & 1RXS J134748.0-490158 \\
HIP 67522 & 13500627-4050090 & G0.5IV & 9.79 & 0.02 & 126 & 30.4 & -3.5 & 3.77 & 0.25 & 21 & 1.2 & 1RXS J135005.7-405001 \\
MML 39 & 13524780-4644092 & G0IV & 9.62 & 0.34 & 145 & 30.8 & -3.3 & 3.78 & 0.44 & 16 & 1.3 & 1RXS J135247.0-464412 \\
MML 40 & 14022072-4144509 & G9IV & 10.71 & 0.90 & 130 & 30.3 & -3.3 & 3.73 & 0.07 & 18 & 1.1 & 1RXS J140220.9-414435 \\
MML 41 & 14090357-4438442 & F9IV & 9.39 & 0.38 & 142 & 30.7 & -3.5 & 3.78 & 0.50 & 15 & 1.3 & 1RXS J140902.6-443838 \\
HIP 70350 & 14233787-4357426 & F7V & 8.13 & 0.12 & 107 & 30.8 & -3.6 & 3.81 & 0.64 & 15 & 1.4 & 1RXS J142338.0-435814 \\
HIP 70376 & 14235639-5029585 & F7V & 9.20 & 0.27 & 122 & 30.4 & -3.6 & 3.81 & 0.39 &  ... &  ... & 1RXS J142356.2-503006 \\
MML 43 & 14270556-4714217 & G7IV & 10.59 & 0.17 & 132 & 30.4 & -3.3 & 3.75 & 0.06 & 23 & 1.1 & 1RXS J142705.3-471420 \\
HIP 70689 & 14273044-5231304 & F2V & 8.53 & 0.00 & 105 & 29.6 & -4.4 & 3.85 & 0.40 &  ... &  ... & 1RXS J142729.5-523141 \\
MML 44 & 14280929-4414175 & G5.5IV & 9.78 & 0.33 & 161 & 30.6 & -3.5 & 3.75 & 0.54 & 9 & 1.4 & 1RXS J142809.6-441438 \\
MML 45 & 14281937-4219341 & G3.5IV & 10.47 & 0.37 & 159 & 30.8 & -3.0 & 3.76 & 0.33 & 16 & 1.3 & 1RXS J142817.6-421958 \\
HIP 70919 & 14301035-4332490 & G8III & 8.88 & 0.00 & 193 & 30.8 & -3.9 & 3.68 & 0.84 & $<$1 & 2.2 & 1RXS J143008.7-433313 \\
HIP 71023 & 14313339-4445019 & F0V & 8.94 & 0.15 & 160 & 30.6 & -3.9 & 3.86 & 0.65 &  ... &  ... & 1RXS J143135.2-444526 \\
HIP 71178 & 14332578-3432376 & G8IVe & 10.18 & 0.35 & 115 & $<$29.8 & $<$-3.9 & 3.74 & 0.16 & 16 & 1.2 & ... \\
MML 46 & 14370422-4145028 & G0.5IV & 9.70 & 0.18 & 156 & 30.7 & -3.4 & 3.78 & 0.59 & 11 & 1.4 & 1RXS J143704.6-414504 \\
MML 47 & 14375022-5457411 & K0+IV & 10.72 & 0.30 & 132 & 30.6 & -3.0 & 3.72 & 0.08 & 13 & 1.2 & 1RXS J143750.9-545708 \\
HIP 71767 & 14404593-4247063 & F3V & 9.02 & 0.20 & 168 & 30.6 & -3.9 & 3.84 & 0.70 & 15 & 1.5 & 1RXS J144044.6-424720 \\
MML 48 & 14413499-4700288 & G4IV & 10.00 & 0.02 & 110 & 30.9 & -2.8 & 3.76 & 0.20 & 20 & 1.2 & 1RXS J144135.3-470039 \\
HIP 72033 & 14440435-4059223 & F7IV/V & 9.17 & 0.23 & 156 & 30.3 & -3.9 & 3.81 & 0.60 & 15 & 1.4 & 1RXS J144405.2-405940 \\
HD 129791B & 14455620-4452346 & K5Ve & 12.93 & 0.26 & 132 & 30.6 & -2.9 & 3.63 & -0.52 & 18 & 0.9 & 1RXS J144556.0-445202 \\
MML 49 & 14473176-4800056 & G2.5IV & 10.72 & 0.43 & 130 & 30.2 & -3.4 & 3.77 & -0.05 &  ... &  ... & 1RXS J144732.2-480019 \\
MML 50 & 14502581-3506486 & K0IV & 10.73 & 0.21 & 190 & 30.6 & -3.3 & 3.72 & 0.52 & 5 & 1.6 & 1RXS J145025.4-350645 \\
MML 51 & 14524198-4141552 & K1IVe & 10.89 & 0.61 & 145 & 30.5 & -3.2 & 3.70 & 0.10 & 9 & 1.3 & 1RXS J145240.7-414206 \\
MML 52 & 14571962-3612274 & G6IV & 10.27 & 0.15 & 132 & 30.3 & -3.4 & 3.75 & 0.18 & 18 & 1.1 & 1RXS J145720.4-361242 \\
MML 53 & 14583769-3540302 & K2-IVe & 10.75 & 0.23 & 136 & 30.3 & -3.4 & 3.69 & 0.21 & 6 & 1.4 & 1RXS J145837.6-354036 \\
MML 54 & 14592275-4013120 & G3IV & 9.71 & 0.40 & 122 & 30.5 & -3.4 & 3.76 & 0.32 & 17 & 1.2 & 1RXS J145923.0-401319 \\
MML 55 & 15005189-4331212 & G9IV & 11.15 & 0.30 & 163 & 30.4 & -3.2 & 3.73 & 0.16 & 14 & 1.2 & 1RXS J150052.5-433107 \\
MML 56 & 15011155-4120406 & G0.5IV & 10.01 & 0.48 & 214 & 31.0 & -3.3 & 3.77 & 0.64 & 10 & 1.4 & 1RXS J150112.0-412040 \\
MML 57 & 15015882-4755464 & G1.5IV & 10.15 & 0.22 & 156 & 30.2 & -3.8 & 3.77 & 0.29 & 19 & 1.2 & 1RXS J150158.5-475559 \\
MML 58 & 15071481-3504595 & G9.5IV & 10.49 & 0.44 & 101 & 30.0 & -3.4 & 3.73 & -0.12 & 28 & 1.0 & 1RXS J150714.5-350500 \\
MML 59 & 15083773-4423170 & G1.5IVe & 10.83 & 0.13 & 180 & 30.9 & -2.9 & 3.77 & 0.29 & 19 & 1.2 & 1RXS J150836.0-442325 \\
MML 60 & 15083849-4400519 & G1.5IV & 10.54 & 0.11 & 134 & 30.4 & -3.3 & 3.77 & 0.20 & 23 & 1.1 & 1RXS J150838.5-440048 \\
MML 61 & 15125018-4508044 & G5IV & 10.71 & 0.27 & 151 & 30.6 & -3.1 & 3.76 & 0.19 & 19 & 1.2 & 1RXS J151250.0-450822 \\
HIP 74501 & 15132923-5543545 & G2IV & 7.47 & 0.00 & 153 & 30.0 & -5.2 & 3.74 & 1.19 & 1 & 2.4 & 1RXS J151330.3-554341 \\
MML 62 & 15180174-5317287 & G7IV & 10.12 & 0.77 & 103 & 30.4 & -3.3 & 3.75 & 0.08 & 22 & 1.1 & 1RXS J151802.0-531719 \\
MML 63 & 15182692-3738021 & G9IV & 11.02 & 0.72 & 127 & 30.4 & -3.0 & 3.73 & 0.05 & 18 & 1.1 & 1RXS J151827.3-373808 \\
MML 64 & 15255964-4501157 & G8IV & 10.90 & 0.19 & 149 & 30.1 & -3.5 & 3.74 & -0.02 & 26 & 1.0 & 1RXS J152600.9-450113 \\
MML 65* & 15293858-3546513 & K0+IV & 10.43 & 0.16 & 135 & 30.3 & -3.4 & 3.72 & 0.17 & 10 & 1.3 & 1RXS J152937.7-354656 \\
HIP 75924 & 15302626-3218122 & G2.5IV & 8.80 & 0.37 & 102 & 30.9 & -2.9 & 3.77 & 0.69 & 5 & 1.7 & 1RXS J153026.1-321815 \\
%

HIP 76084 & 15322013-3108337 & F2V & 8.62 & 0.16 & 149 & 29.8 & -4.6 & 3.85 & 0.73 & 15 & 1.5 & 1RXS J153218.2-310828 \\
MML 66 & 15370214-3136398 & G6IV & 9.99 & 0.43 & 135 & 30.9 & -3.0 & 3.75 & 0.42 & 11 & 1.3 & 1RXS J153701.9-313647 \\
HIP 76472 & 15370466-4009221 & G1IV & 9.39 & 0.29 & 138 & 30.8 & -3.3 & 3.77 & 0.60 & 11 & 1.4 & 1RXS J153706.0-400929 \\
MML 67 & 15371129-4015566 & G8.5IVe & 10.43 & 0.40 & 164 & 30.7 & -3.2 & 3.74 & 0.25 & 11 & 1.3 & 1RXS J153711.6-401608 \\
MML 68 & 15384306-4411474 & G8.5IV & 10.28 & 0.62 & 124 & 30.3 & -3.4 & 3.74 & 0.08 & 19 & 1.1 & 1RXS J153843.1-441149 \\
MML 69 & 15392440-2710218 & G5IV & 9.57 & 0.28 & 127 & 30.7 & -3.3 & 3.76 & 0.46 & 11 & 1.3 & 1RXS J153924.0-271035 \\
MML 70* & 15440376-3311110 & K0IVe & 10.87 & 0.47 & 124 & 30.2 & -3.3 & 3.72 & 0.01 & 18 & 1.1 & 1RXS J154404.1-331120 \\
HIP 77081* & 15442105-3318549 & G7.5IV & 9.69 & 0.15 & 131 & $<$29.9 & $<$-4.1 & 3.74 & 0.35 & 11 & 1.3 & ... \\
HIP 77135* & 15445769-3411535 & G4IV & 9.88 & 0.08 & 139 & 30.5 & -3.4 & 3.76 & 0.42 & 12 & 1.3 & 1RXS J154458.0-341143 \\
HIP 77144 & 15450184-4050310 & G0IV & 9.46 & 0.08 & 125 & 30.6 & -3.4 & 3.78 & 0.36 & 20 & 1.2 & 1RXS J154502.0-405043 \\
MML 71 & 15455225-4222163 & K2-IVe & 10.50 & 0.80 & 130 & 30.9 & -2.9 & 3.69 & 0.12 & 8 & 1.3 & 1RXS J154552.7-422227 \\
MML 72 & 15465179-4919048 & G7.5IV & 10.18 & 0.29 & 132 & 30.5 & -3.3 & 3.74 & 0.21 & 15 & 1.2 & 1RXS J154651.5-491922 \\
HIP 77524* & 15494499-3925089 & K1-IVe & 10.64 & -0.03 & 151 & 30.5 & -3.3 & 3.71 & 0.23 & 9 & 1.3 & 1RXS J154944.7-392509 \\
HIP 77656 & 15511373-4218513 & G5IV & 9.58 & 0.52 & 130 & 30.3 & -3.8 & 3.76 & 0.41 & 12 & 1.3 & 1RXS J155113.5-421858 \\
MML 73* & 15565905-3933430 & G9.5IV & 10.81 & 0.35 & 174 & 30.5 & -3.4 & 3.73 & 0.21 & 11 & 1.3 & 1RXS J155659.0-393400 \\
HD 143099* & 15595826-3824317 & G0V & 9.33 & 0.00 & 142 & 30.4 & -3.7 & 3.78 & 0.48 & 15 & 1.3 & 1RXS J155958.3-382352 \\
MML 74 & 16010792-3254526 & G0IV & 9.50 & 0.15 & 132 & 30.6 & -3.4 & 3.78 & 0.33 & 20 & 1.2 & 1RXS J160108.0-325455 \\
MML 75 & 16010896-3320141 & G5IV & 10.88 & 0.16 & 176 & 30.6 & -3.1 & 3.76 & 0.41 & 11 & 1.4 & 1RXS J160108.9-332021 \\
MML 76 & 16034536-4355492 & G9.5IV & 9.64 & 0.51 & 178 & 31.2 & -3.1 & 3.73 & 0.75 & 4 & 1.8 & 1RXS J160345.8-435544 \\
MML 77* & 16035250-3939013 & K2-IVe & 11.01 & 0.18 & 140 & 30.4 & -3.2 & 3.69 & 0.09 & 7 & 1.4 & 1RXS J160352.0-393901 \\
HD 143939B* & 16044404-3926117 & K3Ve & 11.80 & 0.00 & 149 & 30.5 & -3.0 & 3.67 & -0.37 & 23 & 0.9 & 1RXS J160444.6-392602 \\
MML 78* & 16054499-3906065 & G6.5IV & 10.53 & 0.19 & 130 & 30.5 & -3.1 & 3.75 & 0.13 & 20 & 1.1 & 1RXS J160545.8-390559 \\
HIP 78881* & 16060937-3802180 & F3V & 8.03 & 0.26 & 128 & 30.7 & -3.8 & 3.84 & 0.88 & 11 & 1.6 & 1RXS J160610.3-380215 \\
HIP 79516 & 16133433-4549035 & F5V & 8.91 & 0.02 & 120 & 30.0 & -4.1 & 3.82 & 0.38 & 20 & 1.3 & 1RXS J161335.9-454901 \\
MML 79 & 16135801-3618133 & G9IVe & 11.14 & 0.31 & 123 & 30.4 & -2.9 & 3.73 & -0.13 & 31 & 1.0 & 1RXS J161357.9-361813 \\
MML 80 & 16145207-5026187 & G9.5IV & 10.41 & 0.39 & 129 & 30.6 & -3.1 & 3.73 & 0.34 & 10 & 1.3 & 1RXS J161451.3-502621 \\
MML 81 & 16183856-3839117 & F9IV & 9.02 & 0.32 & 108 & 30.3 & -3.8 & 3.78 & 0.33 & 20 & 1.2 & 1RXS J161839.0-383927 \\
MML 82 & 16211219-4030204 & G8IV & 10.57 & 0.69 & 156 & 30.5 & -3.3 & 3.74 & 0.23 & 13 & 1.3 & 1RXS J162112.0-403032 \\
MML 83 & 16232955-3958008 & G3IV & 10.64 & 0.95 & 169 & 30.4 & -3.4 & 3.76 & 0.11 & 26 & 1.1 & 1RXS J162330.1-395806 \\
MML 84 & 16273054-3749215 & G9.5IV & 10.96 & 0.50 & 180 & 30.5 & -3.3 & 3.73 & 0.21 & 11 & 1.3 & 1RXS J162730.0-374929 \\
HIP 80636 & 16275233-3547003 & G0.5IV & 9.37 & 0.39 & 152 & 30.9 & -3.3 & 3.77 & 0.65 & 10 & 1.4 & 1RXS J162752.8-354702 \\
MML 85 & 16314204-3505171 & G7.5IV & 10.64 & 0.33 & 143 & 30.4 & -3.3 & 3.74 & 0.05 & 22 & 1.1 & 1RXS J163143.7-350521 \\
MML 86 & 16353598-3326347 & K2-IV & 11.00 & 0.34 & 192 & 30.9 & -3.0 & 3.69 & 0.35 & 4 & 1.5 & 1RXS J163533.9-332631 \\
HIP 81380 & 16371286-3900381 & G0IV & 9.82 & 0.33 & 200 & 30.1 & -3.8 & 3.78 & 0.74 & 5 & 1.7 & 1RXS J163713.7-390104 \\
HIP 81447 & 16380553-3401106 & G0.5IV & 9.08 & 0.03 & 172 & 30.0 & -4.3 & 3.78 & 0.78 & 5 & 1.7 & 1RXS J163805.3-340110 \\
MML 87 & 16395929-3924592 & G4IV & 10.59 & 0.27 & 216 & 30.7 & -3.3 & 3.76 & 0.74 & 5 & 1.7 & 1RXS J163958.7-392457 \\
MML 88 & 16422399-4003296 & G1IV & 9.62 & 0.74 & 199 & 30.7 & -3.7 & 3.77 & 0.92 & 4 & 1.9 & 1RXS J164224.5-400329 \\
HD 151868 & 16514560-3803088 & F6V & 9.37 & 0.00 & 192 & $<$30.3 & $<$-4.0 & 3.80 & 0.60 & 15 & 1.4 & ... \\
HIP 82569 & 16524171-3845372 & F3V & 8.85 & 0.13 & 185 & 30.5 & -3.9 & 3.84 & 0.82 & 12 & 1.5 & 1RXS J165242.7-384534 \\
HIP 82747 & 16544485-3653185 & F5IVe & 9.21 & 0.52 & 144 & $<$30.0 & $<$-4.3 & 3.81 & 0.76 & 15 & 1.4 & ... \\
HIP 83159 & 16594248-3726168 & F5V & 9.02 & 0.00 & 152 & 29.9 & -4.3 & 3.82 & 0.54 & 19 & 1.4 & 1RXS J165943.1-372614 \\

\label{table_ucl_list}
\end{longtable}
}

\clearpage

{\footnotesize
\begin{longtable}{@{}l@{\hskip10pt}ll@{\hskip10pt}r@{\hskip10pt}r@{\hskip10pt}r@{\hskip10pt}r@{\hskip10pt}r@{\hskip10pt}r@{\hskip10pt}rrrl@{}}

\caption[]{Low mass members of LCC. The name (first column) reflects the mode
of selection: HIP stars from {\it Hipparcos} survey by
\citet{deZeeuw99}, MML stars from \citet{Mamajek02}, SACY stars were
selected from \citet{Torres06}, PF96 stars are from \citet{Park96},
and HD stars from \citet{Thackeray66} and \citet{Lindroos86}. The
columns are the same as in Table \ref{table_ucl_list}. $V$ magnitudes
are usually accurate to $<$0.1 mag, except when followed by a ":"
($\sim$0.5 mag uncertainties).}\\
\hline
Name & 2MASS J & SpT & $V$ & A$_V$ & Dist & log\,$L_X$ & log\,$\frac{L_X}{L_{\odot}}$ & log\,$T_{eff}$ & log\,$\frac{L}{L_{\odot}}$ & age & $M_{\star}$ & X-ray\\
     &         &     & [mag] & [mag]   & [pc]   & [erg/s]      &                              & [K]              &                            & [Myr] & [$M_{\odot}$] & counterpart\\
\hline
\endfirsthead

\caption[]{ -- continued}\\
\hline
Name & 2MASS J & SpT & $V$ & A$_V$ & Dist & log\,$L_X$ & log\,$\frac{L_X}{L_{\odot}}$ & log\,$T_{eff}$ & log\,$\frac{L}{L_{\odot}}$ & age & $M_{\star}$ & X-ray\\
     &         &     & [mag] & [mag]   & [pc]   & [erg/s]      &                              & [K]              &                            & [Myr] & [$M_{\odot}$] & counterpart\\
\hline
\endhead

\hline
\endfoot

\hline
\endlastfoot

SACY 606   & 10065573-6352086 & K0V(e)  & 11.05 & 0.71 & 173  & 30.5    & -3.4    & 3.720 &  0.31 &  6 & 1.6 & 1RXS J100659.0-635212 \\
SACY 653   & 10494839-6446284 & G9Ve    & 11.73 & 0.21 & 195  & 30.5    & -3.2    & 3.733 &  0.09 & 17 & 1.2 & 1RXS J104949.5-644613 \\
MML 1      & 10574936-6913599 & K1+IV   & 10.39 & 0.35 & 102  & 30.3    & -3.3    & 3.699 &  0.02 & 11 & 1.2 & 1RXS J105751.2-691402 \\
SACY 671   & 11080791-6341469 & M0Ve    & 11.86 & 0.00 & 113  & 30.0    & -3.2    & 3.585 & -0.34 &  3 & 0.7 & 1RXS J110808.9-634125 \\
HIP 55334  & 11195276-7037065 & F2V     &  8.14 & 0.24 &  87  & 29.8    & -4.3    & 3.838 &  0.60 & 27 & 1.4 & 1RXS J111948.2-703711 \\
TWA 12     & 11210549-3845163 & M2e     & 13.6: & 0.45 & 109  & 30.1    & -3.4    & 3.550 & -0.39 &  2 & 0.6 & 1RXS J112105.2-384529 \\
SACY 681   & 11275535-6626046 & K1V(e)  & 10.82 & 0.57 & 109  & 30.3    & -3.2    & 3.706 & -0.06 & 16 & 1.1 & 1RXS J112755.0-662558 \\
MML 2      & 11320835-5803199 & G7IV    &  9.92 & 0.68 &  93  & 29.9    & -3.7    & 3.747 &  0.19 & 17 & 1.2 & 1RXS J113209.3-580319 \\
SACY 684   & 11350376-4850219 & G7V     & 10.28 & 0.00 & 153  & 30.5    & -3.2    & 3.751 &  0.06 & 25 & 1.1 & 1RXS J113501.5-485011 \\
HIP 56673  & 11371464-6940272 & F5IV    &  6.62 & 0.22 & 106  & 30.8    & -4.2    & 3.809 &  1.40 &  2 & 2.4 & 1RXS J113714.2-694025 \\
TWA 19B    & 11472064-4953042 & K7e     & 11.6: & 0.77 & 113  & $<$29.8 & $<$-3.8 & 3.605 & -0.20 &  2 & 0.8 & ... \\
HIP 57524  & 11472454-4953029 & F9IV    &  9.07 & 0.30 & 113  & 30.8    & -3.3    & 3.783 &  0.44 & 17 & 1.3 & 1RXS J114724.3-495250 \\
SACY 695   & 11515049-6407278 & K1V     & 11.99 & 0.33 & 174  & 30.3    & -3.1    & 3.706 & -0.19 & 24 & 1.0 & 1RXS J115149.1-640705 \\
HIP 57950  & 11530799-5643381 & F2IV/V  &  8.26 & 0.17 &  99  & 29.9    & -4.3    & 3.838 &  0.64 & 18 & 1.4 & 1RXS J115308.5-564317 \\
SACY 699   & 11554295-5637314 & M0Ve    & 11.69 & 0.00 &  87  & 30.3    & -2.8    & 3.585 & -0.55 &  7 & 0.7 & 1RXS J115544.5-563739 \\
HIP 58167  & 11554354-5410506 & F3IV    &  8.30 & 0.06 & 103  & 29.7    & -4.5    & 3.829 &  0.63 & 15 & 1.4 & 1RXS J115543.4-541049 \\
SACY 700   & 11555771-5254008 & K4V     & 11.00 & 0.01 & 109  & 29.9    & -3.5    & 3.662 & -0.26 & 13 & 1.1 & 1RXS J115554.5-525332 \\
SACY 706   & 11594608-6101132 & K4V(e)  & 11.36 & 0.01 & 133  & 30.1    & -3.3    & 3.662 & -0.23 & 12 & 1.1 & 1RXS J115946.5-610111 \\
HIP 58528  & 12000940-5707021 & F5V     &  8.54 & 0.05 & 100  & 29.5    & -4.6    & 3.809 &  0.51 & 17 & 1.3 & 1RXS J120009.5-570646 \\
SACY 708   & 12041439-6418516 & G8V     &  9.93 & 0.00 & 135  & 30.8    & -3.0    & 3.742 &  0.16 & 17 & 1.2 & 1RXS J120413.3-641837 \\
MML 3      & 12044888-6409555 & G1IV    &  9.41 & 0.60 & 120  & 30.8    & -3.2    & 3.773 &  0.44 & 15 & 1.3 & 1RXS J120448.2-640942 \\
HIP 58996  & 12054748-5100121 & G1IV    &  8.89 & 0.14 & 102  & 30.5    & -3.6    & 3.772 &  0.43 & 15 & 1.3 & 1RXS J120547.8-510007 \\
MML 4      & 12061352-5702168 & G4IV    & 10.69 & 1.45 & 154  & 30.6    & -3.1    & 3.760 &  0.15 & 23 & 1.1 & 1RXS J120613.9-570215 \\
SACY 713   & 12063292-4247508 & K0V     & 10.66 & 0.31 &  86  & 29.9    & -3.4    & 3.720 & -0.30 & 42 & 0.9 & 1RXS J120632.7-424750 \\
SACY 715   & 12074236-6227282 & K3Ve    & 10.90 & 0.41 & 120  & 30.3    & -3.4    & 3.675 &  0.06 &  6 & 1.3 & 1RXS J120741.4-622720 \\
MML 5      & 12094184-5854450 & K0IVe   & 10.09 & 0.28 & 111  & 30.5    & -3.2    & 3.720 &  0.20 & 12 & 1.2 & 1RXS J120941.5-585440 \\
MML 6      & 12113142-5816533 & G9IV    & 10.19 & 0.28 & 108  & 30.5    & -3.2    & 3.729 &  0.05 & 18 & 1.1 & 1RXS J121131.9-581651 \\
MML 7      & 12113815-7110360 & G3.5IV  &  9.15 & 0.48 &  99  & 30.5    & -3.5    & 3.762 &  0.39 & 14 & 1.3 & 1RXS J121137.3-711032 \\
SACY 724   & 12120804-6554549 & K3Ve    & 11.25 & 0.53 & 102  & 30.0    & -3.4    & 3.675 & -0.23 & 15 & 1.1 & 1RXS J121206.3-655456 \\
SACY 725   & 12121119-4950081 & K2Ve    & 11.37 & 0.58 & 113  & 30.1    & -3.3    & 3.690 & -0.24 & 21 & 1.0 & 1RXS J121210.7-494955 \\
MML 8      & 12123577-5520273 & K0+IV   & 10.48 & 0.48 & 108  & 30.3    & -3.2    & 3.717 &  0.00 & 17 & 1.1 & 1RXS J121236.4-552037 \\
SACY 727   & 12124890-6230317 & K7Ve    & 11.47 & 0.43 & 101  & 30.7    & -2.6    & 3.609 & -0.31 &  4 & 0.8 & 1RXS J121248.7-623027 \\
HIP 59603  & 12132235-5653356 & F2V     &  8.56 & 0.31 & 115  & 29.8    & -4.5    & 3.838 &  0.71 & 14 & 1.5 & 1RXS J121321.6-565323 \\
SACY 728   & 12135700-6255129 & K4Ve    & 11.58 & 0.38 & 122  & 30.4    & -3.0    & 3.662 & -0.17 &  9 & 1.1 & 1RXS J121356.3-625508 \\
MML 9      & 12143410-5110124 & G9IV    & 10.28 & 0.0: & 106  & 30.4    & -3.2    & 3.733 &  0.01 & 21 & 1.0 & 1RXS J121434.2-511004 \\
HIP 59716  & 12145071-5547235 & F5V     &  8.45 & 0.11 & 104  & 30.9    & -3.3    & 3.809 &  0.60 & 15 & 1.4 & 1RXS J121452.4-554704 \\
MML 10     & 12145229-5547037 & G6IV    &  9.64 & 0.87 & 103  & 30.9    & -2.9    & 3.750 &  0.28 & 14 & 1.2 & 1RXS J121452.4-554704 \\
HIP 59764  & 12151855-6325301 & F8/G0V: &  8.43 & 0.25 & 109  & 30.8    & -3.5    & 3.786 &  0.72 & 10 & 1.5 & 1RXS J121518.6-632517 \\
HIP 59781  & 12152822-6232207 & F8/G0V  &  9.12 & 0.30 & 102  & 29.9    & -4.1    & 3.786 &  0.40 & 19 & 1.2 & 1RXS J121529.1-623209 \\
SACY 738   & 12160114-5614068 & K5Ve    & 11.22 & 0.64 &  91  & 30.3    & -3.0    & 3.638 & -0.29 &  8 & 1.0 & 1RXS J121601.8-561405 \\
HIP 59854  & 12162783-5008356 & G1IV    &  9.34 & 0.37 & 130  & 30.9    & -3.2    & 3.772 &  0.49 & 13 & 1.4 & 1RXS J121627.9-500829 \\
SACY 740   & 12163007-6711477 & K4IVe   & 11.56 & 0.58 & 107  & 30.3    & -3.2    & 3.662 & -0.08 &  7 & 1.2 & 1RXS J121630.6-671146 \\
MML 11     & 12182762-5943128 & K1.5IVe & 10.73 & 0.12 &  96  & 30.3    & -3.1    & 3.695 &  0.03 & 10 & 1.2 & 1RXS J121828.6-594307 \\
MML 12     & 12185802-5737191 & G9.5IV  &  9.87 & 0.49 & 106  & 30.6    & -3.2    & 3.724 &  0.32 & 10 & 1.3 & 1RXS J121858.2-573713 \\
MML 13     & 12192161-6454101 & K1-IV   & 10.11 & 0.25 & 103  & 30.5    & -3.2    & 3.707 &  0.26 &  8 & 1.3 & 1RXS J121919.4-645406 \\
HIP 60205  & 12204420-5215249 & F5      & 10.06 & 0.19 & 147  & 29.8    & -4.0    & 3.809 &  0.28 & ... & 1.4: & 1RXS J122047.8-521509 \\
SACY 750   & 12205449-6457242 & K4Ve    & 11.00 & 0.00 &  94  & 29.9    & -3.4    & 3.662 & -0.22 & 11 & 1.1 & 1RXS J122050.8-645724 \\
SACY 751   & 12210808-5212226 & K4Ve    & 11.85 & 0.40 & 106  & 30.0    & -3.2    & 3.662 & -0.41 & 22 & 0.9 & 1RXS J122108.0-521217 \\
MML 14     & 12211648-5317450 & G1.5IV  &  9.34 & 0.44 & 107  & 30.7    & -3.3    & 3.770 &  0.31 & 18 & 1.2 & 1RXS J122116.7-531747 \\
MML 15     & 12215566-4946125 & G5.5IV  & 10.02 & 0.14 & 102  & 30.2    & -3.4    & 3.754 &  0.08 & 25 & 1.1 & 1RXS J122155.9-494609 \\
MML 16     & 12220430-4841248 & K0IVe   & 10.50 & 0.21 & 126  & 30.4    & -3.2    & 3.721 &  0.10 & 13 & 1.2 & 1RXS J122204.0-484118 \\
HIP 60348  & 12222484-5101343 & F5V     &  8.80 & 0.06 & 104  & 29.8    & -4.2    & 3.809 &  0.44 & 19 & 1.3 & 1RXS J122226.1-510120 \\
MML 17     & 12223322-5333489 & G0IV    &  9.42 & 0.41 & 124  & 30.3    & -3.7    & 3.778 &  0.41 & 17 & 1.3 & 1RXS J122233.4-533347 \\
MML 18     & 12234012-5616325 & K0+IV   & 10.85 & 0.44 & 112  & 30.0    & -3.4    & 3.715 & -0.09 & 21 & 1.1 & 1RXS J122339.9-561628 \\
SACY 759   & 12242065-5443540 & G5V     & 10.37 & 1.14 & 145  & 30.7    & -3.5    & 3.761 &  0.56 &  7 & 1.8 & 1RXS J122421.0-544343 \\
HIP 60567  & 12245491-5200157 & F6/F7V  &  9.77 & 0.19 & 136  & 30.1    & -3.8    & 3.801 &  0.34 & 23 & 1.3 & 1RXS J122452.5-520014 \\
SACY 762   & 12264842-5215070 & K5Ve    & 11.66 & 0.47 &  92  & 29.9    & -3.2    & 3.638 & -0.42 & 14 & 1.0 & 1RXS J122648.5-521453 \\

SACY 764   & 12282540-6320589 & G7V     &  9.25 & 0.32 & 109  & 30.5    & -3.5    & 3.751 &  0.41 &  9 & 1.5 & 1RXS J122823.4-632100 \\
HIP 60885  & 12284005-5527193 & G0IV    &  8.89 & 0.16 & 105  & 30.3    & -3.8    & 3.778 &  0.46 & 15 & 1.3 & 1RXS J122840.3-552707 \\
HIP 60913  & 12290224-6455006 & G4.5IV  &  9.04 & 0.21 & 102  & 29.6    & -4.4    & 3.758 &  0.43 & 12 & 1.3 & 1RXS J122858.3-645448 \\
SACY 766   & 12302957-5222269 & K3V(e)  & 12.04 & 0.85 & 103  & 29.9    & -3.3    & 3.675 & -0.41 & 28 & 0.9 & 1RXS J123031.1-522221 \\
SACY 768   & 12333381-5714066 & K1V(e)  & 10.92 & 0.69 &  92  & 29.9    & -3.5    & 3.706 & -0.16 & 22 & 1.0 & 1RXS J123332.4-571345 \\
SACY 769   & 12361767-5042421 & K4Ve    & 11.39 & 0.00 & 105  & 29.9    & -3.3    & 3.662 & -0.45 & 25 & 0.9 & 1RXS J123620.3-504238 \\
MML 19     & 12363895-6344436 & K0+III  &  9.87 & 0.39 & 103  & 30.4    & -3.4    & 3.717 &  0.25 & 10 & 1.3 & 1RXS J123637.5-634446 \\
MML 20     & 12365895-5412178 & K0IV    & 10.40 & 0.11 & 116  & 30.3    & -3.4    & 3.721 &  0.04 & 16 & 1.2 & 1RXS J123657.4-541217 \\
SACY 773   & 12383556-5916438 & K3Ve    & 11.62 & 0.55 &  87  & 29.9    & -3.2    & 3.675 & -0.50 & 38 & 0.8 & 1RXS J123834.9-591645 \\
MML 21     & 12393796-5731406 & G7.5IV  & 10.12 & 0.30 &  98  & 30.4    & -3.1    & 3.742 & -0.06 & 30 & 1.0 & 1RXS J123938.4-573141 \\
SACY 779   & 12404664-5211046 & K2V(e)  & 11.91 & 0.34 & 135  & 30.1    & -3.2    & 3.690 & -0.26 & 23 & 1.0 & 1RXS J124046.9-521108 \\
MML 22     & 12411820-5825558 & G1.5IV  &  9.93 & 0.67 &  97  & 30.2    & -3.4    & 3.771 &  0.14 & 26 & 1.1 & 1RXS J124118.5-582556 \\
SACY 782   & 12442412-5855216 & K2Ve    & 10.26 & 0.48 & 106  & 30.3    & -3.4    & 3.690 &  0.15 &  6 & 1.4 & 1RXS J124423.9-585428 \\
HIP 62171  & 12442659-5420480 & F3V     &  8.90 & 0.25 & 113  & 29.6    & -4.6    & 3.829 &  0.54 & 26 & 1.4 & 1RXS J124427.5-542043 \\
MML 23     & 12443482-6331463 & K1IVe   & 10.78 & 0.27 & 125  & 30.6    & -3.0    & 3.704 &  0.25 &  8 & 1.3 & 1RXS J124432.6-633139 \\
MML 24     & 12450674-4742580 & G8.5IV  & 10.40 & 0.38 & 120  & 30.6    & -3.1    & 3.738 &  0.13 & 17 & 1.2 & 1RXS J124506.9-474254 \\
SACY 787   & 12454884-5410583 & K2V(e)  & 11.40 & 0.60 & 129  & 30.2    & -3.4    & 3.690 & -0.04 & 11 & 1.2 & 1RXS J124547.1-541104 \\
PF96 1     & 12464097-5931429 & M3e     & 13.9: & 0.13 & 110: & 30.0    & -2.8    & 3.532 & -0.85 &  4 & 0.4 & 1WGA J1246.6-5931 \\
HIP 62427  & 12473870-5824567 & F8      &  9.28 & 0.00 & 134  & 30.5    & -3.5    & 3.792 &  0.39 & 19 & 1.3 & 1RXS J124742.1-582544 \\
HIP 62431  & 12474180-5825558 & F0      &  8.02 & 0.34 & 140  & 30.5    & -4.2    & 3.857 &  1.10 &  5 & 1.9 & 1RXS J124742.1-582544 \\
SACY 789   & 12474824-5431308 & M0Ve    & 11.78 & 0.03 &  96  & 30.2    & -3.1    & 3.585 & -0.26 &  2 & 0.7 & 1RXS J124747.8-543141 \\
HIP 62445  & 12475186-5126382 & G4.5IVe &  9.52 & 0.66 & 130  & 30.7    & -3.4    & 3.758 &  0.62 &  5 & 1.7 & 1RXS J124751.7-512638 \\
MML 25     & 12480778-4439167 & G6IVe   &  9.73 & 0.18 &  91  & 30.6    & -3.1    & 3.751 &  0.14 & 20 & 1.1 & 1RXS J124807.6-443913 \\
PF96 3A    & 12483152-5944493 & K5e     & 11.5: & 0.63 & 110: & 29.9:   & -3.5:   & 3.644 & -0.10 &  5 & 1.1 & 1WGA J1248.5-5944 \\
PF96 3B    & 12483152-5944493 & K5e     & 11.5: & 0.63 & 110: & 29.9:   & -3.5:   & 3.644 & -0.10 &  5 & 1.1 & 1WGA J1248.5-5944 \\
MML 26     & 12484818-5635378 & G5IV    & 10.22 & 0.53 & 132  & 30.4    & -3.4    & 3.755 &  0.14 & 22 & 1.1 & 1RXS J124847.4-563525 \\
PF96 4     & 12485496-5949476 & M4e     & 12.7: & 0.07 & 110: & 29.8    & -2.9    & 3.517 & -0.92 &  3 & 0.3 & 1WGA J1248.9-5949 \\
PF96 6     & 12492437-5913112 & M1e     & 12.8: & 0.10 & 110: & 29.8    & -3.2    & 3.564 & -0.60 &  5 & 0.6 & 1WGA J1249.3-5913 \\
HIP 62657  & 12501971-4951488 & F5/F6V  &  8.91 & 0.09 & 107  & 29.7    & -4.3    & 3.806 &  0.43 & 19 & 1.3 & 1RXS J125020.5-495144 \\
SACY 797   & 12505143-5156353 & K5Ve    & 11.67 & 0.33 & 109  & 29.9    & -3.4    & 3.638 & -0.33 & 10 & 1.0 & 1RXS J125051.3-515655 \\
SACY 799   & 12560830-6926539 & K7Ve    & 11.80 & 0.62 & 118  & 30.8    & -2.7    & 3.609 & -0.05 &  2 & 0.8 & 1RXS J125608.8-692652 \\
SACY 800   & 12560940-6127256 & K0Ve    &  9.62 & 0.27 &  75  & 30.7    & -3.0    & 3.720 &  0.05 & 15 & 1.2 & 1RXS J125609.4-612724 \\
HIP 63272  & 12575777-5236546 & F3IV/V  &  8.40 & 0.00 & 113  & 29.6    & -4.6    & 3.829 &  0.63 & 15 & 1.4 & 1RXS J125757.2-523659 \\
MML 27     & 12582559-7028490 & K0+IV   &  9.91 & 0.46 &  85  & 30.3    & -3.3    & 3.714 & -0.01 & 16 & 1.1 & 1RXS J125824.6-702848 \\
MML 28     & 13015069-5304581 & K2-IV   & 11.08 & 0.55 & 108  & 30.0    & -3.3    & 3.690 & -0.37 & 32 & 0.9 & 1RXS J130153.7-530446 \\
MML 29     & 13023752-5459370 & G1IV    & 10.28 & 0.38 & 156  & 30.8    & -3.1    & 3.772 &  0.35 & 18 & 1.2 & 1RXS J130237.2-545933 \\
HIP 63847  & 13050530-6413552 & G3IV    &  9.18 & 0.33 &  98  & 30.4    & -3.5    & 3.764 &  0.36 & 15 & 1.3 & 2RXP J130506.7-641346 \\
HD 113703B & 13061785-4827456 & K0Ve    & 10.8: & 0.00 & 118  & 30.5    & -2.9    & 3.717 & -0.05 & 19 & 1.1 & 1RXS J130618.4-482744 \\
HIP 63975  & 13063577-4602018 & F3/F5V  &  8.08 & 0.00 & 111  & 29.4    & -4.8    & 3.819 &  0.70 & 12 & 1.4 & 1RXS J130630.7-460215 \\
MML 30     & 13064012-5159386 & K0IVe   & 10.53 & 0.17 & 112  & 30.2    & -3.4    & 3.722 & -0.05 & 21 & 1.1 & 1RXS J130638.5-515948 \\
SACY 808   & 13065439-4541313 & K5Ve    & 12.03 & 0.53 & 124  & 30.2    & -3.1    & 3.638 & -0.28 &  8 & 1.0 & 1RXS J130655.0-454125 \\
HD 113791B & 13065720-4954265 & F7V     & 10.90 & 0.03 & 145  & 29.9    & -3.5    & 3.798 &  0.49 & 17 & 1.3 & 1RXS J130659.7-495405 \\
HIP 64044  & 13073350-5254198 & F5V     &  8.83 & 0.27 & 111  & 30.3    & -3.8    & 3.809 &  0.57 & 16 & 1.4 & 1RXS J130733.2-525421 \\
SACY 814   & 13130714-4537438 & K5Ve    & 11.63 & 0.52 & 101  & 29.9    & -3.4    & 3.638 & -0.23 &  6 & 1.1 & 1RXS J131306.7-453740 \\
SACY 815   & 13132810-6000445 & G3V     & 10.01 & 0.20 & 157  & 30.6    & -3.4    & 3.766 &  0.40 & 14 & 1.3 & 1RXS J131327.3-600032 \\
MML 31     & 13142382-5054018 & G5.5IV  & 10.39 & 0.27 & 130  & 30.8    & -2.9    & 3.754 &  0.25 & 16 & 1.2 & 1RXS J131424.3-505402 \\
MML 32     & 13175694-5317562 & G1IV    & 10.41 & 0.22 & 167  & 30.7    & -3.2    & 3.774 &  0.43 & 15 & 1.3 & 1RXS J131754.9-531758 \\
TWA 17     & 13204539-4611377 & K7e     & 13.1: & 0.67 & 142  & 30.2    & -3.0    & 3.635 & -0.38 & 11 & 1.0 & 1RXS J132046.5-461139 \\
TWA 18     & 13213722-4421518 & M0.5e   & 13.7: & 0.16 & 119  & 30.0    & -2.9    & 3.574 & -0.63 &  6 & 0.6 & 1RXS J132137.0-442133 \\
MML 33     & 13220446-4503231 & G0IV    & 10.02 & 0.22 & 140  & 30.3    & -3.6    & 3.779 &  0.18 & 25 & 1.1 & 1RXS J132204.7-450312 \\
MML 34     & 13220753-6938121 & K1IVe   & 10.44 & 0.17 &  86  & 30.2    & -3.2    & 3.702 &  0.00 & 13 & 1.2 & 1RXS J132207.2-693812 \\
SACY 822   & 13233587-4718467 & K3Ve    & 11.19 & 1.23 & 111  & 30.0    & -3.6    & 3.733 &  0.06 & 19 & 1.1 & 1RXS J132336.3-471844 \\
HIP 65517  & 13254783-4814577 & G1.5IV  &  9.76 & 0.14 &  95  & 30.4    & -3.3    & 3.770 &  0.02 & 35 & 1.1 & 1RXS J132548.2-481451 \\
SACY 828   & 13270594-4856180 & K3Ve    & 10.68 & 0.51 & 104  & 30.4    & -3.2    & 3.675 &  0.01 &  7 & 1.3 & 1RXS J132706.3-485617 \\
HIP 66001  & 13315360-5113330 & G2.5IV  &  9.84 & 0.11 & 127  & 30.6    & -3.2    & 3.766 &  0.33 & 17 & 1.2 & 1RXS J133152.6-511335 \\
MML 35     & 13342026-5240360 & G1IV    &  9.29 & 0.45 & 105  & 30.5    & -3.4    & 3.774 &  0.32 & 19 & 1.2 & 1RXS J133420.0-524032 \\
SACY 835   & 13343188-4209305 & K2IVe   &  9.64 & 0.49 &  91  & 30.2    & -3.2    & 3.690 & -0.15 & 16 & 1.1 & 1RXS J133432.2-420929 \\
SACY 841   & 13402554-4633514 & K3Ve    & 11.37 & 0.68 & 143  & 30.8    & -2.9    & 3.675 &  0.08 &  5 & 1.4 & 1RXS J134025.6-463323 \\
HIP 66941  & 13430870-6907393 & G0.5IV  &  7.57 & 0.38 & 107  & 31.1    & -3.5    & 3.775 &  1.09 &  4 & 2.0 & 1RXS J134306.8-690754 \\
MML 37     & 13432853-5436434 & G2IV    &  9.32 & 0.28 &  85  & 30.6    & -3.2    & 3.769 &  0.13 & 26 & 1.1 & 1RXS J134332.7-543638 \\
SACY 848   & 13444279-6347495 & K4Ve    & 11.04 & 0.50 &  84  & 30.2    & -3.2    & 3.662 & -0.24 & 12 & 1.1 & 1RXS J134442.5-634758 \\
SACY 849   & 13455599-5222255 & K3Ve    & 11.34 & 0.49 & 121  & 30.4    & -3.1    & 3.675 & -0.13 & 11 & 1.2 & 1RXS J134555.2-522215 \\
SACY 857   & 13540743-6733449 & G6V     & 10.93 & 0.56 & 142  & 30.2    & -3.4    & 3.756 &  0.05 & 27 & 1.1 & 1RXS J135404.0-673334 \\
MML 42     & 14160567-6917359 & G1IV    & 10.14 & 0.62 & 123  & 30.1    & -3.7    & 3.773 &  0.19 & 24 & 1.1 & 1RXS J141605.3-691756 \\
\label{table_lcc_list}
\end{longtable}
}
\end{landscape}

\end{document}